\documentclass[a4paper,11pt,headings=big,DIV=12]{scrartcl}
\pdfoutput=1
\usepackage{amsmath,amssymb}
\usepackage{color}
\usepackage[dvipsnames, table]{xcolor}
\definecolor{blue}{rgb}{0,0,0.5}
\usepackage[colorlinks,linkcolor=blue,urlcolor=blue,citecolor=blue]{hyperref}
\usepackage{graphicx}
\graphicspath{{../plots/}}
\addtokomafont{disposition}{\rmfamily\boldmath}
\usepackage{cite}
\usepackage{xspace}
\usepackage{relsize}
\usepackage{bm}
\usepackage[utf8]{inputenc}
\usepackage{tabularx}
\usepackage{booktabs}
\usepackage{enumerate}
\usepackage{placeins}

\allowdisplaybreaks

\DeclareOldFontCommand{\rm}{\normalfont\rmfamily}{\mathrm}
\DeclareOldFontCommand{\sf}{\normalfont\sffamily}{\mathsf}
\DeclareOldFontCommand{\tt}{\normalfont\ttfamily}{\mathtt}
\DeclareOldFontCommand{\bf}{\normalfont\bfseries}{\mathbf}
\DeclareOldFontCommand{\it}{\normalfont\itshape}{\mathit}
\DeclareOldFontCommand{\sl}{\normalfont\slshape}{\@nomath\sl}
\DeclareOldFontCommand{\sc}{\normalfont\scshape}{\@nomath\sc}

\newcommand{\smelli}{\texttt{smelli}}
\newcommand{\flavio}{\texttt{flavio}}
\newcommand{\wilson}{\texttt{wilson}}

\begin{document}

\begin{flushright}
{\tt LAPTH-024/19}
\vspace{-0.2cm}
\end{flushright}

\begin{center}
\vspace*{1cm}
{\LARGE\bfseries \boldmath
$B$-decay discrepancies after Moriond 2019
}\\[0.8 cm]
{\textsc{
Jason Aebischer$^a$, Wolfgang Altmannshofer$^b$, Diego Guadagnoli$^c$,\\ M\'eril Reboud$^c$, Peter Stangl$^c$, David M.\ Straub$^a$
}\\[1 cm]
\small
$^a$ Excellence Cluster Universe, Boltzmannstr.~2, 85748~Garching, Germany \\
$^b$ Santa Cruz Institute for Particle Physics, University of California, Santa Cruz, CA~95064, USA \\
$^c$ Laboratoire d’Annecy-le-Vieux de Physique Th\'eorique, UMR5108, CNRS,\\ 9 Chemin de Bellevue, B.P. 110, F-74941, Annecy-le-Vieux Cedex, France
}
\\[1 cm]
\footnotesize
E-mail:
\texttt{jason.aebischer@tum.de},
\texttt{waltmann@ucsc.edu},
\texttt{diego.guadagnoli@lapth.cnrs.fr},
\texttt{meril.reboud@lapth.cnrs.fr},
\texttt{peter.stangl@lapth.cnrs.fr},
\texttt{david.straub@tum.de}
\\[1 cm]
\end{center}

\begin{abstract}\noindent

\noindent Following the updated measurement of the lepton flavour universality (LFU) ratio $R_K$ in $B\to K\ell\ell$ decays by LHCb, as well as a number of further measurements, e.g. $R_{K^*}$ by Belle and $B_s \to \mu \mu$ by ATLAS, we analyse the global status of new physics in $b\to s$ transitions in the weak effective theory at the $b$-quark scale, in the Standard Model effective theory above the electroweak scale, and in simplified models of new physics.
We find that the data continues to strongly prefer a solution with new physics in semi-leptonic Wilson coefficients. A purely muonic contribution to the combination $C_9 = -C_{10}$, well suited to UV-complete interpretations, is now favoured with respect to a muonic contribution to $C_9$ only. An even better fit is obtained by allowing an additional LFU shift in $C_9$. Such a shift can be renormalization-group induced from four-fermion operators above the electroweak scale, in particular from semi-tauonic operators, able to account for the potential discrepancies in $b \to c$ transitions. This scenario is naturally realized in the simplified $U_1$ leptoquark model. We also analyse simplified models where a LFU effect in $b\to s\ell\ell$ is induced radiatively from four-quark operators and show that such a setup is on the brink of exclusion by LHC di-jet resonance searches.
\end{abstract}

\section{Introduction}

In recent years, several deviations from Standard Model (SM) expectations
have been building up in $B$-decay measurements. While each of them could be
a first sign of physics beyond the SM, statistical fluctuations or underestimated
experimental or theoretical systematic uncertainties cannot be excluded at
present. These deviations -- or ``anomalies'' -- can be grouped into four
categories that have very different experimental and theoretical challenges:

\begin{enumerate}[(\em i)]
  \item \label{anom:br_bsmumu} Apparent suppression of various branching ratios of exclusive decays
  based on the $b\to s\mu\mu$ flavour-changing neutral current (FCNC) transition \cite{Aaij:2014pli,Aaij:2015esa}.
  The uncertainties are dominated by the limited knowledge of the $B$ to light
  meson hadronic form factors
  \cite{Straub:2015ica,Horgan:2015vla,Gubernari:2018wyi}.
  \item \label{anom:ang_bsmumu} Deviations from SM expectations in $B\to K^*\mu^+\mu^-$ angular observables \cite{Aaij:2015oid,ATLAS-CONF-2017-023,CMS-PAS-BPH-15-008,Khachatryan:2015isa}
  (also based on the $b\to s\mu\mu$ transition), where form factor uncertainties
  are much smaller than for the branching ratios, but hadronic uncertainties
  are nevertheless significant
  \cite{Khodjamirian:2010vf,Bobeth:2017vxj}.
  \item \label{anom:lfu_bsll} Apparent deviations from $\mu$-$e$ universality
  in $b\to s\ell\ell$ transitions in the processes $B\to K\ell\ell$ and
  $B\to K^*\ell\ell$ (via the $\mu/e$ ratios $R_K$ \cite{Aaij:2014ora} and $R_{K^*}$ \cite{Aaij:2017vbb}, respectively).
  Here the theoretical uncertainties are negligible \cite{Bordone:2016gaq} and
  the sensitivity is only limited by statistics at present.
  \item \label{anom:lfu_bclnu} Apparent deviations from $\tau$-$\mu$ and $\tau$-$e$ universality
  in $b\to c\ell\nu$ transitions \cite{Lees:2012xj,Lees:2013uzd,Huschle:2015rga,Sato:2016svk,Hirose:2016wfn,Aaij:2015yra,Aaij:2017uff}. Uncertainties are dominated by statistics,
  with non-negligible experimental systematics but small theoretical uncertainties \cite{Lattice:2015rga,Na:2015kha,Bernlochner:2017jka,Bigi:2017jbd}.
  (Note that $e$-$\mu$ universality
  in $b\to c\ell\nu$ transitions is tested to hold at the percent level \cite{Abdesselam:2017kjf,Abdesselam:2018nnh,Jung:2018lfu}.)
\end{enumerate}
While the deviations in {\em (\ref{anom:br_bsmumu})} and {\em (\ref{anom:ang_bsmumu})} could be alleviated by more conservative
assumptions on the hadronic uncertainties, it is tantalizing
that a simple description in terms of a
single non-standard Wilson coefficient of a semi-muonic operator like
$(\bar s \gamma^\rho P_Lb)(\bar \mu \gamma_\rho \mu)$
or
$(\bar s \gamma^\rho P_Lb)(\bar \mu \gamma_\rho P_L\mu)$
leads to a consistent description of {\em (\ref{anom:br_bsmumu})}, {\em (\ref{anom:ang_bsmumu})}, and {\em (\ref{anom:lfu_bsll})},
with a best-fit point that improves the fit to the data by more than five
standard deviations compared to the SM (for a single degree of freedom) \cite{Altmannshofer:2017fio,Altmannshofer:2017yso,Capdevila:2017bsm,Geng:2017svp,Ciuchini:2017mik,Hurth:2017hxg}.
Moreover, it was shown that simplified models with a single
tree-level mediator can not only explain {\em (\ref{anom:br_bsmumu})}, {\em (\ref{anom:ang_bsmumu})}, and {\em (\ref{anom:lfu_bsll})}, but even all four categories
of deviations simultaneously without violating any other existing constraints \cite{Alonso:2015sja,Greljo:2015mma,Barbieri:2015yvd,Calibbi:2015kma,Bauer:2015knc,Fajfer:2015ycq}.

Taken together, these observations explain the buzz of activity around these
deviations and the anticipation of improved measurements of the theoretically
clean ratios $R_K$ and $R_{K^*}$.
The purpose of this article is to examine the status of the tensions after inclusion of a number of updated or newly available measurements, in particular:

\begin{itemize}

\item The new measurement of $R_K$ by the LHCb collaboration combining Run-1 data with 2~fb$^{-1}$ of Run-2 data (corresponding to about one third of the full Run-2 data set). The updated measurement finds \cite{Aaij:2019wad}\footnote{%
In our numerical analysis, we use the full one-dimensional numerical likelihood provided in \cite{Aaij:2019wad},
which is markedly non-Gaussian, rather than symmetrizing the uncertainties in \eqref{eq:RKmeas}.
\label{fn:rk}}
\begin{equation}
 R_K = \frac{\text{BR}(B \to K \mu\mu)}{\text{BR}(B \to K ee)} = 0.846{\phantom{.}}^{+0.060}_{-0.054} {\phantom{.}}^{+0.016}_{-0.014} \,, \qquad \text{for}~ 1.1\,\text{GeV}^2 < q^2 < 6\,\text{GeV}^2 \,,
\label{eq:RKmeas}
\end{equation}
where the first uncertainty is statistical and the second systematic, and $q^2$ is the dilepton invariant mass squared. The SM predicts lepton flavour universality, i.e. $R_K^\text{SM}$ is unity with uncertainties that are well below the current experimental sensitivities.
While the updated experimental value is closer to the SM prediction than the Run-1 result~\cite{Aaij:2014ora}, the reduced experimental uncertainties imply a tension between theory and experiment at the level of $2.5\sigma$, which is comparable to the situation before the update.

\item The new, preliminary measurement of $R_{K^*}$ by Belle \cite{Abdesselam:2019wac}. Averaged over $B^\pm$ and $B^0$ decays, the measured $R_{K^*}$ values at low and high $q^2$ are
\begin{equation}
 R_{K^*} = \frac{\text{BR}(B \to K^* \mu\mu)}{\text{BR}(B \to K^* ee)} = \begin{cases} 0.90^{+0.27}_{-0.21}\pm0.10 \,, \qquad \text{for}~ 0.1\,\text{GeV}^2 < q^2 < 8\,\text{GeV}^2 \,, \\ 1.18^{+0.52}_{-0.32}\pm0.10 \,, \qquad \text{for}~ 15\,\text{GeV}^2 < q^2 < 19\,\text{GeV}^2 \,. \end{cases}
\end{equation}
Given their sizable uncertainties, these values are compatible with both the SM predictions ($R_{K^*}^\text{SM}$ approximately unity) and previous results on $R_{K^*}$ from LHCb \cite{Aaij:2017vbb}
\begin{equation}
 R_{K^*} = \frac{\text{BR}(B \to K^* \mu\mu)}{\text{BR}(B \to K^* ee)} = \begin{cases} 0.66^{+0.11}_{-0.07}\pm0.03 \,, \qquad \text{for}~ 0.045\,\text{GeV}^2 < q^2 < 1.1\,\text{GeV}^2 \,, \\ 0.69^{+0.11}_{-0.07}\pm0.05 \,, \qquad \text{for}~ 1.1\,\text{GeV}^2 < q^2 < 6\,\text{GeV}^2 \,, \end{cases}
\end{equation}
that are in tension with the SM predictions by $\sim 2.5\sigma$ in both $q^2$ bins.

\item One further, important piece of information included in our study is the 2018 measurement of $B_s \to \mu \mu$ by the ATLAS collaboration \cite{Aaboud:2018mst}, that we combine with the existing measurements by CMS and LHCb \cite{Chatrchyan:2013bka,CMS:2014xfa,Aaij:2017vad}.

\end{itemize}

\noindent In this paper we will explore the implications of all these, as well as other  data, to be described in fuller detail in the next section,
in the context of global fits to model-independent new physics scenarios, identify those that lead to a good description of the data, and discuss possible realizations in terms of simplified new-physics models.

Our numerical analysis is entirely based on open-source software,
notably the global likelihood in Wilson coefficient space provided
by the \smelli{} package \cite{Aebischer:2018iyb}, built on \flavio{} \cite{Straub:2018kue} and \wilson{} \cite{Aebischer:2018bkb}.
As such, our analysis is easily reproducible and modifiable.

The rest of this work is organized as follows.
\begin{itemize}
  \item In Section~\ref{sec:setup}, we describe our statistical approach
  and the experimental measurements we employ in our numerical analysis.
  \item In Section~\ref{sec:model-indep}, we perform a model-independent
  global analysis of $b\to s\ell\ell$ transitions, first in the weak effective
  theory (WET) below the electroweak (EW) scale, then in the SM
  effective field theory (SMEFT) above the EW scale, which allows us to extend
  the discussion to the charged-current deviations and to incorporate constraints
  from electroweak precision tests and other precision measurements.
  \item In Section~\ref{sec:models}, we discuss a number of specific
  simplified new-physics (NP) models that are favoured by the current data,
  assuming the deviations to be due to NP.
  \item Section~\ref{sec:concl} contains our conclusions.
\end{itemize}

\section{Setup}\label{sec:setup}

Our numerical analysis is based on a global likelihood function
in the space of the Wilson coefficients of the WET valid below the EW scale,
or the SMEFT valid above it. Theoretical uncertainties (for observables
where they cannot be neglected) are treated by computing a covariance
matrix of theoretical uncertainties within the SM and combining
it with the experimental uncertainties (approximated as Gaussian).
The main assumption in this approach is that the sizes of
theory uncertainties are weakly dependent on NP, which we checked for the
observables included. This approach was first applied to $b\to s\ell\ell$
transitions in \cite{Altmannshofer:2014rta}.
The theoretical uncertainties in exclusive $B$-decay observables stem mainly from hadronic form factors, which we take from \cite{Straub:2015ica} for $B$ to light vector meson transitions and from \cite{Gubernari:2018wyi} for $B\to K$,
as well as unknown non-factorizable effects that are parametrized as in \cite{Altmannshofer:2014rta,Straub:2015ica,Straub:2018kue} (and are compatible with more sophisticated approaches \cite{Khodjamirian:2010vf,Bobeth:2017vxj}).
Additional parametric uncertainties (e.g. from CKM matrix elements) are based on \flavio{} v1.3 with default settings \cite{Straub:2018kue}.
For more details on the statistical approach and the
list of observables and measurements included, we refer the reader to \cite{Aebischer:2018iyb}.

Here we highlight the changes in observables
sensitive to $b\to s$ transitions
included with respect
to the recent global analyses \cite{Altmannshofer:2017fio,Altmannshofer:2017yso} by some of us.
\begin{itemize}
  \item We include the LHCb update of $R_K$ \cite{Aaij:2019wad}
  (cf. footnote~\ref{fn:rk}) and the new, preliminary measurement of $R_{K^*}$ by Belle \cite{Abdesselam:2019wac}.
  The Belle results are available for various $q^2$-bin choices, separately for $B^\pm$ and $B^0$ decays and in an isospin averaged form.
  In our numerical analysis we use the $0.1\,$GeV$^2 < q^2 < 8\,$GeV$^2$ and $15\,$GeV$^2 < q^2 < 19\,$GeV$^2$ bins, separately for $B^\pm$ and $B^0$ decays.
  \item We include the new ATLAS measurement of $B_s\to\mu^+\mu^-$ \cite{Aaboud:2018mst},
  that we combine with the CMS and LHCb measurements~\cite{Chatrchyan:2013bka,CMS:2014xfa,Aaij:2017vad}. This combination is
  discussed in detail in appendix~\ref{app:bsmumu}. Our combination is in slight tension with the SM prediction of BR$(B_s \to \mu^+\mu^-)$ by approximately $2\sigma$.
  \item We include the updated LHCb measurement of forward-backward
  asymmetries in $\Lambda_b\to \Lambda\ell^+\ell^-$ \cite{Aaij:2018gwm}
  as well as its branching ratio \cite{Aaij:2015xza}. For the theory predictions of the baryonic decay we follow \cite{Detmold:2016pkz,Meinel:2016grj}.
  \item Here we are working with the global likelihood described in~\cite{Aebischer:2018iyb} (i.e.\ including as many
  observables sensitive to the Wilson coefficients as possible), while in
  \cite{Altmannshofer:2017fio} we focused on observables sensitive to the $b\to s\ell\ell$
  transition only. This means e.g.\ that we also include
  all the observables sensitive to the $b\to s\gamma,g$ dipole transitions
  studied in \cite{Paul:2016urs}.
  In addition, the global likelihood also includes observables that do not directly depend on the Wilson coefficients of interest but whose theory uncertainties are strongly correlated with those of the directly dependent observables.
  This is in particular relevant for the $b\to s\mu\mu$ observables.
  In our figures, we indicate the set of observables consisting of $b\to s\mu\mu$, $b\to s\gamma,g$, and other correlated observables as ``$b\to s\mu\mu$ \& corr. obs.''.

  Like in the previous analysis~\cite{Altmannshofer:2017yso} by some of us, we again include the LFU differences of angular observables\footnote{%
$D_{P_{4,5}^\prime} = P_{4,5}^\prime(B \to K^* \mu^+\mu^-) - P_{4,5}^\prime(B \to K^* e^+e^-)$.
  The observables
$P_{4,5}^\prime$ are defined in~\cite{Descotes-Genon:2013vna}.} $D_{P_{4}^\prime}$ and $D_{P_{5}^\prime}$.
To this end, we have added them to the global likelihood in version 1.3.0 of the \texttt{smelli} package.
\end{itemize}

\section{Model-independent numerical analysis}\label{sec:model-indep}

Having at hand the global likelihood in the space of NP Wilson coefficients, $L(\vec{C})$, we perform a
model-independent numerical analysis by studying it in simple one- and two-coefficient scenarios.
This analysis proceeds in two steps:
\begin{enumerate}
  \item We first investigate the Wilson coefficients of the weak effective theory at the $b$-quark mass scale.
  This analysis can be seen as an update of earlier analyses (see e.g.~\cite{Altmannshofer:2017fio,Altmannshofer:2017yso,Capdevila:2017bsm,Geng:2017svp,Ciuchini:2017mik,Hurth:2017hxg})
  and is completely general, barring new particles lighter than the $b$ quark (see e.g.~\cite{Sala:2017ihs,Ghosh:2017ber,Datta:2017ezo,Altmannshofer:2017bsz}).
  \item Next, we embed these results into the SMEFT at a scale $\Lambda$ above the electroweak scale. This is based on the additional assumptions that there are no new particles beneath $\Lambda$ and that EW symmetry breaking is approximately linear (see e.g.\ \cite{Cata:2015lta}).
  This allows us to correlate NP effects in $b\to s\ell\ell$ model-independently with other sectors like
  EW precision tests or $b\to c$ transitions
  (cf.~\cite{Celis:2017doq,Camargo-Molina:2018cwu,Aebischer:2018iyb,Hurth:2019ula}).
\end{enumerate}

\subsection{$b\to s\ell\ell$ observables in the WET}\label{sec:wet}

We start by investigating the constraints on NP contributions
to the $|\Delta B|=|\Delta S|=1$
Wilson coefficients of the WET at the $b$-quark scale $\mu_b\approx m_b$
that we take to be $4.8\,\text{GeV}$.
We work with the effective Hamiltonian
\begin{equation}
\mathcal H_\text{eff}^{bs\ell\ell}
= \mathcal H_\text{eff, SM}^{bs\ell\ell}
+ \mathcal H_\text{eff, NP}^{bs\ell\ell} \,,
\end{equation}
where the first term contains the SM contributions to the Wilson coefficients.
The second term reads
\begin{equation}
  \mathcal H_\text{eff, NP}^{bs\ell\ell}
  = - \mathcal{N} \bigg(
  C^{bs}_7 O^{bs}_7 + C^{\prime bs}_7 O^{\prime bs}_7
  \\
  +
  \sum_{\ell=e,\mu}
  \sum_{i=9,10,S,P}
  \left(C^{bs\ell\ell}_i O^{bs\ell\ell}_i + C^{\prime bs\ell\ell}_i O^{\prime bs\ell\ell}_i\right)
  \bigg)
  +
   \text{h.c.}\,,
\end{equation}
with the normalization factor
\begin{equation}
  \mathcal{N} = \frac{4G_F}{\sqrt{2}} V_{tb}V_{ts}^* \frac{e^2}{16\pi^2} \,.
  \label{eq:norm}
\end{equation}
The dipole operators are given by\footnote{The sign of the dipole coefficients $C_7^{(\prime)}$ are fixed by our convention for the covariant derivative $D_\mu\psi=\partial_\mu+ieQ_\psi A_\mu+ig_sT^AG^A_\mu$.}
\begin{align}
O^{bs}_7 &= \frac{m_b}{e} (\bar{s} \sigma_{\mu\nu} P_{R} b) F^{\mu\nu}\,,
&
O^{\prime bs}_7 &= \frac{m_b}{e} (\bar{s} \sigma_{\mu\nu} P_{L} b) F^{\mu\nu}\,,
\end{align}
where $\sigma^{\mu\nu}=\frac{i}{2}[\gamma^\mu,\gamma^\nu]$,
and the semi-leptonic operators
\begin{align}
O_9^{bs\ell\ell} &=
(\bar{s} \gamma_{\mu} P_{L} b)(\bar{\ell} \gamma^\mu \ell)\,,
&
O_9^{\prime bs\ell\ell} &=
(\bar{s} \gamma_{\mu} P_{R} b)(\bar{\ell} \gamma^\mu \ell)\,,\label{eq:O9}
\\
O_{10}^{bs\ell\ell} &=
(\bar{s} \gamma_{\mu} P_{L} b)( \bar{\ell} \gamma^\mu \gamma_5 \ell)\,,
&
O_{10}^{\prime bs\ell\ell} &=
(\bar{s} \gamma_{\mu} P_{R} b)( \bar{\ell} \gamma^\mu \gamma_5 \ell)\,,\label{eq:O10}
\\
O_{S}^{bs\ell\ell} &= m_b
(\bar{s} P_{R} b)( \bar{\ell}  \ell)\,,
&
O_{S}^{\prime bs\ell\ell} &= m_b
(\bar{s}  P_{L} b)( \bar{\ell}  \ell)\,,\label{eq:OS}
\\
O_{P}^{bs\ell\ell} &= m_b
(\bar{s} P_{R} b)( \bar{\ell} \gamma_5 \ell)\,,
&
O_{P}^{\prime bs\ell\ell} &= m_b
(\bar{s}  P_{L} b)( \bar{\ell} \gamma_5 \ell)\,.\label{eq:OP}
\end{align}
We have omitted from $\mathcal H_\text{eff, NP}^{bs\ell\ell}$
semi-leptonic tensor operators, which are not generated
at dimension 6 in
theories that have SMEFT as EW-scale limit,
as well as chromomagnetic and four-quark operators.
Even though the latter can contribute via one-loop matrix elements to
$b\to s\ell\ell$ processes, their dominant effects typically stem from
renormalization
group (RG) evolution above the scale $\mu_b$, and we will discuss these effects
in the SMEFT framework in the next section.
For the same reason, we have constrained the sum over lepton flavours to $e$
and $\mu$: semi-tauonic WET operators can contribute via QED RG mixing, but their
direct matrix elements are subleading \cite{Bobeth:2011st}.

\subsubsection{Scenarios with a single Wilson coefficient}\label{sec:1d}

\begin{table}[tbp]
\centering
\renewcommand{\arraystretch}{1.5}
\rowcolors{2}{gray!15}{white}
\addtolength{\tabcolsep}{4pt} 
\begin{tabularx}{\textwidth}{ccccX}
\toprule
\rowcolor{white}
Coeff.  & best fit & $1\sigma$ & $2\sigma$ & pull\\
\hline
$C_9^{bs\mu\mu}                 $ & $-0.97$ & [$-1.12$, $-0.81$] & [$-1.27$, $-0.65$] & $5.9\sigma$ \\
$C_9^{\prime bs\mu\mu}          $ & $+0.14$ & [$-0.03$, $+0.32$] & [$-0.20$, $+0.51$] & $0.8\sigma$ \\
$C_{10}^{bs\mu\mu}              $ & $+0.75$ & [$+0.62$, $+0.89$] & [$+0.48$, $+1.03$] & $5.7\sigma$ \\
$C_{10}^{\prime bs\mu\mu}       $ & $-0.24$ & [$-0.36$, $-0.12$] & [$-0.49$, $+0.00$] & $2.0\sigma$ \\
$C_9^{bs\mu\mu}=C_{10}^{bs\mu\mu}$ & $+0.20$ & [$+0.06$, $+0.36$] & [$-0.09$, $+0.52$] & $1.4\sigma$ \\
$C_9^{bs\mu\mu}=-C_{10}^{bs\mu\mu}$ & $-0.53$ & [$-0.61$, $-0.45$] & [$-0.69$, $-0.37$] & $6.6\sigma$ \\
\hline
$C_9^{bsee}                     $ & $+0.93$ & [$+0.66$, $+1.17$] & [$+0.40$, $+1.42$] & $3.5\sigma$ \\
$C_9^{\prime bsee}              $ & $+0.39$ & [$+0.05$, $+0.65$] & [$-0.27$, $+0.95$] & $1.2\sigma$ \\
$C_{10}^{bsee}                  $ & $-0.83$ & [$-1.05$, $-0.60$] & [$-1.28$, $-0.37$] & $3.6\sigma$ \\
$C_{10}^{\prime bsee}           $ & $-0.27$ & [$-0.57$, $-0.02$] & [$-0.84$, $+0.26$] & $1.1\sigma$ \\
$C_9^{bsee}=C_{10}^{bsee}       $ & $-1.49$ & [$-1.79$, $-1.18$] & [$-2.05$, $-0.79$] & $3.2\sigma$ \\
$C_9^{bsee}=-C_{10}^{bsee}      $ & $+0.47$ & [$+0.33$, $+0.59$] & [$+0.20$, $+0.73$] & $3.5\sigma$ \\
\hline
$\left(C_S^{bs\mu\mu}=-C_P^{bs\mu\mu}\right)\times\text{GeV}$ & $-0.006$ & [$-0.009$, $-0.003$] & [$-0.014$, $-0.001$] & $2.8\sigma$ \\
$\left(C_S^{\prime bs\mu\mu}=C_P^{\prime bs\mu\mu}\right)\times\text{GeV}$ & $-0.006$ & [$-0.009$, $-0.003$] & [$-0.014$, $-0.001$] & $2.8\sigma$ \\

\bottomrule
\end{tabularx}
\addtolength{\tabcolsep}{-4pt} 
\caption{Best-fit values,  1 and $2\sigma$ ranges, and pulls (cf.~Eq.~(\ref{fn:pull})) between the best-fit point and the
SM point for scenarios with NP in a single Wilson coefficient
(or Wilson coefficient combination).
For the scalar Wilson coefficients, we show the SM-like solution, while also a sign-flipped solution is allowed, see \cite{Altmannshofer:2017wqy}.
}
\label{tab:1d}
\end{table}

We now consider the global likelihood in the space of the above Wilson coefficients.
We start with scenarios where only a single NP Wilson coefficient
(or a single linear combination motivated by UV scenarios) is nonzero.
The best-fit values, 1 and $2\sigma$ ranges, and pulls
for several such scenarios are listed in
Table~\ref{tab:1d}.
For the 1D scenarios, the pull in $\sigma$ is defined as
\begin{equation}
 \label{fn:pull}
\text{pull} = \sqrt{\Delta\chi^2}\,, \qquad \text{where} ~ -\frac{1}{2}\Delta\chi^2=\ln L(\vec 0) - \ln L(\vec C_\text{best fit}) \,.
\end{equation}
We make the following observations.
\begin{itemize}
  \item Like in previous analyses, two scenarios stand out, namely a shift to $C_9^{bs\mu\mu}$ by approximately $- 25\%$ of its SM value ($C_9^\text{SM}(\mu_b) \simeq 4.1$), or a shift to the combination $C_9^{bs\mu\mu} = -C_{10}^{bs\mu\mu}$ by approximately $- 15\%$ of its SM value. However, at variance with previous analyses, it is the second scenario, rather than the first one, to have the largest pull. Given our assumptions about hadronic uncertainties, the pull exceeds $6\sigma$.
  The main reason why the combination $C_9^{bs\mu\mu} = -C_{10}^{bs\mu\mu}$ performs better is the $\sim2\sigma$ tension in BR$(B_s \to \mu\mu)$, which remains unresolved in the $C_9^{bs\mu\mu}$ scenario.
  We will comment further on this finding in appendix \ref{app:nonzero_C10}.
  \item New physics in $C_{10}^{bs\mu\mu}$ alone also improves the agreement between theory and data considerably. However, tensions in $B \to K^* \mu\mu$ angular observables remain in this scenario.
  \item Muonic scenarios with right-handed currents on the quark side, $C_9^{\prime bs\mu\mu}$ and $C_{10}^{\prime bs\mu\mu}$, or the lepton side, $C_9^{bs\mu\mu} = C_{10}^{bs\mu\mu}$, do not lead to a good description of the data.
  \item Scenarios with NP in $bsee$ Wilson coefficients only, while able to accommodate the discrepancies in $R_{K^{(*)}}$, do not help for the rest of the data. Given the pulls in Table \ref{tab:1d}, such scenarios are, all in all, less convincing.
\end{itemize}

The scalar Wilson coefficients $C_{S/P}^{bs\mu\mu}$ and $C_{S/P}^{\prime bs\mu\mu}$ are strongly constrained by the $B_s \to \mu\mu$ decay.
In theories that have SMEFT as their EW-scale limit,
they satisfy the constraint~\cite{Alonso:2014csa}\footnote{Discussions of the case where the relations (\ref{eq:CSCP}) are violated can be found in \cite{Cata:2015lta,Arbey:2018ics}. For a detailed numerical study, including also tensor operators, see Ref. \cite{Beaujean:2015gba}.
}
\begin{align} \label{eq:CSCP}
  C_S^{bs\mu\mu} &= - C_P^{bs\mu\mu} \,,
&
  C_S^{\prime bs\mu\mu} &= C_P^{\prime bs\mu\mu} \,.
\end{align}
In this case, they cannot lead to significant modifications in semi-leptonic $b \to s \mu \mu$ transitions \cite{Altmannshofer:2017wqy}.
However, the preference of the combination discussed in appendix~\ref{app:bsmumu} for a suppressed $B_s\to\mu^+\mu^-$ branching ratio
means that a destructive interference of these Wilson coefficients with the SM contribution to the leptonic decay can lead to a moderate improvement of the likelihood.

\subsubsection{Scenarios with a pair of Wilson coefficients}\label{sec:2d}

\begin{figure}[tbp]
\centering
\includegraphics[width=0.5\textwidth]{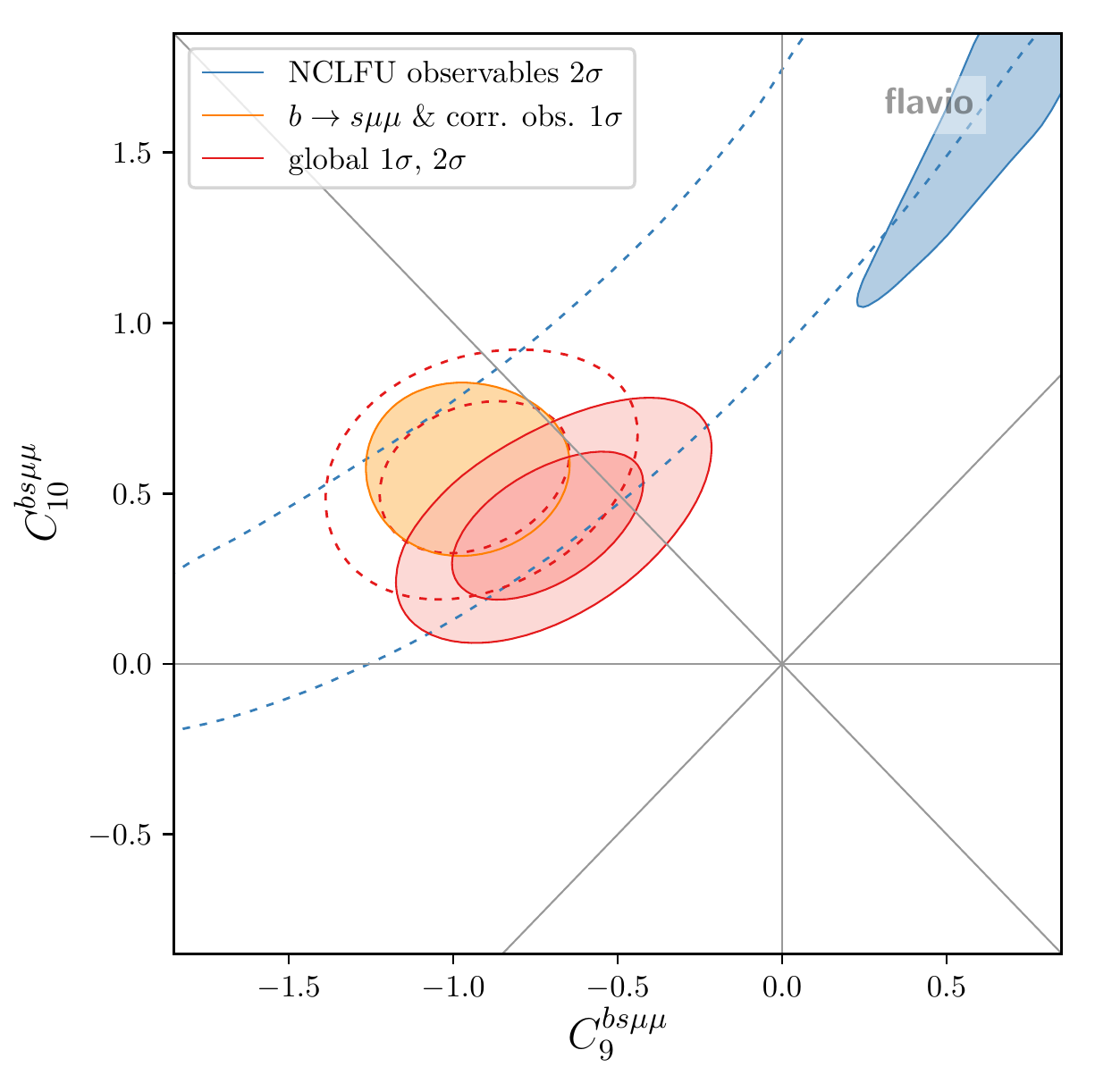}%
\includegraphics[width=0.5\textwidth]{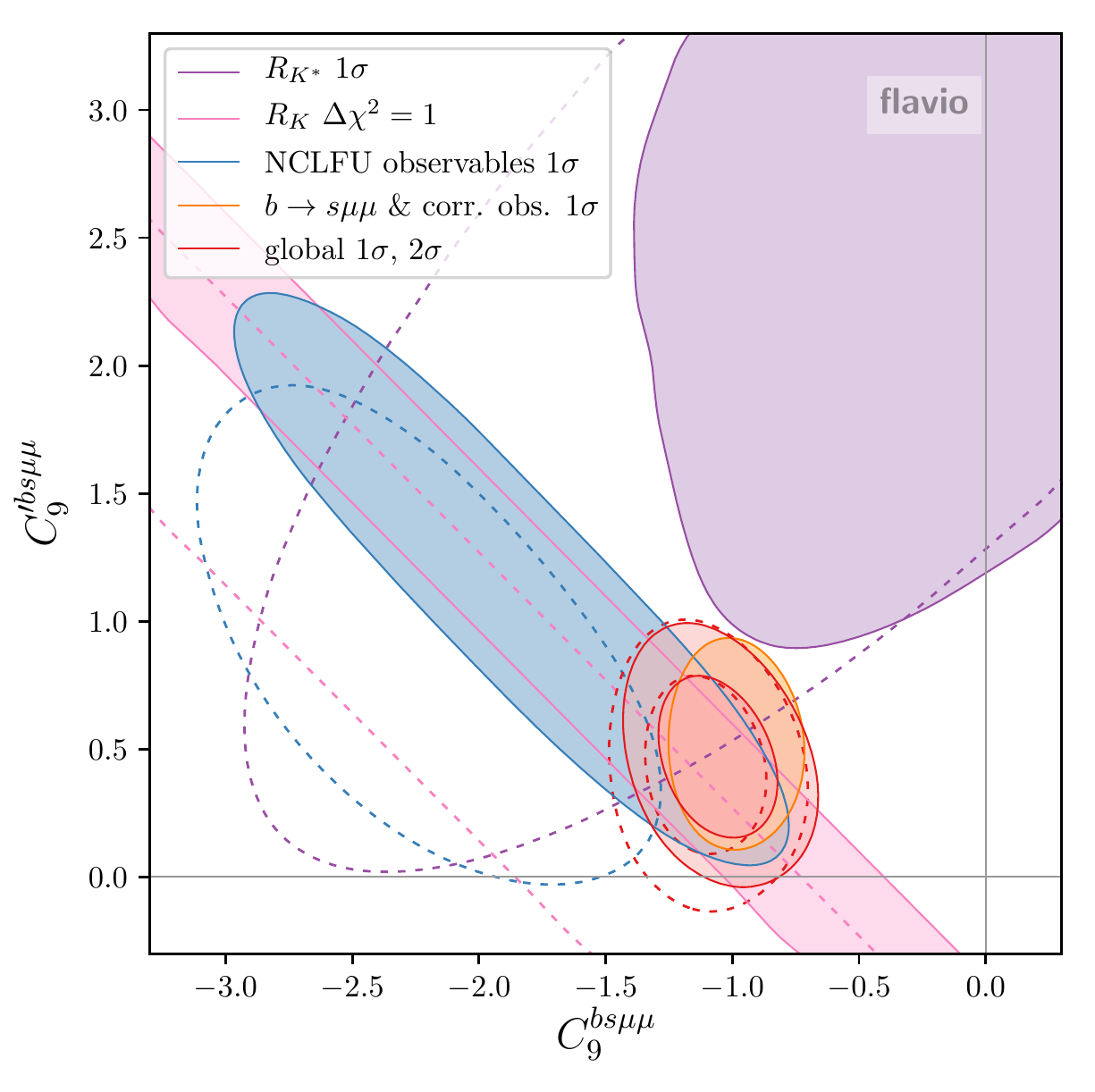}%
\caption{Likelihood contours of the global fit and several fits to subsets of observables (see text for details) in the plane of the WET Wilson coefficients $C_9^{bs\mu\mu}$ and $C_{10}^{bs\mu\mu}$ (left), and $C_9^{bs\mu\mu}$ and $C_{9}^{\prime bs\mu\mu}$ (right). Solid (dashed) contours include (exclude) the Moriond-2019 results for $R_K$ and $R_{K^*}$.
As $R_{K}$ only constrains a single combination of Wilson coefficients in the right plot, its 1$\sigma$~contour corresponds to $\Delta \chi^2 = 1$.
  For the other fits, 1 and 2$\sigma$~contours correspond to $\Delta \chi^2 \approx 2.3$ and $6.2$, respectively.}
\label{fig:C9C10}
\end{figure}

Next, we consider the likelihood in the space of pairs of Wilson coefficients. The results in Table~\ref{tab:1d} suggest that NP in both $C_9^{bs\mu\mu}$ and $C_{10}^{bs\mu\mu}$ ought to give an excellent fit to the data. The left plot of Fig.~\ref{fig:C9C10} shows the best fit regions in the $C_9^{bs\mu\mu}$\,-\,$C_{10}^{bs\mu\mu}$ plane.
The orange regions correspond to the $1\sigma$ constraints from $b \to s \mu\mu$ observables (including $B_s\to\mu^+\mu^-$) and observables whose uncertainties are correlated with those of the $b \to s \mu\mu$ observables (cf.\ last point in Sec.~\ref{sec:setup}).
In blue we show regions corresponding to the $1\sigma$ (right plot) and $2\sigma$ (left plot) constraints from the {\em neutral-current LFU} (NCLFU) observables $R_K$, $R_{K^*}$, $D_{P^\prime_{4}}$, and $D_{P^\prime_{5}}$.
In the right plot, the $1\sigma$ constraints from only $R_K$ (purple) and only $R_{K^*}$ (pink) are shown.
The combined $1$ and $2\sigma$ region is shown in red. The dotted contours indicate the situation without the Moriond-2019 results for $R_K$ and $R_{K^*}$.
The best fit point $C_9^{bs\mu\mu} \simeq -0.73$ and $C_{10}^{bs\mu\mu} \simeq 0.40$ has a $\sqrt{\Delta \chi^2} = 6.6$, which, corrected for the two degrees of freedom, corresponds to a pull of $6.3\sigma$.
In this scenario a slight tension between $R_K$ and $R_{K^*}$ remains, as it predicts $R_K \simeq R_{K^*}$ while the data seems to indicate $R_K > R_{K^*}$.
In addition, there is also a slight tension between the fit to NCLFU observables and the fit to $b \to s \mu\mu$ ones, especially in the $C_9^{bs\mu\mu}$ direction.

Overall, we find a similarly good fit of the data in a scenario with NP in $C_9^{bs\mu\mu}$ and $C_{9}^{\prime bs\mu\mu}$. The scenario is shown in the right plot of Fig.~\ref{fig:C9C10}. The best fit values for the Wilson coefficients are $C_9^{bs\mu\mu} \simeq -1.06$ and $C_{9}^{\prime bs\mu\mu} \simeq 0.47$. The $\sqrt{\Delta \chi^2} = 6.4$ corresponds to a pull of $6.0\sigma$.
Interestingly, in this scenario a non-zero $C_{9}^{\prime bs\mu\mu}$ is preferred at the $2\sigma$ level. The right-handed quark current allows one to accommodate the current experimental results for the LFU ratios, $R_K > R_{K^*}$. This scenario cannot address the tension in BR$(B_s \to \mu^+\mu^-)$. It predicts BR$(B_s \to \mu^+\mu^-) =$ BR$(B_s \to \mu^+\mu^-)_\text{SM}$.

Other two-coefficient scenarios (including dipole coefficients, scalar coefficients, and electron specific semileptonic coefficients) are discussed in appendix \ref{app:2d}.

\subsubsection{Universal vs. non-universal Wilson coefficients}\label{sec:universal}

\begin{figure}[tbp]
\centering
\includegraphics[width=0.5\textwidth]{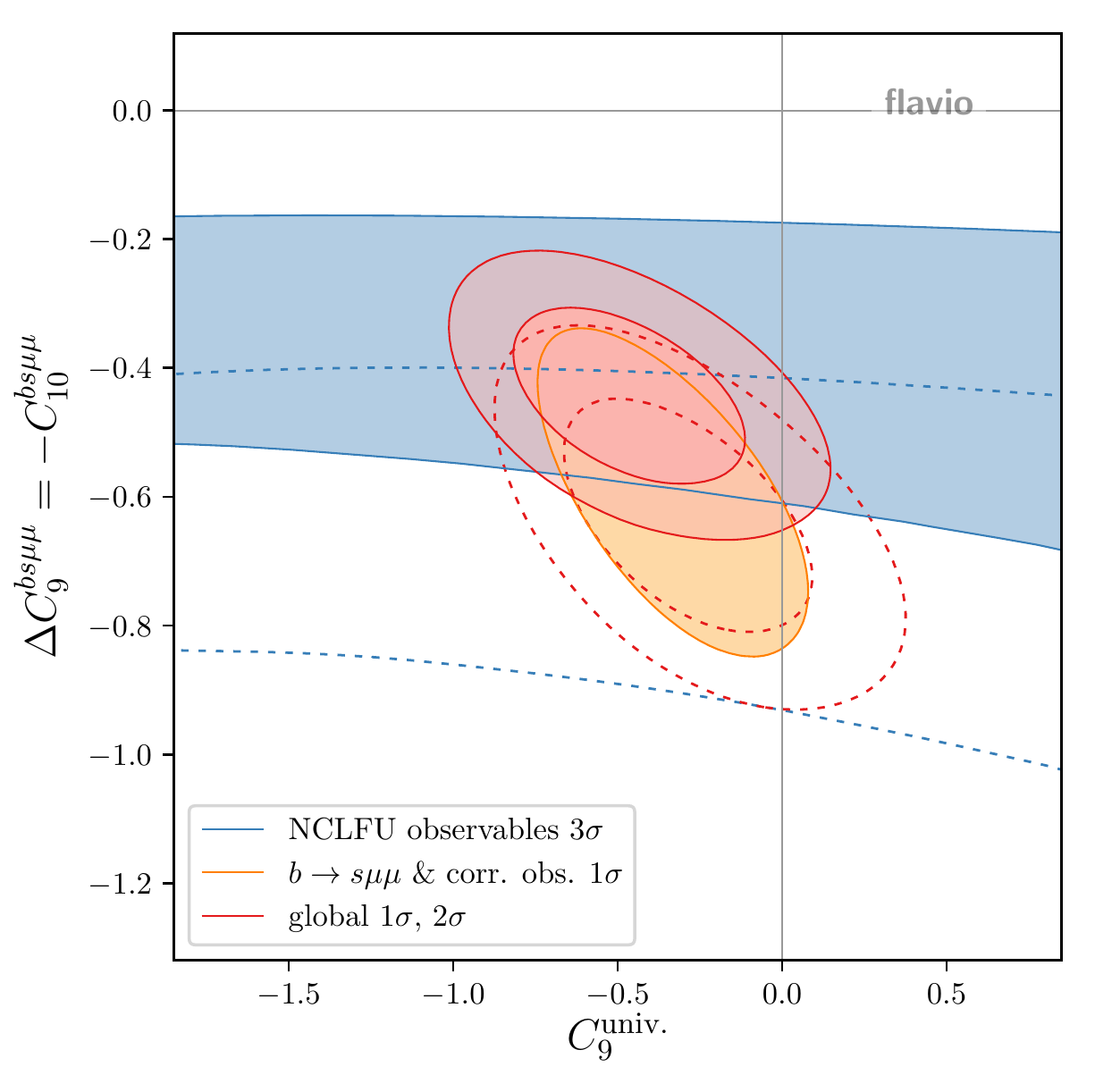}%
\caption{Likelihood contours from NCLFU observables ($R_{K^{(*)}}$ and $D_{P^\prime_{4,5}}$), $b\to s\mu\mu$ observables, and the global fit in the plane of a lepton flavour universal contribution to $C_9^{\rm univ.}$ $\equiv C_9^{bs\ell \ell}, \forall \ell,$ and a muon specific contribution to the linear combination $C_9=-C_{10}$ (see text for details). Solid (dashed) contours include (exclude) the Moriond-2019 results for $R_K$ and $R_{K^*}$.}
\label{fig:C9u}
\end{figure}

In view of the updated $R_{K^{(*)}}$ measurements, which are closer to the SM prediction than the Run-1 results, our fit in $C_9^{bs\mu\mu}$ and $C_{10}^{bs\mu\mu}$ shows a tension between the fit to NCLFU observables and the fit to $b \to s \mu\mu$ ones, especially in the $C_9^{bs\mu\mu}$ direction.
Therefore, it is interesting to investigate whether lepton flavour universal new physics that mostly affects $b \to s \mu\mu$ observables but none of the NCLFU observables is preferred by the global analysis.
In Fig.~\ref{fig:C9u}
we show the likelihood in the space of a LFU contribution to $C_9$
vs.\ a purely muonic contribution to the linear combination $C_9=-C_{10}$, i.e. we consider a two-parameter scenario where the total NP Wilson
coefficients are given by\footnote{Such decomposition was adopted for the first time in \cite{Alguero:2018nvb}.}
\begin{align}
C_9^{bs\mu\mu} &= \Delta C_9^{bs\mu\mu} + C_9^\text{univ.}
\,,\\
C_9^{bsee} &= C_9^{bs\tau\tau} = C_9^\text{univ.}
\,,\\
C_{10}^{bs\mu\mu} &= -\Delta C_{9}^{bs\mu\mu}
\,,\\
C_{10}^{bsee} &= C_{10}^{bs\tau\tau} = 0
\,.
\end{align}
The best fit values in this scenario are $C_9^{\rm univ.} = -0.49$ and $\Delta C_9^{bs\mu\mu} = -0.44$ with a $\sqrt{\Delta \chi^2} = 6.8$ that corresponds to a pull of $6.5\sigma$. The updated values of $R_{K^{(*)}}$ favour a non zero lepton flavour universal contribution to $C_9$ in this scenario.

In the following Sections \ref{sec:SMEFT}, \ref{sec:semitauonic}, and \ref{sec:fourquark} we will discuss how such a lepton-flavour universal
NP effect in $C_9$ can arise from RG effects.

\subsection{The global picture in the SMEFT}\label{sec:SMEFT}

We next discuss the interpretation of the above results within the SMEFT.
In contrast to the discussion in WET at the $b$-quark scale, more Wilson coefficients
become relevant in SMEFT due to RG mixing above \cite{Jenkins:2013zja,Jenkins:2013wua,Alonso:2013hga} and below \cite{Aebischer:2017gaw,Jenkins:2017dyc} the EW scale.
Due to the pattern of Wilson coefficients preferred by the global fit, we focus
on SMEFT Wilson coefficients that either contribute to the
semimuonic Wilson coefficients in the form $C_9^{bs\mu\mu}=-C_{10}^{bs\mu\mu}$
or induce a LFU effect in $C_9^{bs\ell\ell}$.
Our notation in the following will be such that $\ell$ refers to leptons below the EW scale and $l$ to the lepton doublets above the EW scale.
Furthermore, we will work in a basis where generation indices for RH quarks are taken to coincide with the mass basis \cite{Aebischer:2017ugx}, which can be done without loss of generality.

The direct matching contributions to $C_{9,10}$ at the EW scale are well known \cite{DAmbrosio:2002vsn,Buras:2014fpa},
\begin{align}\label{eq:matching1}
2\mathcal{N} \, C_{9}^{bs\ell_i\ell_i} &=
[C_{qe}]_{23ii}
+
[C_{lq}^{(1)}]_{ii23}
+
[C_{lq}^{(3)}]_{ii23}
- \zeta c_Z
\,,\\ \label{eq:matching2}
2\mathcal{N} \, C_{10}^{bs\ell_i\ell_i} &=
[C_{qe}]_{23ii}
-
[C_{lq}^{(1)}]_{ii23}
-
[C_{lq}^{(3)}]_{ii23}
+ c_Z
\,,
\end{align}
where the normalization factor $\mathcal{N}$ is defined in~(\ref{eq:norm}), the $Z$ penguin coefficient $c_Z$ is
\begin{equation}
c_Z = [C_{\varphi q}^{(1)}]_{23}+[C_{\varphi q}^{(3)}]_{23}\,,
\end{equation}
and $\zeta=1-4s_w^2\approx 0.08$ is the accidentally suppressed vector coupling of the $Z$ to charged leptons.
Using the notation of~\cite{Grzadkowski:2010es}, the corresponding operators are given by
\begin{align} \label{eq:SMEFTops1}
 &[O_{qe}]_{23ii} = (\bar q_2 \gamma_\mu q_3)(\bar e_i \gamma^\mu e_i) \,,  \\ \label{eq:SMEFTops2}
 &[O_{lq}^{(1)}]_{ii23} = (\bar l_i \gamma_\mu l_i)(\bar q_2 \gamma^\mu q_3)  \,, &[O_{lq}^{(3)}]_{ii23} = (\bar l_i \gamma_\mu \tau^I l_i)(\bar q_2 \gamma^\mu \tau^I  q_3) \,, \\
 &[O_{\varphi q}^{(1)}]_{23} = (\varphi^\dagger i\overset\leftrightarrow D_\mu \varphi)(\bar q_2 \gamma^\mu q_3)  \,, &[O_{\varphi q}^{(3)}]_{23} = (\varphi^\dagger i\overset\leftrightarrow {D}{}_\mu^I \varphi)(\bar q_2 \gamma^\mu \tau^I q_3)\,,
\end{align}
where $q_i$, and $l_i$ are the left-handed $SU(2)_L$ doublet quarks and leptons and $e_i$ are the right-handed lepton singlets, $\varphi$ is the SM Higgs doublet, and $\tau^I$ are the Pauli matrices.

The equations (\ref{eq:matching1}) and (\ref{eq:matching2}) highlight the well-known fact that a LFU contribution
to $C_{9,10}$ induced by the SMEFT coefficients $[C_{\varphi q}^{(1,3)}]_{23}$
(yielding a flavour-changing $\bar sbZ$ coupling) is not preferred by the data since
it leads to $|C_{10}^{bs\ell_i\ell_i}|\gg |C_{9}^{bs\ell_i\ell_i}|$.
Likewise, the coefficient $[C_{qe}]_{ii23}$ alone leads to
$C_{9}^{bs\ell_i\ell_i}=C_{10}^{bs\ell_i\ell_i}$ that is in poor agreement with the data as well.
Thus, if the dominant NP effect in $C_{9,10}^{bs\mu\mu}$ does \emph{not} stem
from an RG effect but a direct matching contribution, it must involve one of the
SMEFT Wilson coefficients $[C_{lq}^{(1,3)}]_{2223}$.

\begin{figure}
\centering
\includegraphics[width=0.45\textwidth]{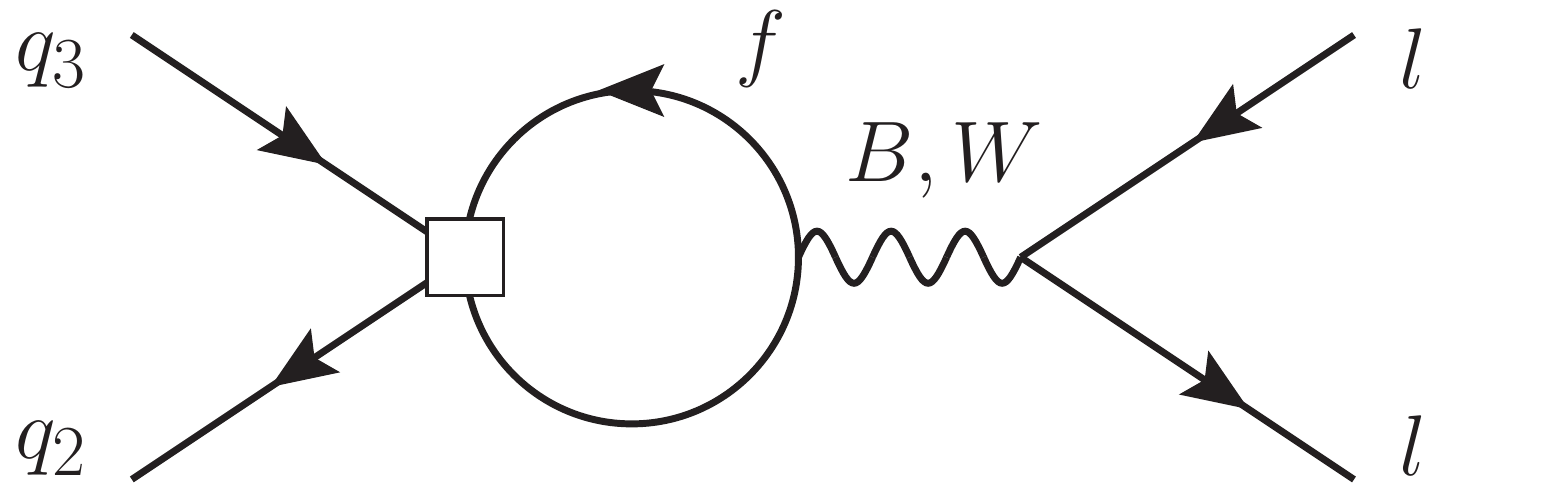}
\hspace{1cm}
\includegraphics[width=0.45\textwidth]{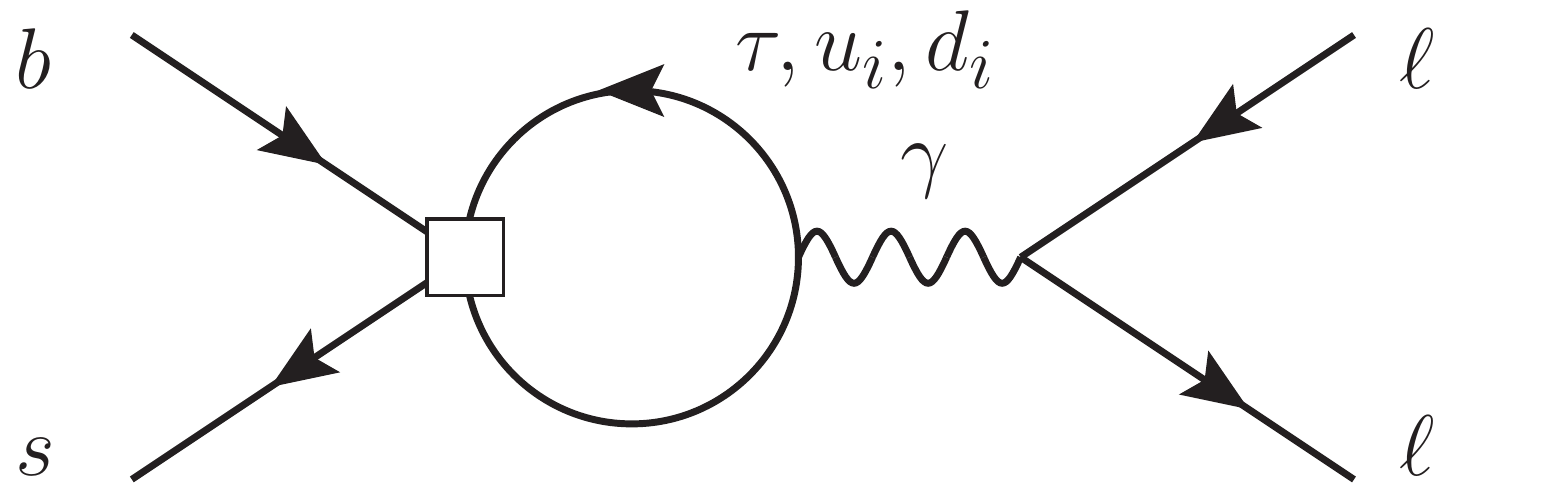}
\caption{Diagrams inducing a contribution to $C_9$ through RG running above (left panel) and below (right panel) the EW scale. A sizeable contribution to $C_9$ is obtained when $f = u_{1,2}, d_{1,2,3}$ or $l_3$, see text for details.}
\label{fig:RGE}
\end{figure}

Apart from the direct matching contributions, additional SMEFT Wilson coefficients can induce a contribution to $C_9$ at the $b$ mass scale through RG evolution above or below the EW scale, as pictured in Fig.~\ref{fig:RGE}. In view of the size of the effect preferred by the data, we can identify three qualitatively different effects that can play a role:
\begin{itemize}
  \item Wilson coefficients $[C_{eu}]_{2233}$ and
  $[C_{lu}]_{2233}$ of the ditop-dimuon operators $[O_{eu}]_{2233} = (\bar e_2 \gamma_\mu e_2)(\bar u_3 \gamma^\mu u_3)$ and
  $[O_{lu}]_{2233} = (\bar l_2 \gamma_\mu l_2)(\bar u_3 \gamma^\mu u_3)$ that induce a contribution to $C_9^{bs\mu\mu}$ from electroweak running above the EW scale. However, this solution is seriously challenged by EW precision tests \cite{Camargo-Molina:2018cwu} and we do not consider it further.

  \item Wilson coefficients $[C_{lq}^{(1,3)}]_{3323}$
  or $[C_{qe}]_{2333}$ of semitauonic operators $[O_{lq}^{(1)}]_{3323}$, $[O_{lq}^{(3)}]_{3323}$,
  or $[O_{qe}]_{2333}$ defined in (\ref{eq:SMEFTops1}) and (\ref{eq:SMEFTops2}),
  that induce a LFU contribution to $C_9^{bs\ell\ell}$
  from gauge-induced running both above and below the EW scale \cite{Bobeth:2011st,Crivellin:2018yvo}.

  \item Four-quark operators (defined below in section \ref{sec:fourquark}) that also induce a LFU contribution to $C_9^{bs\ell\ell}$ analogously to the semitauonic ones \cite{Jager:2017gal}.
\end{itemize}
The case of semitauonic operators is particularly interesting as it potentially allows for a simultaneous explanation of the anomalies in neutral-current $b\to s$ transitions and in charged-current $b\to c$ transitions involving taus \cite{Crivellin:2018yvo,Aebischer:2018iyb}. We now discuss these two possibilities in turn, from a model-independent point of view. Specific realizations in terms of simplified models will be discussed in Section~\ref{sec:models}.

\subsection{Semi-tauonic operators}\label{sec:semitauonic}

\begin{figure}[tbp]
\centering
\includegraphics[width=0.5\textwidth]{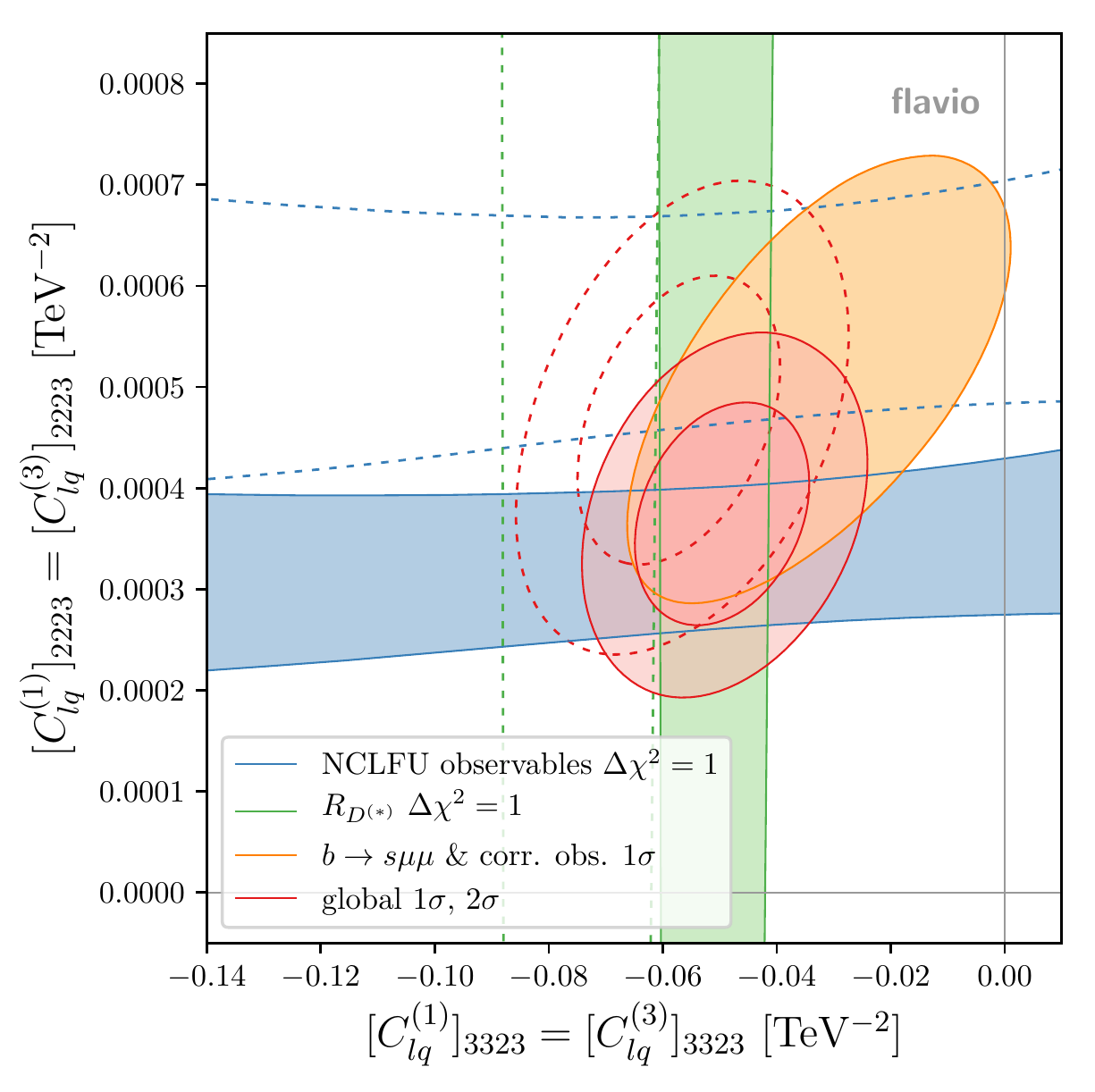}%
\caption{Likelihood contours from $R_{D^{(*)}}$, NCLFU observables ($R_{K^{(*)}}$ and $D_{P^\prime_{4,5}}$), and $b\to s\mu\mu$ observables in the space
  of the two SMEFT Wilson coefficients  $[C_{lq}^{(1)}]_{3323}=[C_{lq}^{(3)}]_{3323}$ and $[C_{lq}^{(1)}]_{2223}=[C_{lq}^{(3)}]_{2223}$ at 2~TeV.
  All other Wilson coefficients are assumed to vanish at 2~TeV.
  Solid (dashed) contours include (exclude) the Moriond-2019 results for $R_K$, $R_{K^*}$, $R_{D}$, and $R_{D^*}$.
  For sets of data that effectively constrain only a single Wilson coefficient (namely $R_{D^{(*)}}$ and NCLFU observables),
  1$\sigma$~contours correspond to $\Delta \chi^2 = 1$.
  For the other data ($b\to s\mu\mu$ and the global likelihood), 1 and 2$\sigma$~contours correspond to $\Delta \chi^2 \approx 2.3$ and $6.2$, respectively.
}
\label{fig:semitauonic}
\end{figure}

Intriguingly, a large value for $[C_{lq}^{(3)}]_{3323}$, that can explain the hints for LFU violation in charged-current $b\to c$ transitions
($R_{D}$ and $R_{D^*}$), also induces a LFU effect in $C_9$ that goes in the right
direction to solve the $b\to s\mu\mu$ anomalies in branching ratios and angular
observables.
An additional contribution to $[C_{lq}^{(a)}]_{2223}$ ($a=1$ or 3) of similar size can accommodate the deviations in $R_K$ and $R_{K^*}$.
Since the linear combination $[C_{lq}^{(1)}]_{3323}-[C_{lq}^{(3)}]_{3323}$ generates a sizable contribution
to $B\to K^{(*)}\nu\bar\nu$ decays \cite{Buras:2014fpa} that is constrained by
$B$-factory searches for these modes, such models are only viable if the semitauonic singlet and triplet Wilson coefficients are approximately equal\footnote{Note
that exact equality is not preserved by the RG evolution in SMEFT.}.

Fig.~\ref{fig:semitauonic} shows the likelihood contributions from
$R_{D^{(*)}}$, NCLFU observables, $b\to s\mu\mu$ observables, and the global likelihood
in the space of the two Wilson coefficients $[C_{lq}^{(1)}]_{3323}=[C_{lq}^{(3)}]_{3323}$ and $[C_{lq}^{(1)}]_{2223}=[C_{lq}^{(3)}]_{2223}$ at the renormalization scale $\mu = 2\,$TeV.
It is interesting to note that before the Moriond 2019 updates (indicated
by the dashed contours), for a purely muonic solution with
$[C_{lq}^{(1,3)}]_{3323}=0$ (corresponding to the vertical axis),
the best-fit values for NCLFU and $b\to s\mu\mu$ data were in perfect agreement with each other (even though $R_{D^{(*)}}$ cannot be explained in this case).
Including the $R_{K^{(*)}}$ updates, the best-fit point of the
NCLFU and $b\to s\mu\mu$ data instead lies in the region with non-zero
semitauonic Wilson coefficients, just as required to explain the
$R_{D^{(*)}}$ anomalies.
In fact, the agreement between the $1\sigma$ regions for $R_{K^{(*)}}$ \& $D_{P^\prime_{4,5}}$, $R_{D^{(*)}}$, and $b\to s\mu\mu$ {\em improves} compared to the case without the $R_{K^{(*)}}$ updates.
We note that a further improvement of the fit is achieved by taking into account the Moriond 2019 update of $R_{D^{(*)}}$ by Belle, which moves the $1\sigma$ region for $R_{D^{(*)}}$ slightly closer to the SM value, exactly to the region where the contours of NCLFU and $b\to s\mu\mu$ observables overlap.
The best fit values in this scenario are $[C_{lq}^{(1,3)}]_{3323} = -5.0\times 10^{-2}$ TeV$^{-2}$ and $[C_{lq}^{(1,3)}]_{2223} = 3.9\times 10^{-4}$ TeV$^{-2}$ with a $\sqrt{\Delta \chi^2} = 8.1$ that corresponds to a pull of $7.8\sigma$.
The pull is considerably larger in the present scenario than in those discussed in section \ref{sec:wet} since it can also explain discrepancies in $b\to c$ transitions.

It is interesting to use the global fit in this scenario as the basis for predictions of several observables that are sensitive to the Wilson coefficients $[C_{lq}^{(1,3)}]_{3323}$ and $[C_{lq}^{(1,3)}]_{2223}$ and are supposed to be measured with higher precision in the near future.
We collect predictions for LFU ratios, angular observables, and branching ratios in $B$ and $B_s$ decays in Table~\ref{tab:prediction}.

\begin{table}[tbp]
\centering
\renewcommand{\arraystretch}{1.5}
\rowcolors{2}{gray!15}{white}
\addtolength{\tabcolsep}{5pt} 
\begin{tabularx}{\textwidth}{l|ll|l}
\toprule
\rowcolor{white}
Observable & $1\sigma$ &$2\sigma$&SM \\
\hline
$R_{K^\ast}^{[0.045,1.1]}$ & $0.88\,_{-0.01}^{+0.01}$ & $[0.86,0.90]$ & $0.926 \pm 0.004$ \\
$R_{K^\ast}^{[1.1,6.0]}$ & $0.81\,_{-0.04}^{+0.04}$ & $[0.73,0.89]$ & $0.9964 \pm 0.0006$ \\
$R_{K^\ast}^{[0.1,8.0]}$ & $0.83\,_{-0.03}^{+0.04}$ & $[0.77,0.90]$ & $0.995 \pm 0.002$ \\
$R_{K^\ast}^{[15,19]}$ & $0.79\,_{-0.04}^{+0.04}$ & $[0.71,0.88]$ & $0.99807 \pm 0.00004$ \\
$R_K^{[1.0,6.0]}$ & $0.80\,_{-0.04}^{+0.04}$ & $[0.71,0.88]$ & $1.0008 \pm 0.0003$ \\
$R_\phi^{[1.0,6.0]}$ & $0.81\,_{-0.04}^{+0.04}$ & $[0.73,0.89]$ & $0.9970 \pm 0.0003$ \\
$\langle P_5^\prime\rangle^{[4.0,6.0]}$ & $-0.58\,_{-0.12}^{+0.13}$ & $[-0.82,-0.33]$ & $-0.763 \pm 0.072$ \\
$R_D$ & $0.34\,_{-0.01}^{+0.01}$ & $[0.32,0.37]$ & $0.303 \pm 0.006$ \\
$R_{D^{\ast}}$ & $0.29\,_{-0.01}^{+0.01}$ & $[0.27,0.31]$ & $0.255 \pm 0.004$ \\
$\overline{\text{BR}}(B_s\to \mu^+\mu^-)$ & $2.98\,_{-0.19}^{+0.20}\times 10^{-9}$ & $[2.60,3.38]$$\,\times 10^{-9}$ & $\left(3.67 \pm 0.16\right) \times 10^{-9}$ \\
$\text{BR}(B^\pm\to K^\pm \tau^+\tau^-)$ & $3.05\,_{-1.06}^{+1.78}\times 10^{-5}$ & $[1.01,6.47]$$\,\times 10^{-5}$ & $\left(1.66 \pm 0.19\right) \times 10^{-7}$ \\
$\overline{\text{BR}}(B_s\to \tau^+\tau^-)$ & $1.41\,_{-0.47}^{+0.80}\times 10^{-4}$ & $[0.52,2.94]$$\,\times 10^{-4}$ & $\left(7.78 \pm 0.33\right) \times 10^{-7}$ \\

\bottomrule
\end{tabularx}
\addtolength{\tabcolsep}{-6pt} 
\caption{Predictions for LFU ratios, angular observables, and branching ratios in $B$ and $B_s$ decays
  from the global likelihood in the space of SMEFT Wilson coefficients
  $[C_{lq}^{(1)}]_{3323}=[C_{lq}^{(3)}]_{3323}$ and $[C_{lq}^{(1)}]_{2223}=[C_{lq}^{(3)}]_{2223}$ (cf.~Fig.~\ref{fig:semitauonic}) and corresponding SM predictions.
}
\label{tab:prediction}
\end{table}

\subsection{Four-quark operators}\label{sec:fourquark}

Four-quark SMEFT operators can induce a LFU contribution to $C_9$ through
gauge-induced RG running above and below the EW scale from diagrams like in Fig.~\ref{fig:RGE}.
Since the flavour-changing quark current in $O_9$ is left-handed, these operators
must contain at least two $q$ fields and the other current must be a $SU(3)_c$ singlet.
Neglecting CKM mixing (i.e.\
to zeroth order in the Cabibbo angle), the following operators
could thus play a role:
\begin{align}
& [O_{qq}^{(1)}]_{23ii} = (\bar q_2 \gamma_\mu q_3)(\bar q_i \gamma^\mu q_i)
\,,& [O_{qq}^{(3)}]_{23ii} = (\bar q_2 \gamma_\mu \tau^I q_3)(\bar q_i \gamma^\mu \tau^I q_i) \,, \\
& [O_{qu}^{(1)}]_{23ii} = (\bar q_2 \gamma_\mu q_3)(\bar u_i \gamma^\mu u_i)
\,,& [O_{qd}^{(1)}]_{23ii} =(\bar q_2 \gamma_\mu q_3)(\bar d_i \gamma^\mu d_i) \,.
\end{align}

Rather than performing a simultaneous global analysis of all of these operators,
we now discuss each of them separately, assuming all the others to be absent,
bearing in mind that cancellations can modify the picture in the general case
(even though the different RG evolution of the operators makes all cancellations
unstable).

\begin{itemize}
  \item $[O_{qq}^{(1)}]_{2333}$ and $[O_{qu}^{(1)}]_{2333}$ induce a LFU
  contribution to $C_{10}$ that is much bigger than the one in $C_9$ through a
  $\bar sbZ$ coupling generated by a top quark loop, so they cannot explain the data.
  \item In the basis where the
  down-type quark mass matrix is diagonal,
  all the $[O_{qq}^{(a)}]_{23ii}$ operators lead to enormous
  NP contributions to CP violation in $D^0$-$\bar D^0$ mixing that are excluded
  by observations. Note that this happens even for real SMEFT Wilson
  coefficients since the CKM rotation between the mass basis for left-handed
  down-type quarks (relevant for $b_L\to s_L$ transitions) and up-type quarks
  (relevant for $u_L\leftrightarrow c_L$) is itself CP violating.
  Working instead in the basis where the up-type quark mass matrix is diagonal\footnote{\label{foot:up-align}Operators and couplings in such up-aligned basis are here and henceforth denoted with a hat.}
  and only using operators $[\hat O_{qq}^{(a)}]_{iijj}$ that do not contribute
  to $D^0$-$\bar D^0$ mixing, it turns out that all these operators either
  generate excessive contributions to CP violation in $K^0$-$\bar K^0$ mixing or do not
  generate an appreciable contribution to $C_9$.
  \item $[O_{qu}^{(1)}]_{2311}$ and $[O_{qd}^{(1)}]_{2311}$ can induce a NP
  contribution to $\epsilon'/\epsilon$ \cite{Aebischer:2018quc,Aebischer:2018csl} (i.e.\ direct CP violation in $K^0\to\pi\pi$)
  through RG running above the EW scale, but for a low enough scale
  they can lead to a visible effect in $C_9$ without violating this bound.
  \item $[O_{qd}^{(1)}]_{2322}$ can induce a NP
  contribution to $\Delta M_s$, the mass difference in the $B_s$-$\bar B_s$ system,
  through RG running above the EW scale, but for a low enough scale
  it can lead to a visible effect in $C_9$ without violating this bound.
  \item An effect in $C_9$ generated by $[O_{qu}^{(1)}]_{2322}$ and $[O_{qd}^{(1)}]_{2333}$
  is not strongly constrained at the level of the EFT.
\end{itemize}

To summarize, barring cancellations,
a LFU contribution to $C_9$ could be generated by the SMEFT
Wilson coefficients
\begin{align}
[C_{qu}^{(1)}]_{2311}
\,,&&
[C_{qu}^{(1)}]_{2322}
\,,&&
[C_{qd}^{(1)}]_{2311}
\,,&&
[C_{qd}^{(1)}]_{2322}
\,,&&
[C_{qd}^{(1)}]_{2333}
\,.
\label{eq:viable4q}
\end{align}
The generic size of these Wilson coefficients required for a visible effect in $C_9$
is in the ballpark of $1/\text{TeV}^2$.

Consequently, realistic model
implementations of such an effect have to rely on tree-level mediators with
sizeable couplings to quarks and masses potentially in the reach of direct production
at LHC. We will discuss such simplified models in Section~\ref{sec:leptophobic}.

\section{Simplified new-physics models}\label{sec:models}

In this section we consider simplified models with a single tree-level mediator multiplet giving rise to the Wilson coefficient patterns that are in agreement with the above findings in the EFT.

In Section~\ref{sec:U1}, we consider the $U_1$ vector leptoquark, transforming as $({\mathbf 3},{\mathbf 1})_{2/3}$ under the SM gauge group, that is known to
be the only viable simultaneous single-mediator explanation of the $R_{K^{(*)}}$ and $R_{D^{(*)}}$
anomalies. Since it generates the semitauonic operators discussed in Section~\ref{sec:semitauonic},
it can also generate a LFU contribution to $C_9$.

In Section~\ref{sec:leptophobic}, we discuss realizations of LFU contributions to $C_9$ via RG effects from four-quark operators as discussed in Section~\ref{sec:fourquark}.
We will show that there is a single viable mediator,
a scalar transforming as $({\mathbf 8},{\mathbf 2})_{1/2}$ under the SM gauge group, and that it is strongly
constrained by LHC di-jet searches.

\subsection{Explaining the data by a single mediator: the $U_1$ leptoquark solution}\label{sec:U1}
The only single mediator that can yield non-zero values for $[C_{lq}^{(1)}]_{3323}=[C_{lq}^{(3)}]_{3323}$ and $[C_{lq}^{(1)}]_{2223}=[C_{lq}^{(3)}]_{2223}$ is the $U_1$ vector leptoquark \cite{Barbieri:2015yvd,Alonso:2015sja,Calibbi:2015kma,Fajfer:2015ycq,Hiller:2016kry,Bhattacharya:2016mcc,Buttazzo:2017ixm,DiLuzio:2017vat,Assad:2017iib,Calibbi:2017qbu,Bordone:2017bld,Barbieri:2017tuq,Greljo:2018tuh,Blanke:2018sro,Kumar:2018kmr,Fornal:2018dqn}, which transforms as $({\mathbf 3},{\mathbf 1})_{2/3}$ under the SM gauge group.
We define its couplings to the left-handed SM fermion doublets $q$ and $l$ as
\begin{equation}
\label{eq:L_U1}
 \mathcal{L}_{U_1}\supset g_{lq}^{ji}\left(\bar{q}^i\gamma^\mu l^j\right) U_\mu + \text{h.c.}
\end{equation}

From the tree-level matching at the scale $\Lambda = M_U$, one finds
\begin{equation}
\label{eq:Clq_matched_w_U1}
 [C_{lq}^{(1)}]_{ijkl}=[C_{lq}^{(3)}]_{ijkl} = -\frac{g_{lq}^{jk}\,g_{lq}^{il*}}{2 M_U^2}.
\end{equation}
Consequently, for a given leptoquark mass, a $\tau$-channel contribution to $R_{D^{(*)}}$ depends only on $g_{lq}^{32}$ and $g_{lq}^{33}$, while a $\mu$-channel contribution to $R_{K^{(*)}}$ depends only on $g_{lq}^{22}$ and $g_{lq}^{23}$. The NLO corrections to such semileptonic operators are known to be of the order $13\%$ \cite{Aebischer:2018acj} and will be neglected in the following.

As has been shown in \cite{Crivellin:2018yvo}, the $U_1$ leptoquark model generates one-loop matching contributions to the electric and chromomagnetic dipole operators in the WET.
They can lead to relevant shifts in the Wilson coefficient $C_7$ at the $b$-quark scale, which are constrained by measurements of $b\to s \gamma$ observables (cf.~\cite{Paul:2016urs}).
In order to be sensitive to this possibly important effect, we will take the one-loop matching contributions to the SMEFT quark-dipole operators into account.
These operators are defined as~\cite{Grzadkowski:2010es}
\begin{align}
 & [O_{dB}]_{ij} = (\bar q_i \sigma^{\mu\nu} d_j) \varphi B_{\mu\nu} \,, & [O_{dW}]_{ij} = (\bar q_i \sigma^{\mu\nu}  d_j)\tau^I \varphi W^I_{\mu\nu}\,, \\
 & [O_{dG}]_{ij} = (\bar q_i \sigma^{\mu\nu} T^A d_j) \varphi G^A_{\mu\nu} \,. &
\end{align}
The matching result depends on the couplings of the $U_1$ vector leptoquark to the SM gauge bosons, which can be written as
\begin{equation}\label{eq:dipole_U1}
  \mathcal{L}_{U_1}\supset
  -\frac{1}{2}U_{\mu\nu}^\dagger U^{\mu\nu}+
  i g_sk_s U_\mu^\dag T^AU_\nu G^{A,\mu\nu}+i g' \frac{2}{3}k_YU_\mu^\dag U_\nu B^{\mu\nu}\,,
\end{equation}
where
\begin{equation}
 U_{\mu\nu}=D_{\mu}U_{\nu}-D_{\nu}U_{\mu}
 \quad
 \text{with}
 \quad
 D_{\mu} = \partial_\mu + i g_s T^A G^A_\mu + i g' \frac{2}{3} B_\mu.
\end{equation}
These couplings are determined by SM gauge invariance except for the two parameters $k_s$ and $k_Y$.
In the following, we make the choice $k_s=k_Y=1$, which leads to a cancellation of divergent tree-level diagrams in $U_1$-gluon and $U_1$-$B$-boson scattering~\cite{Ferrara:1992yc} and further avoids logarithmically divergent contributions to the dipole operators \cite{Barbieri:2015yvd}, making them finite and gauge independent.
We note that $k_s=k_Y=1$ is automatically satisfied in any model in which the $U_1$ leptoquark stems from the spontaneous breaking of a gauge symmetry but can also be realized for a composite $U_1$~\cite{Barbieri:2016las}.

\begin{figure}
\centering
\includegraphics[width=0.35\textwidth]{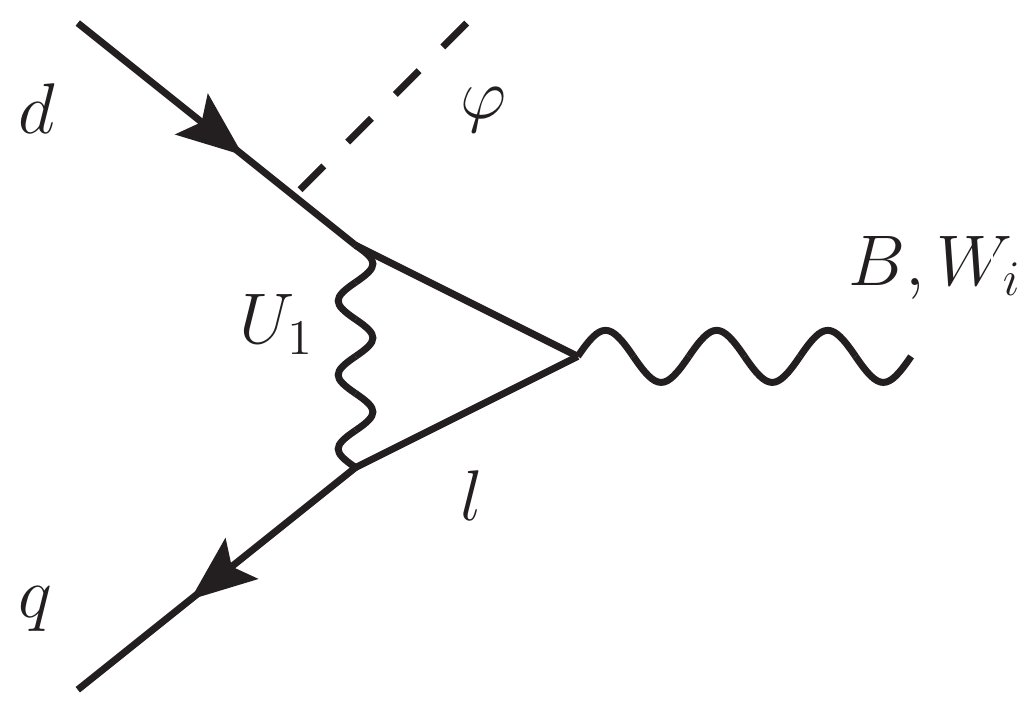}
\hspace{1.5cm}
\includegraphics[width=0.35\textwidth]{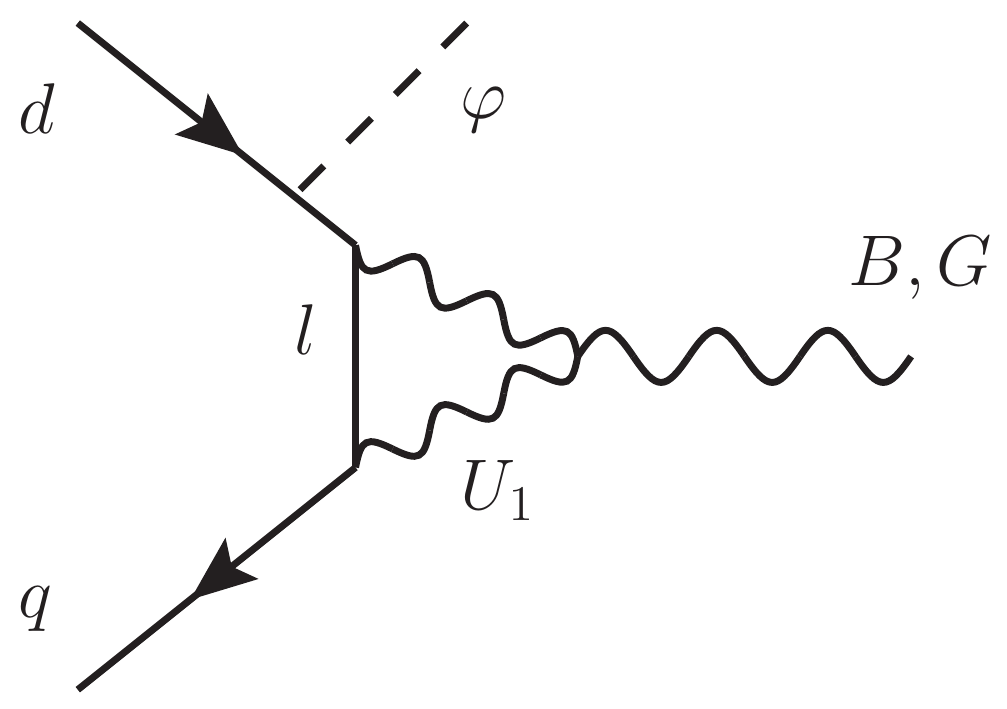}
\caption{Diagrams contributing to the matching of the $U_1$ leptoquark model onto the SMEFT operators $O_{dB}, O_{dW}$ and $O_{dG}$.}
\label{fig:U1matching}
\end{figure}

We perform the one-loop matching at the scale $\Lambda=M_U$ by computing the diagrams shown in Fig.~\ref{fig:U1matching}.
Working in the basis in which the down-type Yukawa matrix is diagonal, and using the conventions mentioned above, we find the Wilson coefficients of the EW dipole operators
\begin{align}
  [C_{dW}]_{23}&=Y_b\frac{g}{16\pi^2}\left(\frac{1}{6}\right)\,\frac{g_{lq}^{i2}\,g_{lq}^{i3*}}{M_U^2}\,,
  \quad&
  [C_{dW}]_{32}&=Y_s\frac{g}{16\pi^2}\left(\frac{1}{6}\right)\,\frac{g_{lq}^{i3}\,g_{lq}^{i2*}}{M_U^2}\,, \\
  [C_{dB}]_{23}&=Y_b\frac{g'}{16\pi^2}\left(-\frac{4}{9}\right)\,\frac{g_{lq}^{i2}\,g_{lq}^{i3*}}{M_U^2}\,,
  \quad&
  [C_{dB}]_{32}&=Y_s\frac{g'}{16\pi^2}\left(-\frac{4}{9}\right)\,\frac{g_{lq}^{i3}\,g_{lq}^{i2*}}{M_U^2}\,,
\end{align}
where $Y_b$ and $Y_s$ denote the Yukawa couplings of the $b$ and $s$ quark respectively and a summation over the lepton index is implied. The Wilson coefficients of the chromomagnetic dipole operators at the matching scale read
\begin{equation}
  [C_{dG}]_{23}=Y_b\frac{g_s}{16\pi^2}\left(-\frac{5}{12}\right)\,\frac{g_{lq}^{i2}\,g_{lq}^{i3*}}{M_U^2}\,,
  \quad\quad
  [C_{dG}]_{32}=Y_s\frac{g_s}{16\pi^2}\left(-\frac{5}{12}\right)\,\frac{g_{lq}^{i3}\,g_{lq}^{i2*}}{M_U^2}\,.
\end{equation}
Using the tree-level matching conditions from SMEFT onto WET \cite{Aebischer:2015fzz,Jenkins:2017jig}, we have checked that these results are consistent with the findings in \cite{Crivellin:2018yvo}.

\subsubsection{$R_{D^{(*)}}$ and indirect constraints}\label{sec:U1_indirect}

\begin{figure}[t]
\centering
\includegraphics[width=0.5\textwidth]{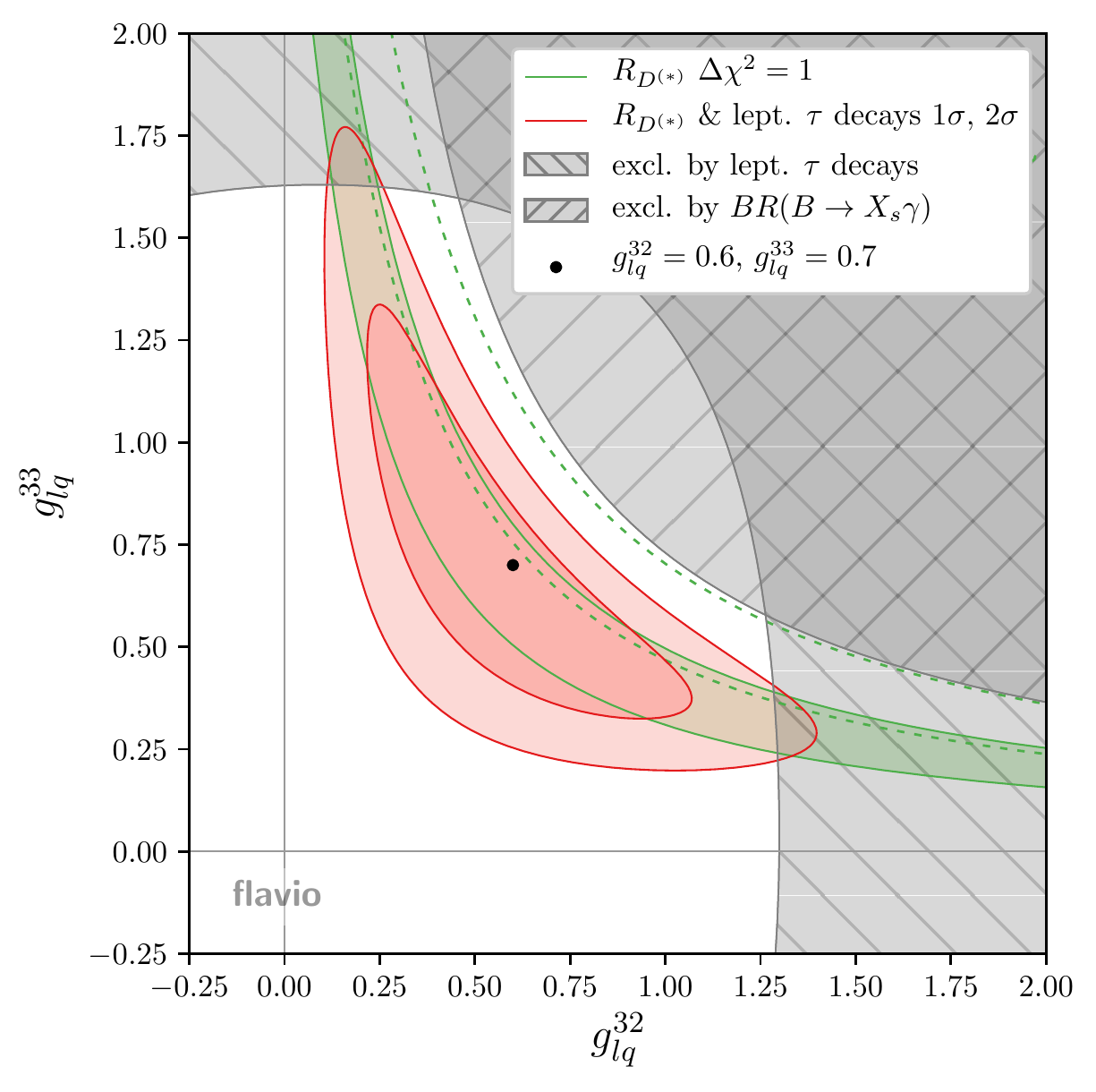}%
\caption{Likelihood contours from different observables in the space of the tauonic $U_1$ leptoquark couplings $g_{lq}^{32}$ and $g_{lq}^{33}$ at 2 TeV.
The grey areas are excluded at the $2\sigma$ level.
  $R_{D^{(*)}}$ data and leptonic $\tau$ decays select a well-defined region in the $g_{lq}^{32}$ versus $g_{lq}^{33}$ plane.
  For $R_{D^{(*)}}$, which only constrain one degree of freedom, 1$\sigma$~contours correspond to $\Delta \chi^2 = 1$, while for others (the global likelihood, leptonic $\tau$ decays, BR$(B\to X_s\gamma)$), 1 and 2$\sigma$~contours correspond to $\Delta \chi^2 \approx 2.3$ and $6.2$, respectively.
}
\label{fig:U1_32_33}
\end{figure}

We perform a fit with fixed $M_U=2$~TeV in the space of tauonic couplings $g_{lq}^{32}$ and $g_{lq}^{33}$, which we take to be real for simplicity.
This allows us to determine the region in which $R_{D^{(*)}}$ can be explained by the semi-tauonic operators discussed in Section~\ref{sec:semitauonic}.
The results of the fit are shown in Fig.~\ref{fig:U1_32_33} and our findings are as follows:
\begin{itemize}
 \item The strongest constraints are due to
 \begin{itemize}
  \item leptonic tau decays $\tau\to\ell\nu\nu$, which receive a contribution due to RG running,\footnote{\label{fn:RGlogs}%
Our analysis includes RG-induced logarithms.
Note that the interactions in Eq.~(\ref{eq:L_U1}) and (\ref{eq:dipole_U1}) provide a simplified model and not a complete UV theory of the $U_1$-leptoquark.
In such a UV theory, it could in principle be possible that the RG-induced logarithms are (partially) canceled by finite terms, which are not present in the simplified model.
Barring cancellations, and in view of the renormalization-scale independence of the full result -- logarithms plus analytic terms -- the contributions from the RG-induced logarithms usually provide a realistic estimate of the size of the effects.}
  \item BR$(B\to X_s\gamma)$, which receives a contribution from the one-loop matching onto dipole operators in SMEFT as discussed above.\footnote{Such contributions are, however, model-dependent. For example, they will be quite different in models with additional vector-like fermions running in the loops \cite{Calibbi:2017qbu,Bordone:2017bld}, as shown explicitly in Ref. \cite{Cornella:2019hct}.}
 \end{itemize}
 This underlines the importance of taking into account loop effects, both in the RG running and in the matching, as emphasized already in~\cite{Feruglio:2016gvd,Feruglio:2017rjo,Crivellin:2018yvo}.
 \item A combined fit to $R_{D^{(*)}}$ and leptonic tau decays selects a well-defined region in the space of $g_{lq}^{32}$ and $g_{lq}^{33}$ in which $R_{D^{(*)}}$ can be explained while satisfying all constraints.
 \item In order to explain $R_{D^{(*)}}$ while at the same time avoiding exclusion at the $2\sigma$ level from leptonic tau decays, a minimal ratio of tauonic couplings $\frac{g_{lq}^{32}}{g_{lq}^{33}} \gtrsim 0.1$ is required (assuming vanishing right-handed couplings), which is compatible with findings in~\cite{Buttazzo:2017ixm}.
 This puts some tension on models based on a $U(2)_q$ flavour symmetry~\cite{Barbieri:2015yvd,Buttazzo:2017ixm,Bordone:2017bld,Greljo:2018tuh}, where the natural expectation for the size of $\frac{g_{lq}^{32}}{g_{lq}^{33}}$ is $|V_{cb}|\approx0.04$.
\end{itemize}
\begin{figure}[t]
\centering
\includegraphics[width=0.5\textwidth]{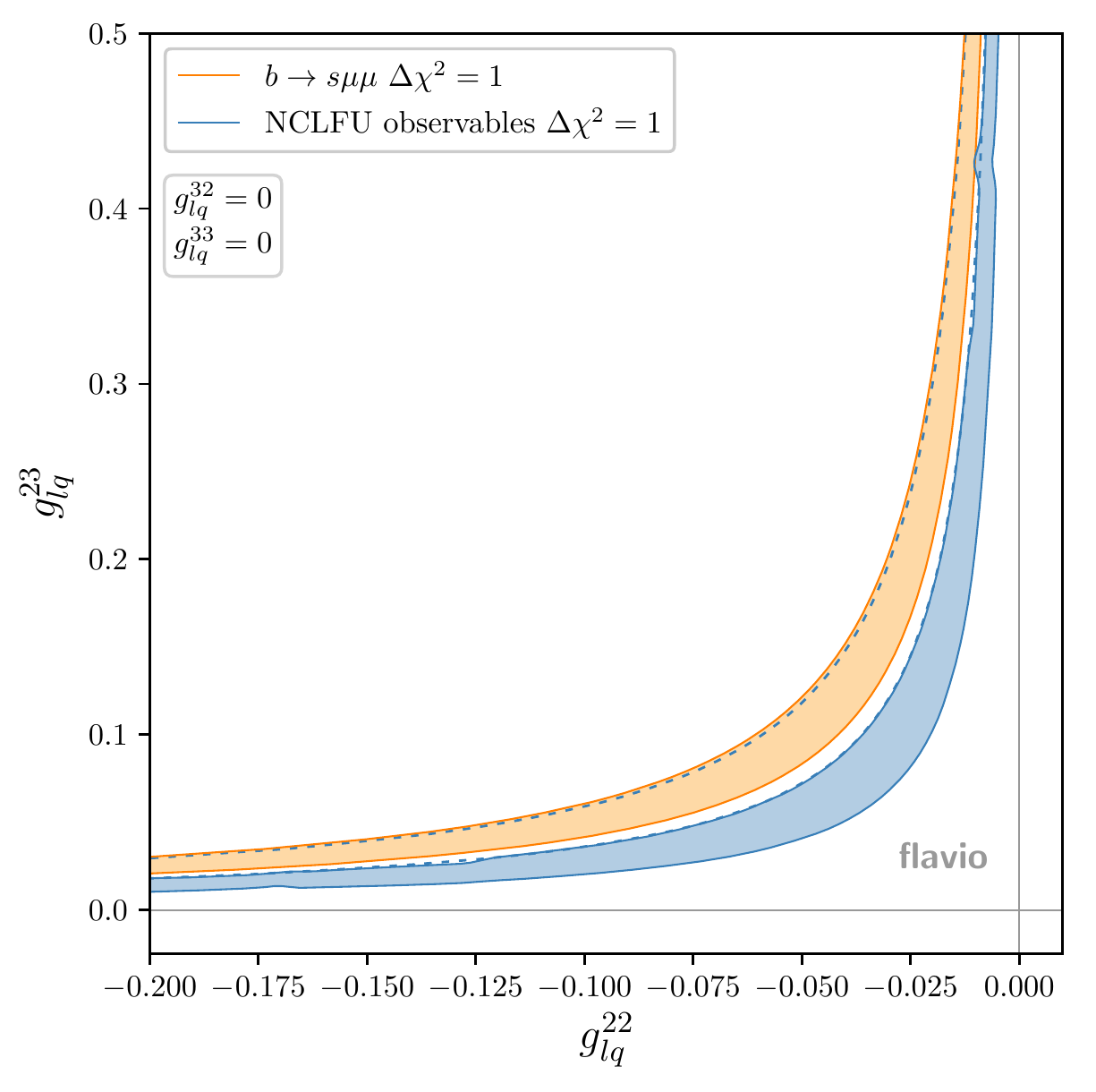}%
\includegraphics[width=0.5\textwidth]{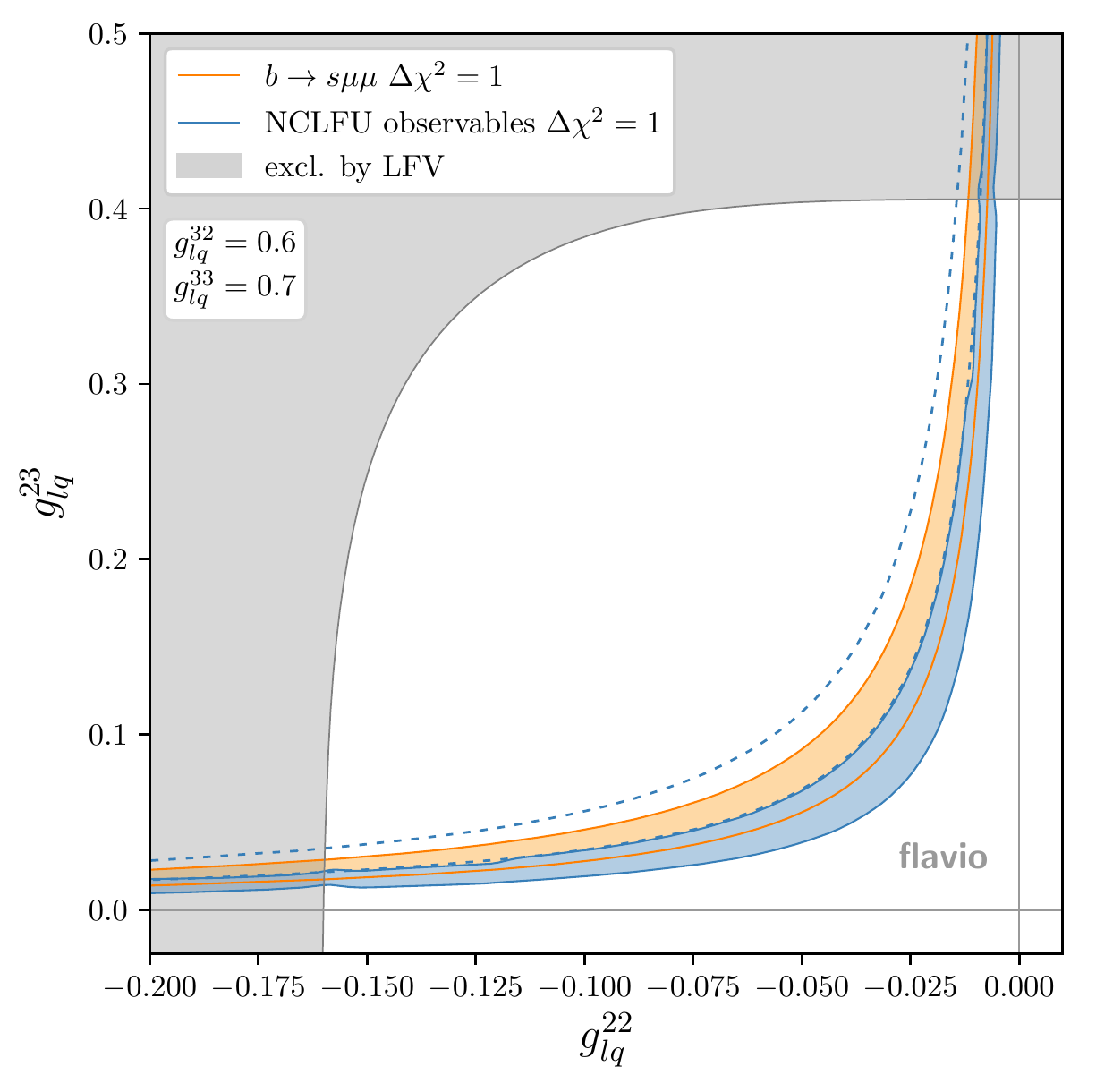}%
\caption{Likelihood contours from different observables in the space of the muonic $U_1$ leptoquark couplings $g_{lq}^{22}$ and $g_{lq}^{23}$ at 2 TeV.
Fits are shown for vanishing tauonic couplings $g_{lq}^{32}=0$, $g_{lq}^{33}=0$ (left) and at the benchmark point $g_{lq}^{32}=0.5$, $g_{lq}^{33}=0.7$ (right).
The grey area is excluded at the $2\sigma$ level.
  For observables that only constrain one degree of freedom (here NCLFU and $b\to s\mu\mu$ observables),
  1$\sigma$~contours correspond to $\Delta \chi^2 = 1$, while for the lepton flavour violating observables, the 2$\sigma$~contour corresponds to $\Delta \chi^2 \approx 6.2$.
}
\label{fig:U1_22_23}
\end{figure}
Based on the above results, we select a benchmark point from the best-fit region in the fit to tauonic couplings,
\begin{equation}
 g_{lq}^{32}=0.6,
 \quad\quad
 g_{lq}^{33}=0.7,
\end{equation}
which is also shown in Fig.~\ref{fig:U1_32_33}.
We then perform two fits in the space of muonic couplings $g_{lq}^{22}$ and $g_{lq}^{23}$ shown in Fig.~\ref{fig:U1_22_23}: one for vanishing tauonic couplings (left panel) and one at the benchmark point $g_{lq}^{32}=0.6$, $g_{lq}^{33}=0.7$ (right panel).
Our findings are as follows:
\begin{itemize}
 \item For vanishing tauonic couplings (left panel of Fig.~\ref{fig:U1_22_23}), the data available before Moriond 2019 leads to a very good agreement between the fits to $b\to s\mu\mu$ (orange contour) and NCLFU observables (dashed blue contour), while the $R_{D^{(*)}}$ measurements cannot be explained in this scenario.
 Taking into account the updated and new measurements of $R_{K^{(*)}}$ presented at Moriond 2019, one finds a slight tension between the fits to $b\to s\mu\mu$ (orange contour) and NCLFU observables (solid blue contour).
 This is analogous to the tension mentioned in sections~\ref{sec:2d} and~\ref{sec:universal}.
 \item The tension disappears if one considers non-zero tauonic couplings that can also explain $R_{D^{(*)}}$, which is exemplified in the right panel of Fig.~\ref{fig:U1_22_23} for the benchmark point $g_{lq}^{32}=0.6$, $g_{lq}^{33}=0.7$.
 As discussed in section~\ref{sec:SMEFT}, the semi-tauonic operators obtained from the tree-level matching (cf.~Eq.~\ref{eq:Clq_matched_w_U1}) induce a lepton-flavour universal contribution to $C_9$, which affects the predictions of $b\to s\mu\mu$ observables and makes the fits to $b\to s\mu\mu$ and NCLFU observables again compatible with each other at the 1$\sigma$ level.
 Consequently, the deviations in neutral current and charged current $B$-decays can all be explained at once.
This very well agrees with our findings in the SMEFT scenario in Section~\ref{sec:semitauonic}.
 \item Given the presence of non-zero values for the tauonic couplings at the benchmark point, the strongest constraint on the muonic couplings $g_{lq}^{22}$ and $g_{lq}^{23}$ is due to LFV observables, in particular $\tau\to\phi\mu$ and $B\to K\tau\mu$.
 The region in the space of muonic couplings that is excluded at the 2$\sigma$ level by these observables is shown in gray in the right panel of Fig.~\ref{fig:U1_22_23}.
\end{itemize}
In conclusion we find that the $U_1$ vector leptoquark can still provide an excellent description of the $B$ anomalies while satisfying all indirect constraints.

\subsubsection{Comparison between indirect and direct constraints}

\begin{figure}[t]
\centering
\includegraphics[width=0.5\textwidth]{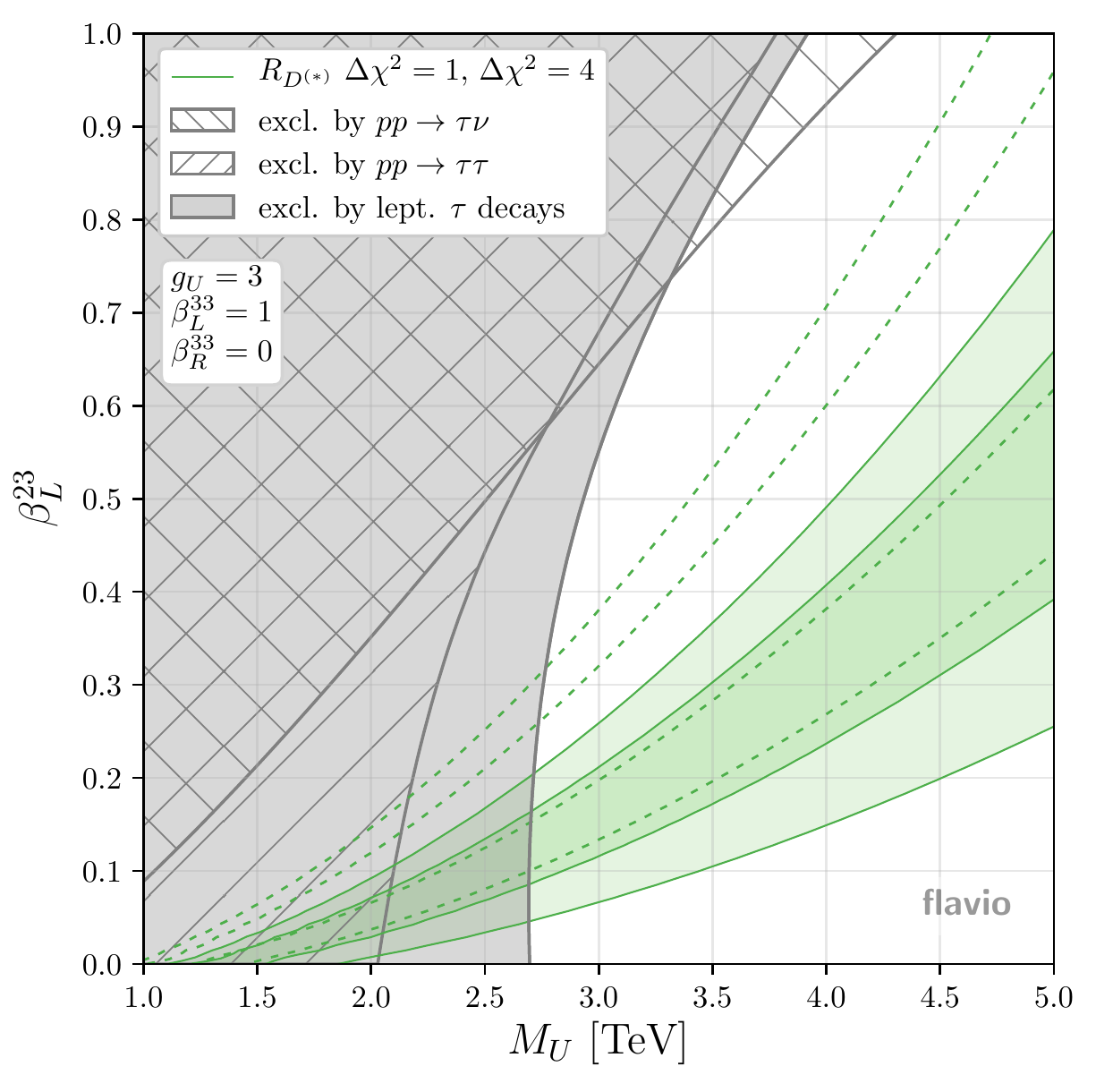}%
\caption{Best fit region in the space of $U_1$ leptoquark mass $M_U$ and coupling $\beta^{23}_L$ (cf.~Eq.~(\ref{eq:glq_betaL})). The green region is the preferred region from $R_{D^{(*)}}$, while the gray shaded area is excluded by leptonic tau decays at the $2\sigma$ level. The hatched areas are excluded by LHC searches recasted in~\cite{Baker:2019sli}.}
\label{fig:U1_mU_betaL23}
\end{figure}

In addition to indirect constraints, high-$p_T$ signatures of models containing a $U_1$ leptoquark have been discussed in detail considering current and future LHC searches \cite{Faroughy:2016osc,Greljo:2017vvb,Diaz:2017lit,Angelescu:2018tyl,Greljo:2018tzh,Baker:2019sli}.
In this section, we compare direct constraints found in the latest study, \cite{Baker:2019sli}, to the strong indirect constraints discussed in Section~\ref{sec:U1_indirect}.
To this end, we adopt the notation of~\cite{Baker:2019sli} and use the parameters $\beta_L^{ij}$ and $g_U$, which are related to our notation by
\begin{equation}
\label{eq:glq_betaL}
  g_{lq}^{ij} = \frac{\beta_L^{ji}\,g_U}{\sqrt{2}}.
\end{equation}
We perform a fit with fixed $g_U=3$, $\beta_L^{33}=1$ (i.e. $g_{lq}^{33}\approx 2$) in the space of $M_U$ and $\beta_L^{23}$.
These values are chosen to allow for a direct comparison with the constraint from $pp\to \tau\tau$ shown in Fig.~1 of~\cite{Baker:2019sli} and $pp\to\tau\nu$ shown in Fig.~6 of~\cite{Baker:2019sli}.
We include both of these direct constraints in our Fig.~\ref{fig:U1_mU_betaL23} as hatched areas.
In addition, we show the results from our fit, namely the constraint from leptonic $\tau$ decays and the region preferred by the $R_{D^{(*)}}$ measurements.
Our findings are as follows:
\begin{itemize}
 \item The indirect constraint from leptonic $\tau$ decays is stronger than the direct constraints in nearly all of the parameter space shown in Fig.~\ref{fig:U1_mU_betaL23}, except for a small region at large $\beta_L^{23}\gtrsim 0.75$, where the constraint from $pp\to\tau\nu$ is the strongest one.
 \item In the region where $R_{D^{(*)}}$ can be explained, the indirect constraint from leptonic $\tau$ decays is considerably stronger than the direct ones.
 \item Small values for $\frac{\beta_L^{23}}{\beta_L^{33}}$ as naturally expected in models based on a $U(2)_q$ flavour symmetry~\cite{Barbieri:2015yvd,Buttazzo:2017ixm,Bordone:2017bld,Greljo:2018tuh} require a relatively small mass $M_U$ to explain $R_{D^{(*)}}$.
 Thus, as also pointed out in \cite{Faroughy:2016osc,Buttazzo:2017ixm}, there is already some tension between this natural expectation and the direct searches.
\end{itemize}
We note that the direct constraints shown in Fig.~\ref{fig:U1_mU_betaL23} depend on the coupling strength $g_U$.
While the assumptions $g_U=3$, $\beta_L^{33}=1$ lead to a lower bound on the leptoquark mass $M_U\gtrsim2.7$~TeV, this bound does not apply to the scenario discussed in Section~\ref{sec:U1_indirect}, which features considerably smaller couplings\footnote{%
The partonic cross section relevant for the direct constraints in Fig.~\ref{fig:U1_mU_betaL23} scales as $\sigma\sim(g_U/M_U)^4$~\cite{Baker:2019sli}.
}.
Latest direct constraints from $U_1$ pair production that are independent of the coupling strength $g_U$ only exclude masses $M_U\lesssim1.5$~TeV~\cite{Angelescu:2018tyl,Baker:2019sli}.
Therefore, the scenario discussed in Section~\ref{sec:U1_indirect} is currently not constrained by direct searches.

\subsection{Lepton flavour universal $C_9$ from leptophobic mediators}\label{sec:leptophobic}

As discussed model-independently in Section~\ref{sec:fourquark},
a lepton flavour universal contribution to $C_9$ can also be induced from
a four-quark operator via RG effects.
However, the four-quark operator would realistically have to be generated by
the tree-level exchange of a resonance with mass not exceeding a few TeV and
$O(1)$ couplings. Such resonance could then be in reach of direct LHC searches,
apart from other indirect constraints.

Since it was shown in Section~\ref{sec:fourquark}
that the only viable operators are of type
$O_{qu}^{(1)}$ and $O_{qd}^{(1)}$,
the conceivable tree-level mediators,
excluding fields that admit baryon number violating diquark couplings\footnote{%
Note that baryon number violating diquark couplings
do not necessarily lead to tree-level proton decay.}, are (see e.g.\ \cite{deBlas:2017xtg})
\begin{itemize}
  \item $({\mathbf 1}, {\mathbf 1})_0$ with spin 1 ($Z'$),
  \item $({\mathbf 8}, {\mathbf 1})_0$ with spin 1 ($G'$),
  \item $({\mathbf 1}, {\mathbf 2})_{1/2}$ with spin 0 ($H'$),
  \item $({\mathbf 8}, {\mathbf 2})_{1/2}$ with spin 0 ($\Phi$).
\end{itemize}
However, it is immediately clear that the spin-1 mediators
$Z'$ and $G'$ are not viable: to generate the
Wilson coefficients $[C_{qu}^{(1)}]_{23ii}$
or $[C_{qu}^{(1)}]_{23ii}$ with sufficient size, they would require a sizeable
flavour-violating coupling to left-handed down-type quarks that would induce excessive contributions to
$B_s$-$\bar B_s$ mixing. Thus the only potentially viable
models are the scalar mediators.

The new scalar bosons have the following Lagrangian couplings to quarks,
\begin{align}
  \mathcal L_{H'} &\supset
  y_{H'qu}^{ij} \,\bar q_i u_j \tilde H'
  -
  y_{H'qd}^{ij} \,\bar q_i  d_j H'
  + \text{h.c.}
  \,,
  \\
  \mathcal L_{\Phi} &\supset
  y_{\Phi qu}^{ij} \,\bar q_i  T^A u_j \tilde \Phi^A
  -
  y_{\Phi qd}^{ij} \,\bar q_i  T^A d_j \Phi^A
  + \text{h.c.}
  \,.&
\end{align}
The SMEFT Wilson coefficients that are generated by a tree-level scalar exchange and can contribute to $C_9$ via RG effects read
\begin{equation}
  [C_{qR}^{(1)}]_{ijkl} =
  \sum_{X=H',\Phi}c_X\, \frac{y_{XqR}^{jk*}y_{XqR}^{il}}{M_{X}^2}\,,
\end{equation}
where $R=u,d$ and $(c_{H'},c_\Phi)= (-1/6, -2/9)$.

Clearly, to generate one of the down-type Wilson coefficients $[C_{qd}^{(1)}]_{23ii}$
in \eqref{eq:viable4q},
at least one flavour-violating coupling in the down-aligned basis has to be present, leading to excessive contributions to $B^0$-$\bar B^0$ or $B_s$-$\bar B_s$ mixing. Thus we assume vanishing down-type couplings $y_{Xqd}^{ij}=0$ in the following.

For the up-type couplings $y_{Xqu}$, it is instead convenient to work in the up-aligned basis (cf. footnote~\ref{foot:up-align} for notation), as setting $\hat y_{Xqu}^{12}=\hat y_{Xqu}^{21}=0$ allows avoiding dangerous contributions to $D^0$-$\bar D^0$ mixing. We then obtain the following non-vanishing matching conditions relevant for RG-induced contributions to $C_9$,
\begin{align}
  [C_{qu}^{(1)}]_{2311} & = c_X\frac{1}{M_X^2}
  \, V_{is}^*V_{jb}  \,\hat y_{Xqu}^{j1*}\hat y_{Xqu}^{i1}
  \approx
  c_X\frac{1}{M_X^2}\left(
  V_{us}^*V_{tb}\,\hat y_{Xqu}^{31*} \hat y_{Xqu}^{11} + V_{ts}^*V_{tb}\,|\hat y_{Xqu}^{31}|^2 + O(\lambda^3)
  \right)
  \,,\label{eq:Cqu2311}\\
  [C_{qu}^{(1)}]_{2322} &= c_X\frac{1}{M_X^2}
  \, V_{is}^*V_{jb}  \,\hat y_{Xqu}^{j2*}\hat y_{Xqu}^{i2}
  \nonumber\\
  &\approx
  c_X\frac{1}{M_X^2}\left(
  V_{cs}^*V_{tb}\,\hat y_{Xqu}^{32*} \hat y_{Xqu}^{22} +
  V_{cs}^*V_{cb}\,|\hat y_{Xqu}^{22}|^2 +
  V_{ts}^*V_{tb}\,|\hat y_{Xqu}^{32}|^2
  + O(\lambda^3)
  \right)
  \,.\label{eq:Cqu2322}
\end{align}
Numerically, it turns out that a visible lepton flavour universal effect in $C_9$ requires $O(1)$ couplings for a scalar mass of the order of 2~TeV even for terms without CKM suppression. Thus, LHC searches for di-jet resonances are sensitive to
the new scalars.
If the NP effect in $C_9$ is generated through the terms in \eqref{eq:Cqu2311}, the scalar has
sizeable couplings to up quarks,
leading to an excessive production cross section at the LHC.
To avoid this, we will further assume $\hat y_{Xqu}^{1i}=\hat y_{Xqu}^{i1}=0$ and use the terms in
\eqref{eq:Cqu2322}.
Production in $pp$ collisions is still possible via charm quarks. The leading-order cross sections of the charged and neutral components read

\begin{align}
\sigma(pp\to X^\pm) &=\tilde c_X\frac{\pi}{12s}
\left[
|\hat y_{Xqu}^{22}|^2 \left(
\mathcal L_{s\bar c} + \mathcal L_{c\bar s}
\right)
+
|\hat y_{Xqu}^{32}|^2 \left(
\mathcal L_{b\bar c} + \mathcal L_{c\bar b}
\right)
\right],
\\
\sigma(pp\to X^0) &= \tilde c_X\frac{\pi}{6s}
|\hat y_{Xqu}^{22}|^2
\mathcal L_{c\bar c}
\,,
\end{align}
where $(\tilde c_{H'},\tilde c_\Phi)=(1,4/3)$, $\sqrt{s}$ being the center of mass energy and $\mathcal L_{ij}$ are
the parton luminosities as defined in~\cite{Stangl:2018kty} and we have neglected
contributions suppressed by CKM factors.

We confront these cross sections with exclusion limits from ATLAS \cite{Aaboud:2019zxd} and CMS \cite{Sirunyan:2018xlo}.
Our procedure to obtain constraints on the scalar model parameters is detailed in appendix~\ref{app:dijet}.
For definiteness, we choose $X=\Phi$ in the following. In fact, according to the results of \cite{Jager:2017gal}, the $H'$ case is considerably more constrained by contributions to $B\to X_s\gamma$ introduced radiatively at the two-loop level.

\begin{figure}
  \includegraphics[width=\textwidth]{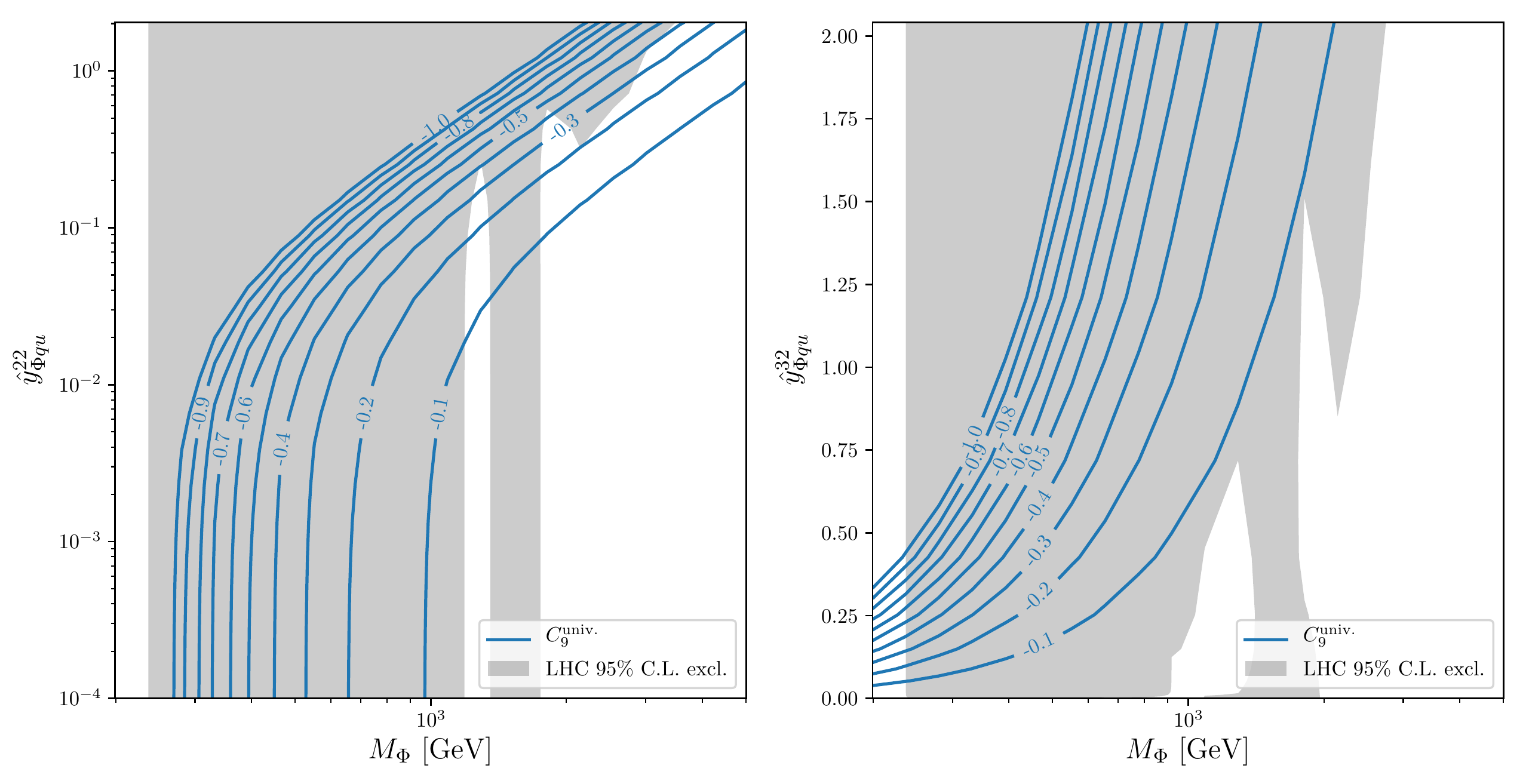}
  \caption{Contours of constant $C_9^\text{univ.}$
  in the colour octet scalar model vs.\ LHC di-jet exclusion for a scenario with $\hat y_{\Phi qu}^{32}=-1$
  and varying $\hat y_{\Phi qu}^{22}$ (left) and for a
  scenario with $\hat y_{\Phi qu}^{22}=0$ and varying
  $\hat y_{\Phi qu}^{32}$ (right).
}
  \label{fig:Phi-exclusion}
\end{figure}

The left plot in Fig.~\ref{fig:Phi-exclusion} shows contours
of $C_9^\text{univ.}$
in the plane of $\hat y_{\Phi qu}^{22}$ vs.\ the color-octet scalar mass for a scenario in which $\hat y_{\Phi qu}^{32}=-1$.
We also show the 95\% C.L.\ exclusion from di-jet
resonance searches at LHC. Obviously, a visible negative
contribution to $C_9^\text{univ.}$ of at most $-0.3$
can only be generated in a thin sliver of parameter space
for masses above 2~TeV and with $\hat y_{\Phi qu}^{22}\sim1$.
This scenario is on the brink of exclusion.
The bending of the $C_9^\text{univ.}$ contours at low
masses in the left plot of Fig.~\ref{fig:Phi-exclusion}
is due to the fact that there is a CKM-suppressed contribution even for $\hat y_{\Phi qu}^{22}=0$ from
the third term in (\ref{eq:Cqu2322}).
To investigate whether such scenario, where production
is only possible through a $b$ quark PDF, is allowed,
in the right plot of Fig.~\ref{fig:Phi-exclusion} we show
the $C_9^\text{univ.}$ contours and the LHC exclusion
for $\hat y_{\Phi qu}^{32}$ vs.\ the color octet scalar mass setting
$\hat y_{\Phi qu}^{22}=0$. Clearly, generating an appreciable
contribution to $C_9^\text{univ.}$ is excluded
by di-jet searches in this scenario.

\section{Conclusions}\label{sec:concl}

Motivated by the updated measurements of the theoretically clean lepton flavour universality tests $R_K$ and $R_{K^*}$ by the LHCb and Belle experiments, as well as by additional measurements, notably of $B_s \to \mu \mu$ by the ATLAS collaboration,
we have updated the global EFT analysis of new physics in $b\to s\ell\ell$ transitions.
A new-physics effect in the semi-muonic Wilson coefficient $C_9^{bs\mu\mu}$
continues to give a much improved fit to the data compared to the SM.
However, compared to previous global analyses, we find that there is now also a preference for a non-zero value of the semi-muonic Wilson coefficient $C_{10}^{bs\mu\mu}$, mostly driven by the global combination of the $B_s\to\mu^+\mu^-$ branching ratio including the ATLAS measurement. The single-coefficient scenario giving the best fit to the data is the one where $C_9^{bs\mu\mu}=-C_{10}^{bs\mu\mu}$, which is known to be well suited to UV-complete interpretations, and indeed is predicted in several new-physics models with tree-level mediators coupling dominantly to left-handed fermions.

We have also studied the possibility of a simultaneous interpretation of the $b\to s\ell\ell$ data and the discrepancies in $b\to c\tau\nu$ transitions in the framework of a global likelihood in SMEFT Wilson coefficient space. We find one especially compelling scenario, characterised by new physics in all-left-handed semitauonic four-fermion operators. These operators can explain directly the discrepancies in $b\to c\tau\nu$ transitions, and, at the same time, radiatively induce a lepton flavour universal contribution to the $b\to s\ell\ell$ Wilson coefficients. An additional nonzero semimuonic Wilson coefficient then allows accommodating the $R_{K^{(*)}}$ discrepancies. Such picture can be quantitatively realized in the context of the $U_1$ leptoquark simplified model, and we find that indeed an excellent description of the data can be obtained, including the deviations in $b\to c\tau\nu$ transitions.

Another logical possibility to generate a lepton flavour universal NP effect in $C_9$ is via RG effects from a four-quark operator. We have investigated this possibility in the SMEFT and in simplified tree-level models. We find that the only potentially viable setup is a colour-octet scalar. Due to its TeV-scale mass and large coupling to quarks, it is strongly constrained by di-jet resonance searches at the LHC and can be tested in the near future.

Our study illustrates how the theoretical picture has evolved as a consequence of crucial measurement updates, and how this picture stays coherent in spite of the numerous constraints. We collect in Table \ref{tab:prediction} a number of predictions directly related to the discussed SMEFT scenario. The situation will only get more exciting due to the host of new analyses using the full Run-2 data set, as well as the Belle-II data set, to which we look forward.

\paragraph{Note added:} The results of this work have first been presented
at the Moriond conference \cite{MoriondTalk} on March 22, 2019.
On the same day, several preprints with overlapping scope
\cite{Alguero:2019ptt,Ciuchini:2019usw,Alok:2019ufo}
have been
submitted to the arXiv and appeared on March 25, one day before our preprint.

\section*{Acknowledgements}
D.S. warmly thanks the organisers of the Rencontres de Moriond 2019 on ``Electroweak Interactions and Unified Theories'' for the opportunity to present the results in this paper prior to their appearance on the arXiv. We also acknowledge useful remarks from Ben Allanach and Sébastien Descotes-Genon.
The  work  of  D. S.  and  J. A.  is supported by the DFG cluster of excellence “Origin and Structure of the Universe”.
The research of W. A. is supported by the National Science Foundation under Grant No. NSF 1912719.
The numerical analysis has been carried out on the computing facilities of the Munich Computational Center for Particle and Astrophysics (C2PAP).

\appendix

\section{Combination of $B_q\to\mu^+\mu^-$ measurements}\label{app:bsmumu}

In this appendix we discuss our procedure of combining the measurements by ATLAS, CMS, and LHCb of the branching ratios of $B^0\to\mu^+\mu^-$ and $B_s\to\mu^+\mu^-$ \cite{Chatrchyan:2013bka,CMS:2014xfa,Aaij:2017vad,Aaboud:2018mst}.

\begin{figure}
\centering
\includegraphics[width=0.65\textwidth]{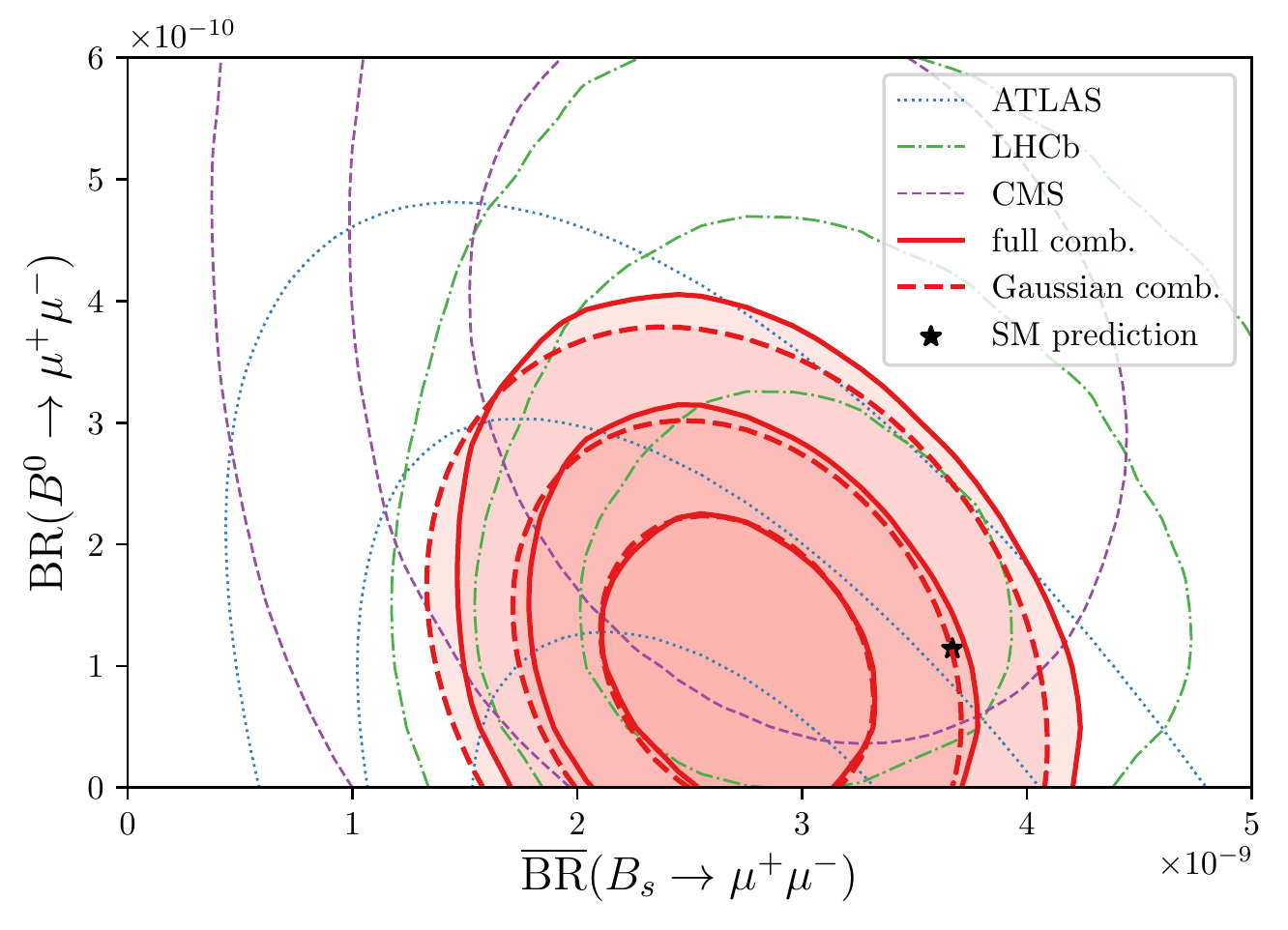}
\caption{Two-dimensional likelihood contours in the space of the $B^0\to\mu^+\mu^-$ and $B_s\to\mu^+\mu^-$ branching ratios from individual measurements (thin contours), the naive combination (thick solid contours), and the Gaussian approximation to it (thick dashed contours), compared to the SM central values.}
\label{fig:Bsmumu}
\end{figure}

In all three cases, the measurements are correlated, since the $B^0$ and $B_s$ have a similar mass, such that the signal regions in dimuon invariant mass squared overlap. ATLAS and CMS provide two-dimensional likelihood contours, from which we interpolate the two-dimensional likelihoods, while LHCb directly provides the two-dimensional likelihood numerically. The three resulting likelihoods are shown as thin lines in Fig.~\ref{fig:Bsmumu} and compared to the SM central values.

Next, we assume the three experiments to be uncorrelated (which we assume to be a good approximation given the dominance of statistical uncertainties) and combine the two-dimensional likelihoods by multiplying them. The resulting contour is also shown in Fig.~\ref{fig:Bsmumu}.

For our global likelihood in Wilson coefficient space, we need to make an additional approximation, namely that the experimental likelihood is approximately Gaussian (see \cite{Aebischer:2018iyb} for a discussion).
Thus we fit a two-dimensional Gaussian to the product likelihood. The resulting contours are shown as thick dashed lines in Fig.~\ref{fig:Bsmumu}.

Since throughout our numerical analysis, we never consider NP effects in $b\to d\mu\mu$ transitions, it is also interesting to compare the combined confidence regions for the $B_s\to\mu^+\mu^-$ branching ratio, fixing $B^0\to\mu^+\mu^-$ to its SM central value or profiling over it. We find
\begin{align}
\overline{\text{BR}}(B_s\to\mu^+\mu^-)
&= (2.67^{+0.45}_{-0.35}) \times 10^{-9}
&& \text{BR}(B^0\to\mu^+\mu^-) \text{ profiled,}
 \\
 \overline{\text{BR}}(B_s\to\mu^+\mu^-)
 &= (2.65^{+0.46}_{-0.33}) \times 10^{-9}
 && \text{BR}(B^0\to\mu^+\mu^-) \text{ SM-like.}
\end{align}
We stress that the similarity of these two numbers is not trivial, as for individual measurements, especially the CMS and ATLAS ones, the best-fit value for $\text{BR}(B^0\to\mu^+\mu^-)$ deviates strongly from the SM prediction.
Conversely, for $B^0\to\mu^+\mu^-$ we get
\begin{align}
{\text{BR}}(B^0\to\mu^+\mu^-)
&= (1.00^{+0.86}_{-0.57}) \times 10^{-10}
&& \overline{\text{BR}}(B_s\to\mu^+\mu^-) \text{ profiled,}
 \\
 {\text{BR}}(B^0\to\mu^+\mu^-)
 &= (0.57^{+0.86}_{-0.36}) \times 10^{-10}
 && \overline{\text{BR}}(B_s\to\mu^+\mu^-) \text{ SM-like.}
\end{align}
The values for the Gaussian approximation only differ from these numbers in a negligible way.

Compared to the SM predictions\footnote{For the SM values, we have
used \flavio{} v1.3 with default settings.
The $B_s\to\mu^+\mu^-$ branching ratio refers to the time-integrated one, see Refs. \cite{Bobeth:2013uxa,Beneke:2017vpq} for state-of-the-art discussions on the SM uncertainty.
The decay constants are taken from the
2019 FLAG average with $2+1+1$ flavours \cite{Aoki:2019cca},
$V_{cb}=0.04221(78)$ from inclusive decays,
and $V_{ub}=0.00373(14)$ from $B\to\pi\ell\nu$.},
\begin{align}
  \overline{\text{BR}}(B_s\to\mu^+\mu^-)_\text{SM}
  &= (3.67\pm0.15) \times 10^{-9},
 \\
 {\text{BR}}(B^0\to\mu^+\mu^-)_\text{SM}
 &= (1.14\pm0.12) \times 10^{-10},
\end{align}
we then find the following one-dimensional pulls\footnote{Here, the ``one-dimensional pull'' is $-2$ times the logarithm of the likelihood ratio at the SM vs.\ the experimental point, after the experimental uncertainties have been convolved with the covariance of the SM uncertainties.}:
\begin{itemize}
  \item if both branching ratios are SM-like, $2.3\sigma$\footnote{Converting the likelihood ratio to a pull with two degrees of freedom, one obtains $1.8\sigma$; this is why the star in Fig.~\ref{fig:Bsmumu} is inside the $2\sigma$ contour.},
  \item if $B_s\to\mu^+\mu^-$ is SM-like and
  $B^0\to\mu^+\mu^-$ profiled over, $2.3\sigma$,
  \item if $B^0\to\mu^+\mu^-$ is SM-like and
  $B_s\to\mu^+\mu^-$ profiled over, $0.2\sigma$.
\end{itemize}

\section{Global likelihood of the Standard Model}

In view of the sizable pulls in various NP scenarios considered in this work, an
interesting question is how good the overall agreement of the SM
with the data is. To this end, we consider the value of the global likelihood at the SM point, $L(\vec{0})$, or
$\chi^2_\text{SM}\equiv -2\ln L(\vec{0})$.
Here, the normalization of the likelihood is such that $\chi^2=0$
corresponds to the case where all measurements are in exact agreement with the theoretical predictions.
By means of Wilks' theorem, this $\chi^2$ value can be translated to a $p$-value, quantifying the goodness of fit.

However, the intepretation of this global $\chi^2$ value is not straightforward
as it is subject to several ambuguities.
\begin{itemize}
  \item Since the global likelihood contains many observables not sensitive to the Wilson coefficients that we consider in our analysis, which focuses on discrepancies in $B$ physics, this $p$-value depends on the number of observables included in the test.
  \item  The likelihood contains measurements that are actually averages (e.g.\ by PDG or HFLAV) of several measurements, such that the number of
  degrees of freedom is reduced and a constant contribution to the $\chi^2$ coming from a tension between the averaged measurements is concealed.
  \item Due to the statistical approach, where theory uncertainties are combined
  with the experimental ones and no explicit nuisance parameters are present,
  the contribution to the absolute $\chi^2$ from different sectors can depend
  sensitively on the central values chosen for the parameters. For instance,
  a lower $V_{cb}$ value would lead to better agreement of $b\to s\ell\ell$
  branching ratios (via lower $|V_{tb}V_{ts}^*|$) but worse agreement with
  $b\to c\ell\nu$ transitions. While this issue does not affect the $\Delta\chi^2$
  used in our numerical analysis, it makes the interpretation of the absolute
  $\chi^2$ for individual sectors difficult.
\end{itemize}

With these caveats in mind, we provide in table \ref{tab:pvalues} the absolute $\chi^2$
values obtained with \smelli{} for various subsets of observables.
The number $N_\text{obs}$ in the first column refers to the number
of \textit{observations}, i.e.\ independent measurements of an observable,
which is greater than or equal to the number of \textit{observables}. In our case, $N_{\rm obs}$ is to be interpreted as the $\chi^2$'s number of degrees of freedom. Through the \texttt{chi2\char`_dict} method introduced in v1.3, it is possible to explore the $\chi^2$ also for different observable sets or parameter inputs.

\begin{table}[tbp]
  \centering
  \begin{tabular}{lccc}
    \toprule
    Observables & $N_\text{obs}$ &  $\chi^2$ & $p$-value [\%]\\
    \midrule
    $b\to s\ell\ell$ & 127 & 126.7 & 49 \\
    $+R_{K^{(*)}}+D^{\mu e}_{P_i'}$ & 138 & 149.9 & 23 \\
    $+B_s\to\mu^+\mu^-+b\to s\gamma$ & 143 & 155.9 & 22 \\
    $+\Delta F=2$ & 149 & 193.4 & 0.8 \\
    $+b\to c\tau\nu$ & 218 & 264.7 & 1.7 \\
    all quark flavour & 258 & 301.9 & 3.1\\
    all low-energy & 276 & 308.6 & 8.6\\
    global & 313 & 361.4 & 3.1\\
    \bottomrule
  \end{tabular}
  \caption{Absolute $\chi^2$ values and $p$-values for the global fit and various subsets of observables.}
  \label{tab:pvalues}
\end{table}

\section{$C_9$ vs. $C_9 = -C_{10}$} \label{app:nonzero_C10}

As already discussed in Sec.~\ref{sec:1d} and summarized in Table \ref{tab:1d}, we find a preference for the $C_9^{bs\mu\mu} = -C_{10}^{bs\mu\mu}$ scenario over the $C_9^{bs\mu\mu}$-only scenario.
Since the opposite result was found in previous analyses, like e.g.~\cite{Altmannshofer:2017fio}, some further comments are in order.
The change in preference is not related to the updated measurements of $R_K$ and $R_{K^*}$ but can be traced back to the inclusion of several observables into the likelihood that were not considered in~\cite{Altmannshofer:2017fio}.\footnote{%
Also other post-Moriond 2019 global fits that appeared slightly before and after the preprint of this paper use different sets of observables and found a preference for the $C_9^{bs\mu\mu}$-only scenario \cite{Alguero:2019ptt,Ciuchini:2019usw,Alok:2019ufo,Kowalska:2019ley,Arbey:2019duh}.
}
Among the newly included observables mentioned in Sec.~\ref{sec:setup}, the following are in particular relevant here.

\begin{figure}
\centering
\includegraphics[width=0.5\textwidth]{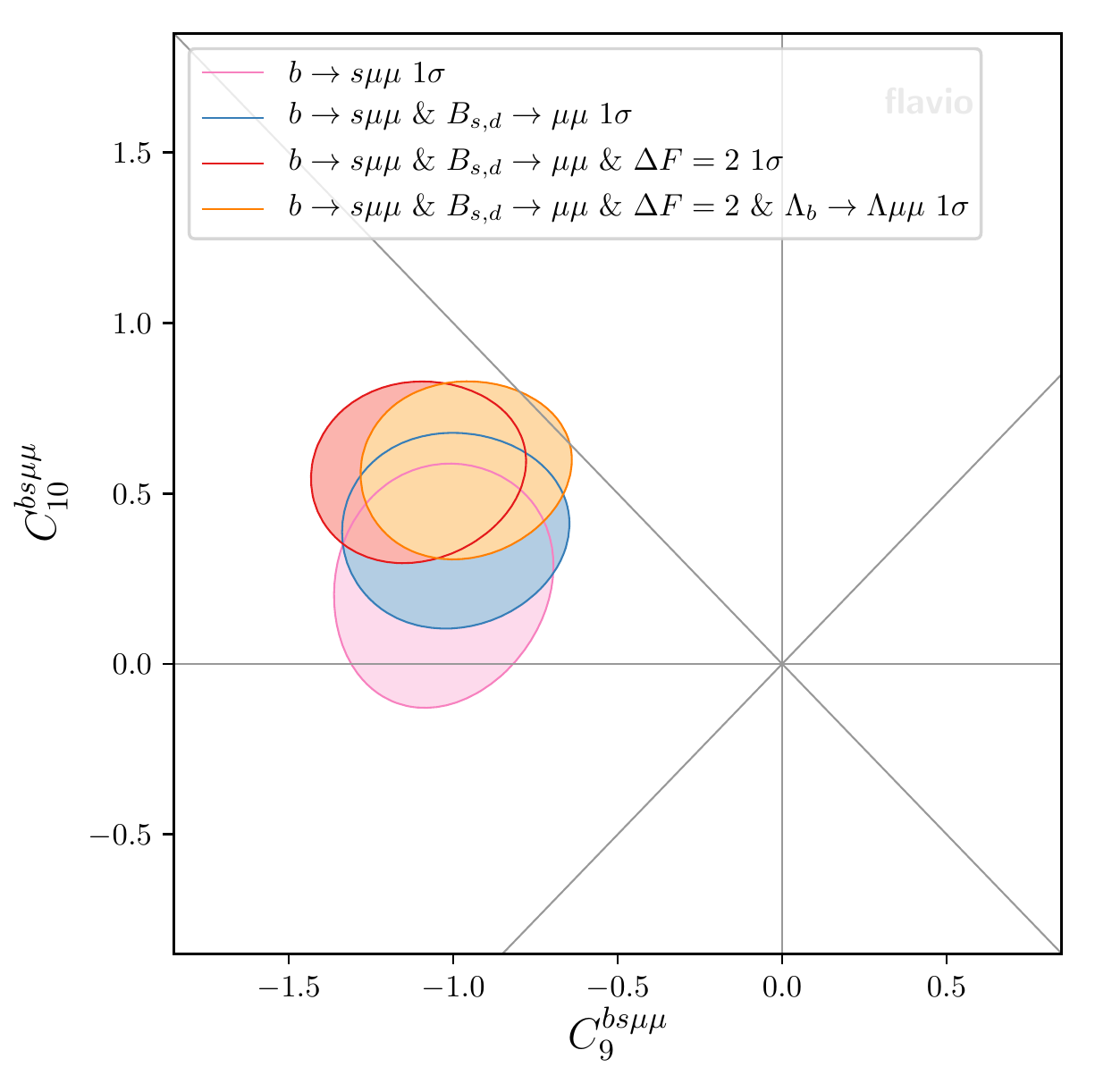}%
\caption{The $1\sigma$ contour for the $b \to s \mu \mu$ likelihood, as a function of the observables detailed in the legend. The last contour in the legend, displayed in yellow, coincides with the corresponding one in Fig. \ref{fig:C9C10} (left).}
\label{fig:nonzero_C10}
\end{figure}

\begin{itemize}

\item The new ATLAS measurement of $\text{BR}(B_{s,d}\to\mu^+\mu^-)$  \cite{Aaboud:2018mst}, which we combine with the measurements by LHCb and CMS \cite{Chatrchyan:2013bka,CMS:2014xfa,Aaij:2017vad} using the procedure detailed in Appendix~\ref{app:bsmumu}.

\item The LHCb measurements of $\Lambda_b\to \Lambda\ell^+\ell^-$ observables \cite{Aaij:2018gwm}.

\item
$\Delta F = 2$ observables,\footnote{%
For the full list of observables, see~\cite{Aebischer:2018iyb}.} most importantly $\varepsilon_K$ and $\Delta M_{s}$.
Like the theoretical predictions for $\text{BR}(B_{s,d}\to\mu^+\mu^-)$, also the predictions for these observables depend on the $B_{s,d}$-meson decay constants $F_{B_{s,d}}$ and the CKM parameters.
As detailed in~\cite{Aebischer:2018iyb}, these nuisance parameters are ``integrated out'' and enter the likelihood in terms of a covariance matrix that contains all experimental and theoretical uncertainties together with their correlations.
Due to these correlations, experimental information on $\Delta F = 2$ observables can reduce the theoretical uncertainties of $\text{BR}(B_{s,d}\to\mu^+\mu^-)$.
Such a reduction of theoretical uncertainties has been discussed explicitly for $\Delta M_{s,d}$ and $\text{BR}(B_{s,d}\to\mu^+\mu^-)$ in models with minimal flavour violation in~\cite{Buras:2003td}.
Via the correlations, $\Delta F = 2$ observables can have an impact on a fit to the Wilson coefficients $C_9^{bs\mu\mu}$ and $C_{10}^{bs\mu\mu}$ even if they do not directly depend on these coefficients themselves.

\end{itemize}
To illustrate the effect of including the above listed observables into the likelihood, we show 1$\sigma$ contours for several fits to subsets of observables in Fig.~\ref{fig:nonzero_C10}.
\begin{itemize}
 \item The pink contour corresponds to a fit including only $b \to s \mu \mu$ observables excluding $\text{BR}(B_{s,d}\to\mu^+\mu^-)$.
 This fit is equivalent to the one shown in~\cite{Altmannshofer:2017fio} and clearly prefers the $C_9^{bs\mu\mu}$-only scenario over the $C_9^{bs\mu\mu} = -C_{10}^{bs\mu\mu}$ one.
 \item The blue contour is obtained from a fit that also includes $\text{BR}(B_{s,d}\to\mu^+\mu^-)$.
 Due to the slight tension between experimental data and SM prediction of $\overline{\text{BR}}(B_s\to\mu^+\mu^-)$ (cf.\ App.~\ref{app:bsmumu}), a non-zero NP contribution to $C_{10}^{bs\mu\mu}$ is preferred.
 In this case, the $C_9^{bs\mu\mu}$-only scenario and the $C_9^{bs\mu\mu} = -C_{10}^{bs\mu\mu}$ scenario perform similarly well and neither of them is clearly preferred.
 \item The red contour corresponds to a fit that additionally includes $\Delta F=2$ observables.
 As described above, taking them into account can have an effect on the uncertainties of $B_s\to\mu\mu$ and we find that this leads to an increased preference for a non-zero NP contribution to $C_{10}^{bs\mu\mu}$.
 In this case, $C_9^{bs\mu\mu} = -C_{10}^{bs\mu\mu}$ is favoured over $C_9^{bs\mu\mu}$-only.
 \item The orange contour is obtained by adding $\Lambda_b\to \Lambda\ell^+\ell^-$ observables to the fit.
 While this does not further increase the preference for a non-zero NP contribution to $C_{10}^{bs\mu\mu}$, this contour now intersects with the $C_9^{bs\mu\mu} = -C_{10}^{bs\mu\mu}$ line, making this scenario clearly preferred over the $C_9^{bs\mu\mu}$-only one.
\end{itemize}
 Other observables that are correlated with the $b \to s \mu \mu$ observables (e.g. $b\to s\gamma$) only have a marginal impact on the fit. Therefore, the orange contour in Fig.~\ref{fig:nonzero_C10} essentially coincides with the orange contour in Fig.~\ref{fig:C9C10} left.

Note that the main effect of the correlations between the $C_{10}^{bs\mu\mu}$-dependent observables and the $\Delta F = 2$ observables vanishes if the theory predictions of the $\Delta F = 2$ observables exactly equal the corresponding experimental values.
This means that neglecting the effect of the correlated $\Delta F = 2$ observables is equivalent to assuming NP contributions that shift all of them exactly to the experimentally measured values.
Rather than relying on such a strong hypothesis, we assume the theory predictions of $\Delta F = 2$ observables to be SM-like.\footnote{%
Many NP models that explain the deviations in rare $B$ decays also predict a NP contribution to the Wilson coefficient $C_{VLL}^{bsbs}$ of the operator $\left(\bar{s}_L \gamma^\mu b_L\right)\left(\bar{s}_L \gamma^\mu b_L\right)$, which modifies the theory prediction of $\Delta M_s$.
A large class of models predicts $\Delta M_s>\Delta M_s^{\rm SM}$~\cite{Blanke:2006yh,Altmannshofer:2007cs,DiLuzio:2017fdq}.
We have explicitly checked that a contribution to $C_{VLL}^{bsbs}$ that increases $\Delta M_s$ up to its 2$\sigma$ experimental bound can only marginally decrease the preference for a non-zero $C_{10}^{bs\mu\mu}$.
On the other hand, any model that predicts $\Delta M_s<\Delta M_s^{\rm SM}$ would further increase this preference.
Assuming SM-like theory predictions for $\Delta F = 2$ observables, the dominant effect of the correlations is due to $\varepsilon_K$.
}

\section{Di-jet bounds on leptophobic mediators} \label{app:dijet}

\begin{figure}
  \centering
  \includegraphics[width=0.9\textwidth]{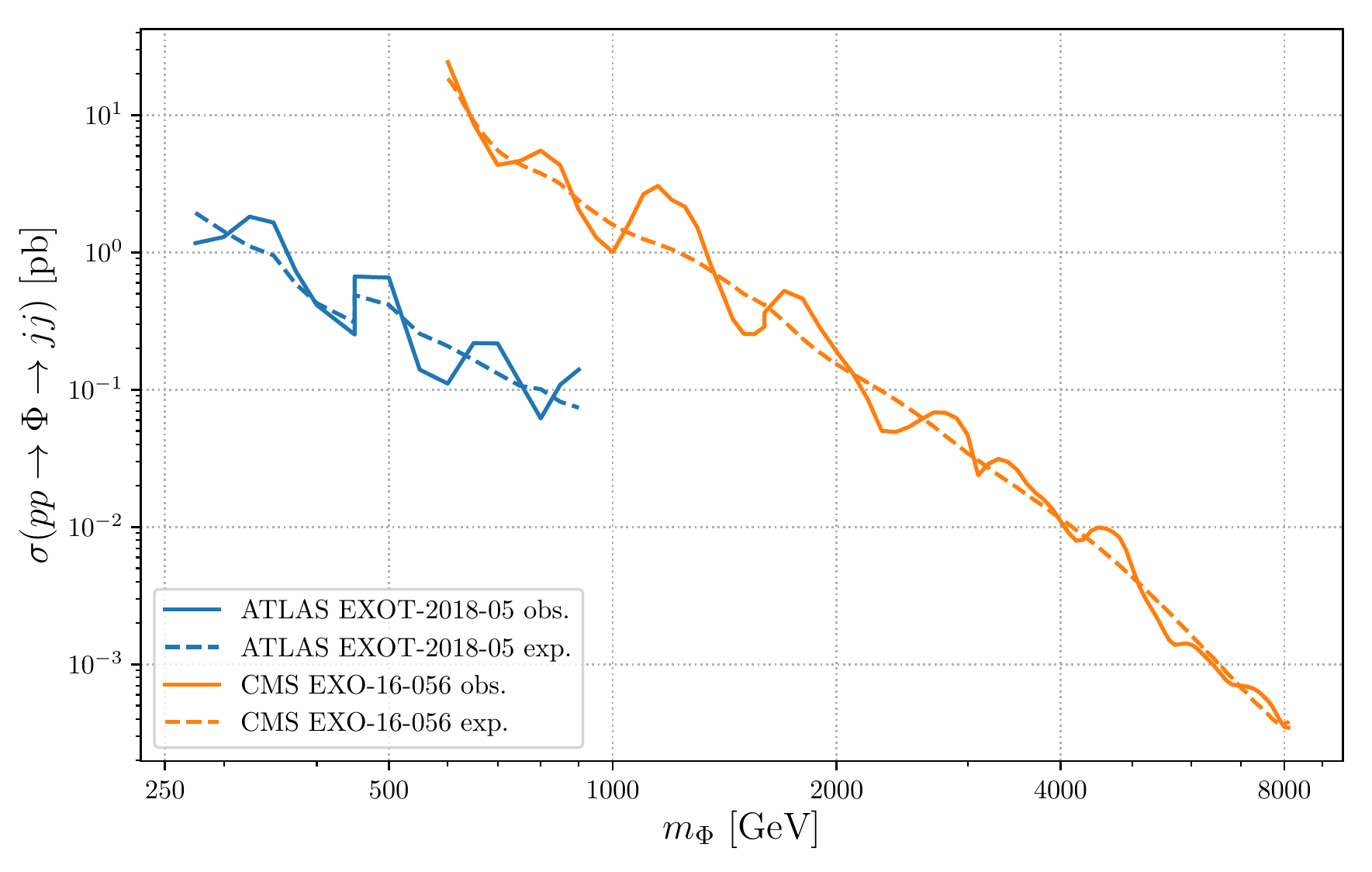}
  \caption{95\% C.L.\ observed and expected limits on the di-jet cross section in 13~TeV $pp$ collisions from
  the two analyses used in Section~\ref{sec:leptophobic}.
  For the assumptions on efficiencies and acceptances see main text.}
  \label{fig:dijets}
\end{figure}

As discussed in Section~\ref{sec:leptophobic},
generating a LFU contribution to $C_9$
radiatively from four-quark operators requires a
fairly light scalar mediator with strong coupling to quarks,
that can lead to a signal in di-jet resonance searches at the LHC. In our numerical analysis, we employ two recent di-jet resonance searches.
\begin{itemize}
  \item A search at $\sqrt{s}=13$~TeV by CMS
  based on 36~fb$^{-1}$ and covering the mass range from 600 to 8100~GeV \cite{Sirunyan:2018xlo}.
  We apply the 95\% C.L.\ constraint on the production cross section times branching ratio times acceptance, assuming 100\% decay into di-jets and a constant acceptance of 57\%.
  \item A search at $\sqrt{s}=13$~TeV by ATLAS
  based on 80~fb$^{-1}$ employing initial state radiation to cover the low-mass range  not covered by CMS \cite{Aaboud:2019zxd}.
  We apply the 95\% C.L.\ constraint on the production cross section times branching ratio times acceptance times efficiency, assuming 100\% decay into di-jets and using the mass-dependent efficiency and acceptance for the $Z'$ model given in the publication.
\end{itemize}
The 95\% C.L.\ bounds under these assumptions are shown in Fig.~\ref{fig:dijets}.

\section{WET and SMEFT scenarios with Wilson coefficient pairs} \label{app:2d}
The aim of this appendix is to present additional projections of the likelihood
onto pairs of Wilson coefficients both in WET and SMEFT.

\begin{figure}[p]
\centering
\includegraphics[width=0.5\textwidth]{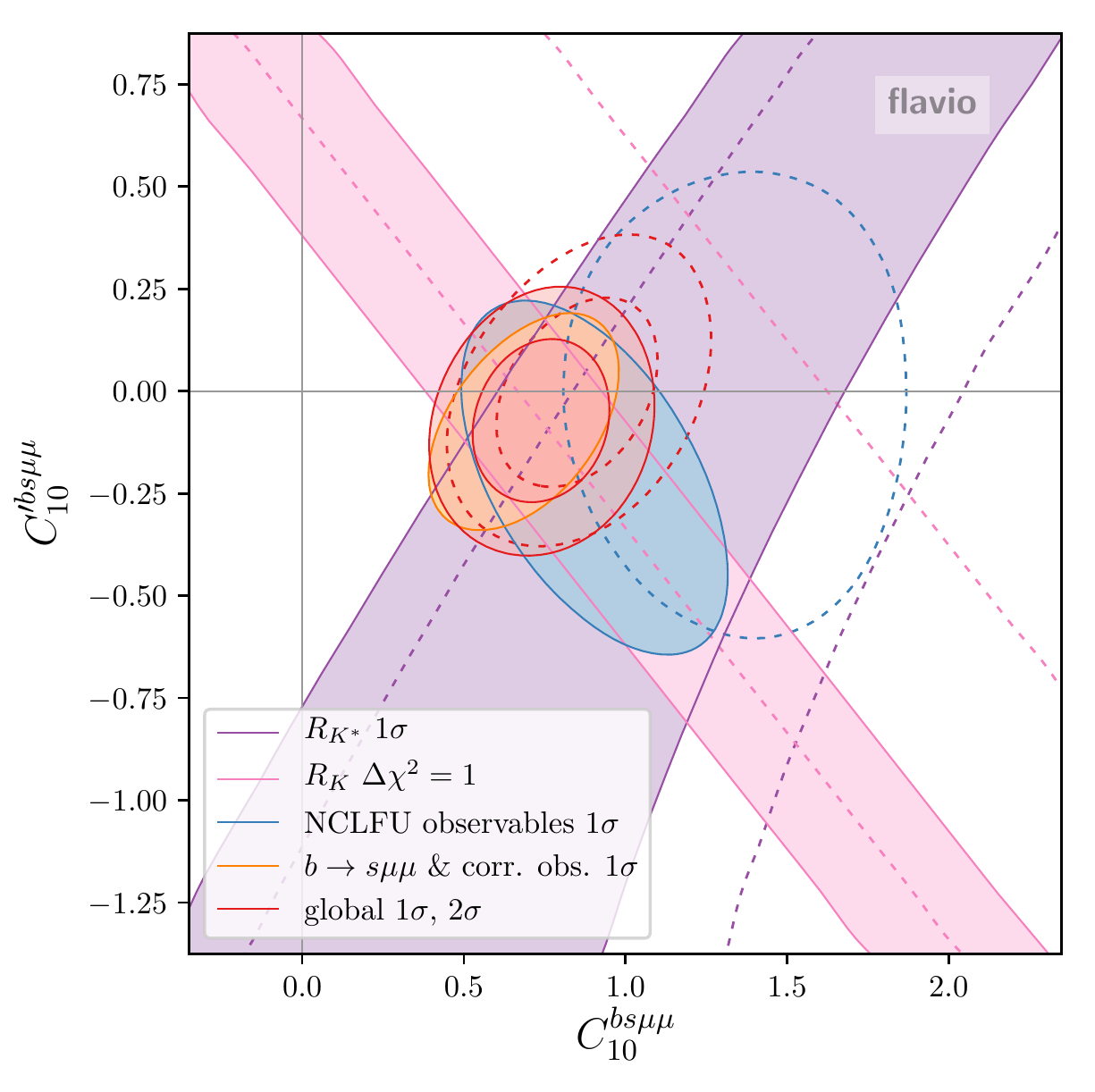}%
\includegraphics[width=0.5\textwidth]{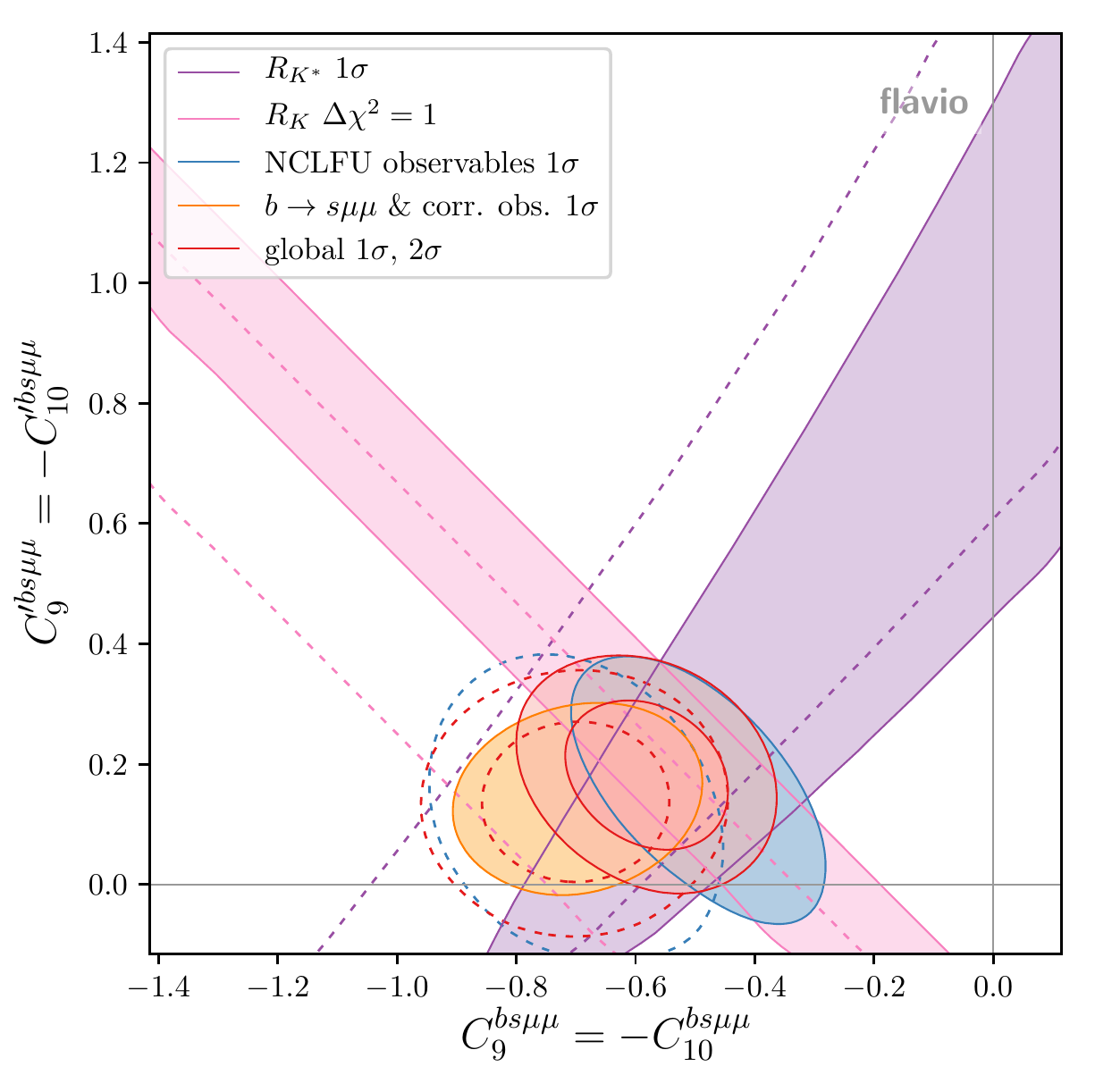}%
\caption{Likelihood contours from NCLFU observables ($R_{K^{(*)}}$ and $D_{P^\prime_{4,5}}$), $b\to s\mu\mu$ observables and the global likelihood
  in the space of $C_{10}^{bs\mu\mu}$ and $C'{}_{10}^{bs\mu\mu}$ (left) and
  in the space of $C_{9}^{bs\mu\mu}=-C_{10}^{bs\mu\mu}$ and $C'{}_{9}^{bs\mu\mu}=-C'{}_{10}^{bs\mu\mu}$ (right).
  All other Wilson coefficients are assumed to vanish.
  Solid (dashed) contours include (exclude) the Moriond-2019 results for $R_{K}$ and $R_{K^{*}}$.}
\label{fig:axial}
\end{figure}

\begin{figure}[p]
\centering
\includegraphics[width=0.5\textwidth]{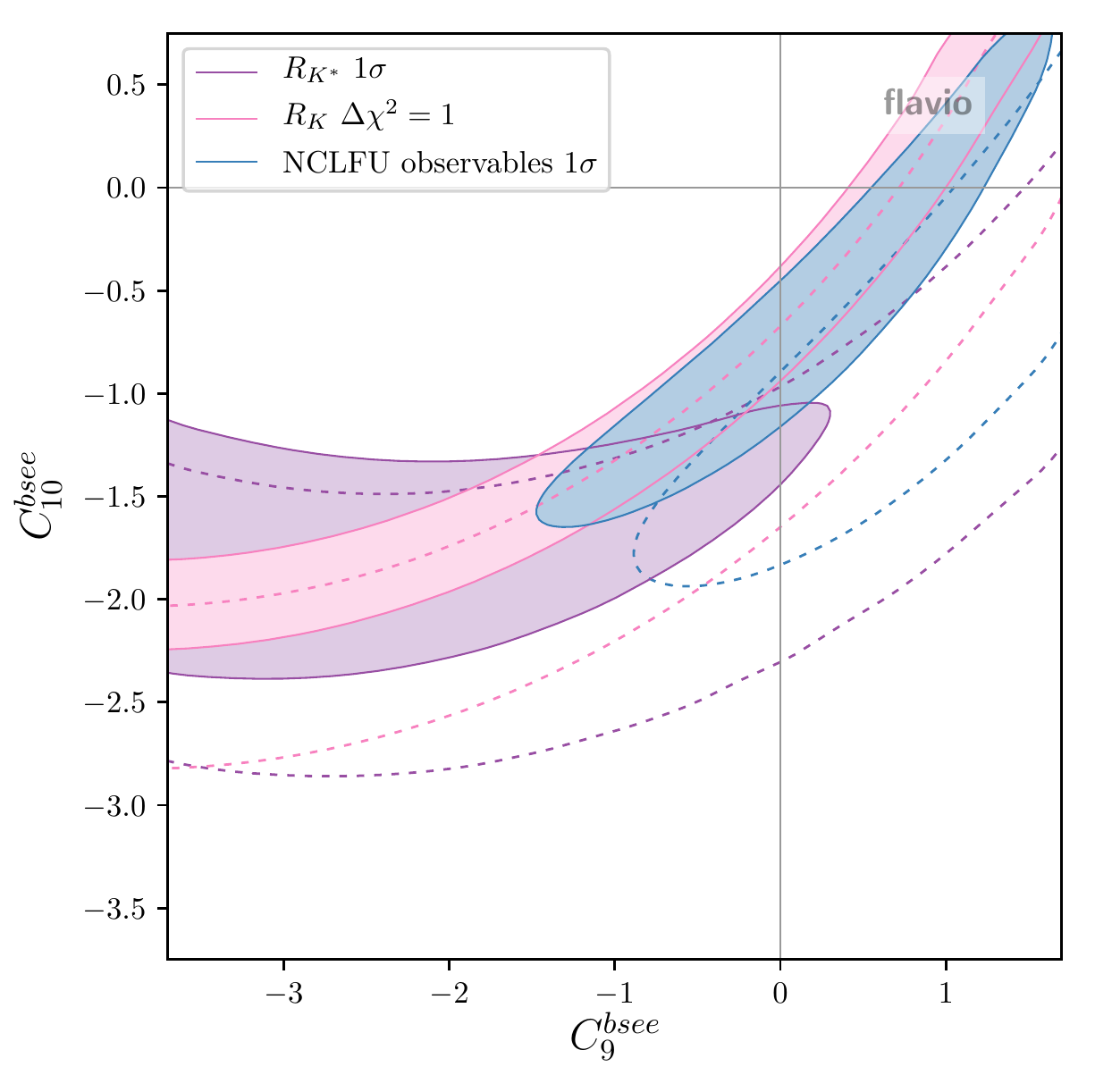}%
\caption{Likelihood contours from NCLFU observables ($R_{K^{(*)}}$ and $D_{P^\prime_{4,5}}$) in the space of $C_{9}^{bsee}$ and $C_{10}^{bsee}$.
  All other Wilson coefficients are assumed to vanish.
  Solid (dashed) contours include (exclude) the Moriond-2019 results for $R_{K}$ and $R_{K^{*}}$.}
\label{fig:bsmumu_C9e_C10e}
\end{figure}

\begin{figure}[p]
\centering
\includegraphics[width=0.5\textwidth]{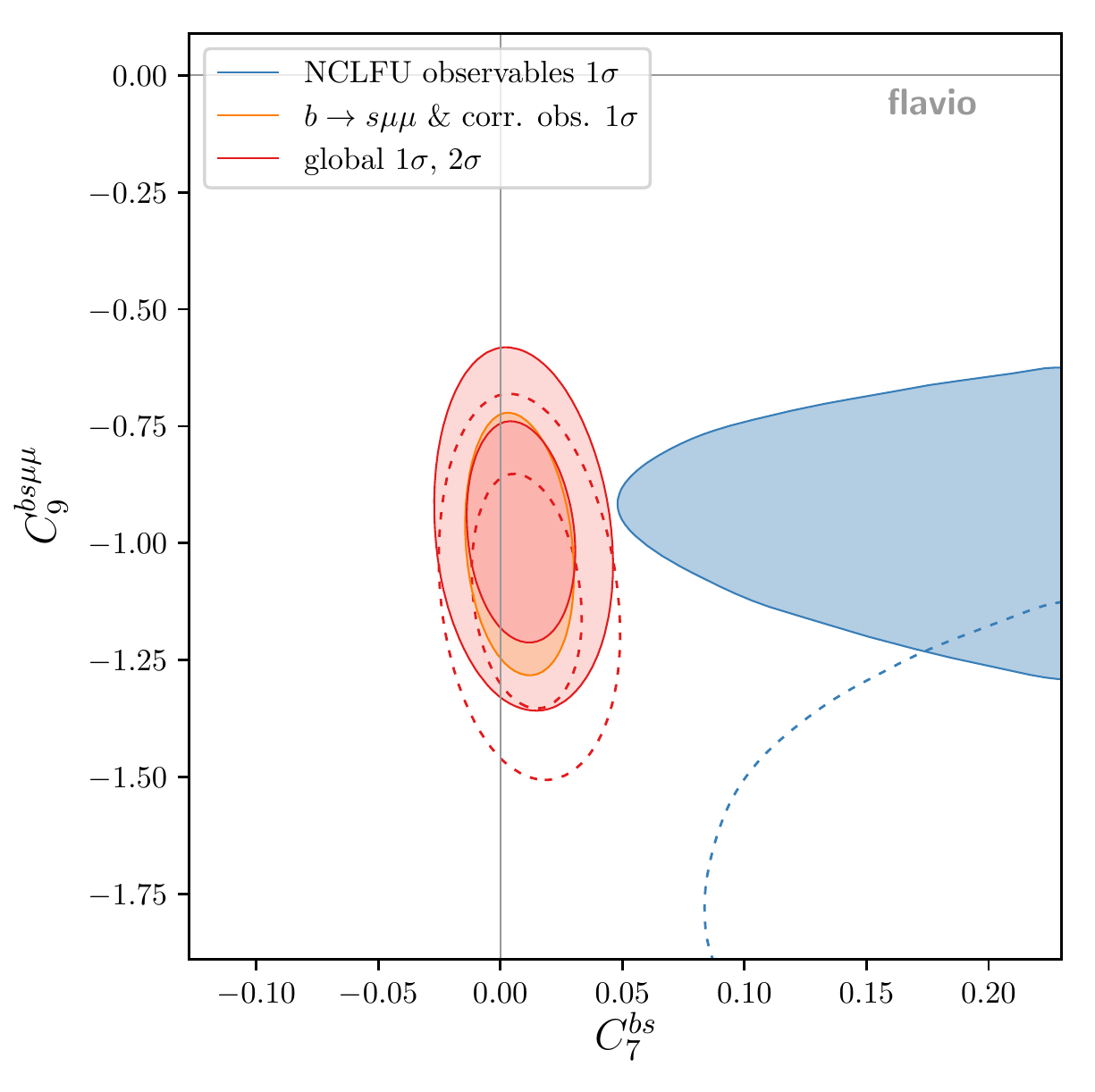}%
\includegraphics[width=0.5\textwidth]{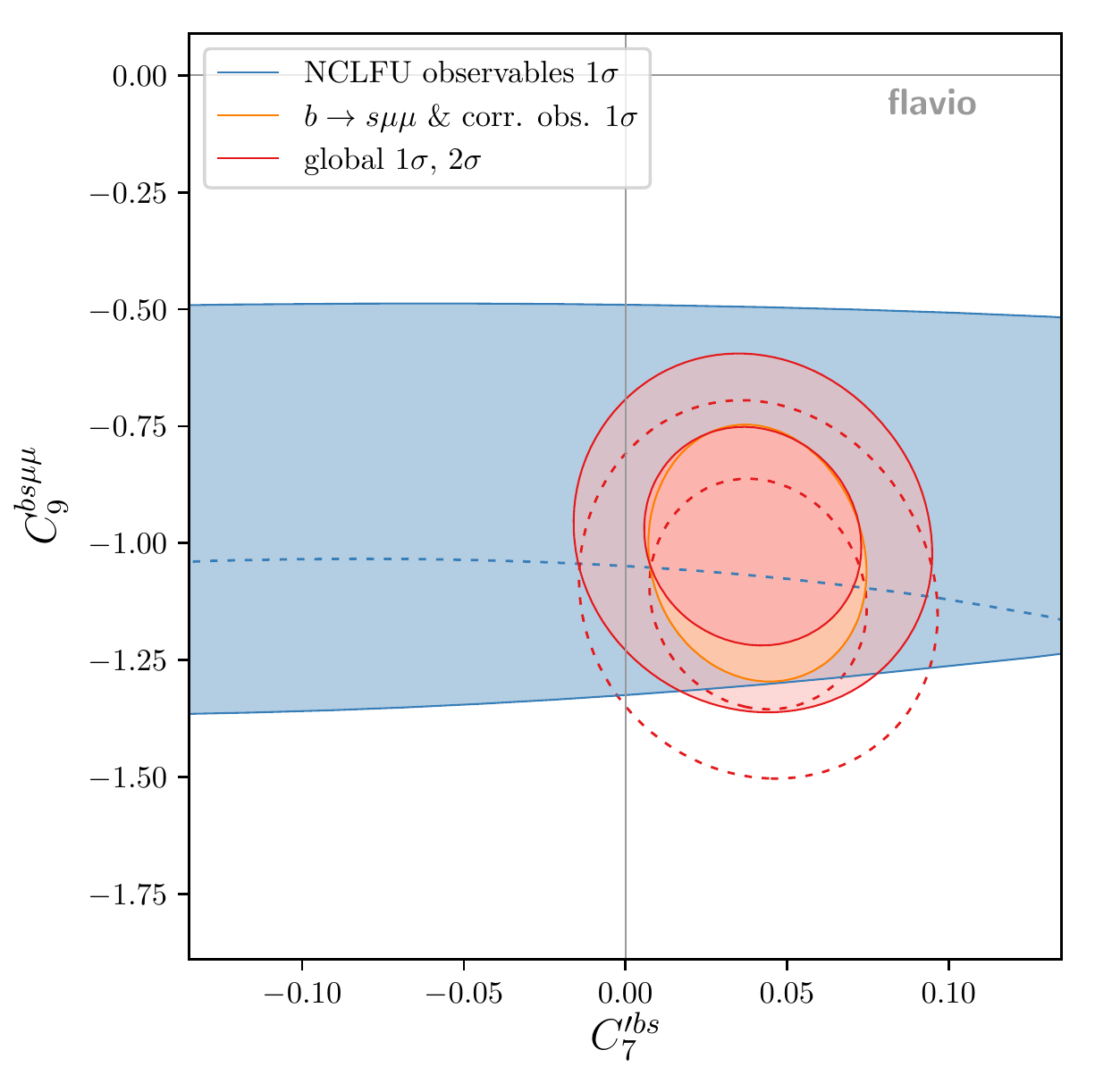}%
\caption{Likelihood contours from NCLFU observables ($R_{K^{(*)}}$ and $D_{P^\prime_{4,5}}$), $b\to s\mu\mu$ observables and the global likelihood
  in the space of $C_{7}^{bs}$ and $C_{9}^{bs\mu\mu}$ (left) and
  in the space of $C'{}_{7}^{bs}$ and $C_{9}^{bs\mu\mu}$ (right).
  All other Wilson coefficients are assumed to vanish.
  Solid (dashed) contours include (exclude) the Moriond-2019 results for $R_{K}$ and $R_{K^{*}}$.}
\label{fig:bsmumu_C7_C9}
\end{figure}

\begin{figure}[p]
\centering
\includegraphics[width=0.5\textwidth]{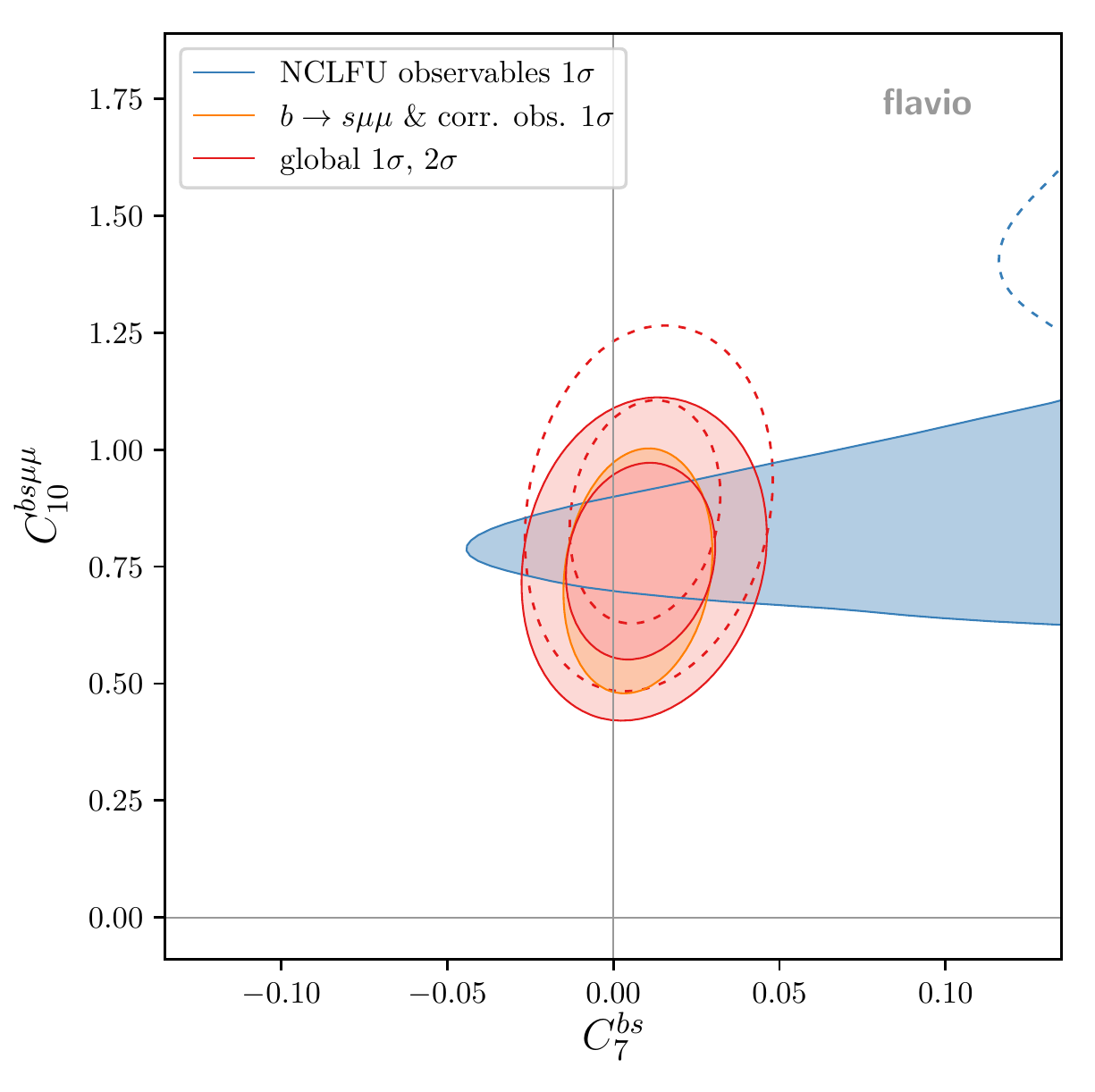}%
\includegraphics[width=0.5\textwidth]{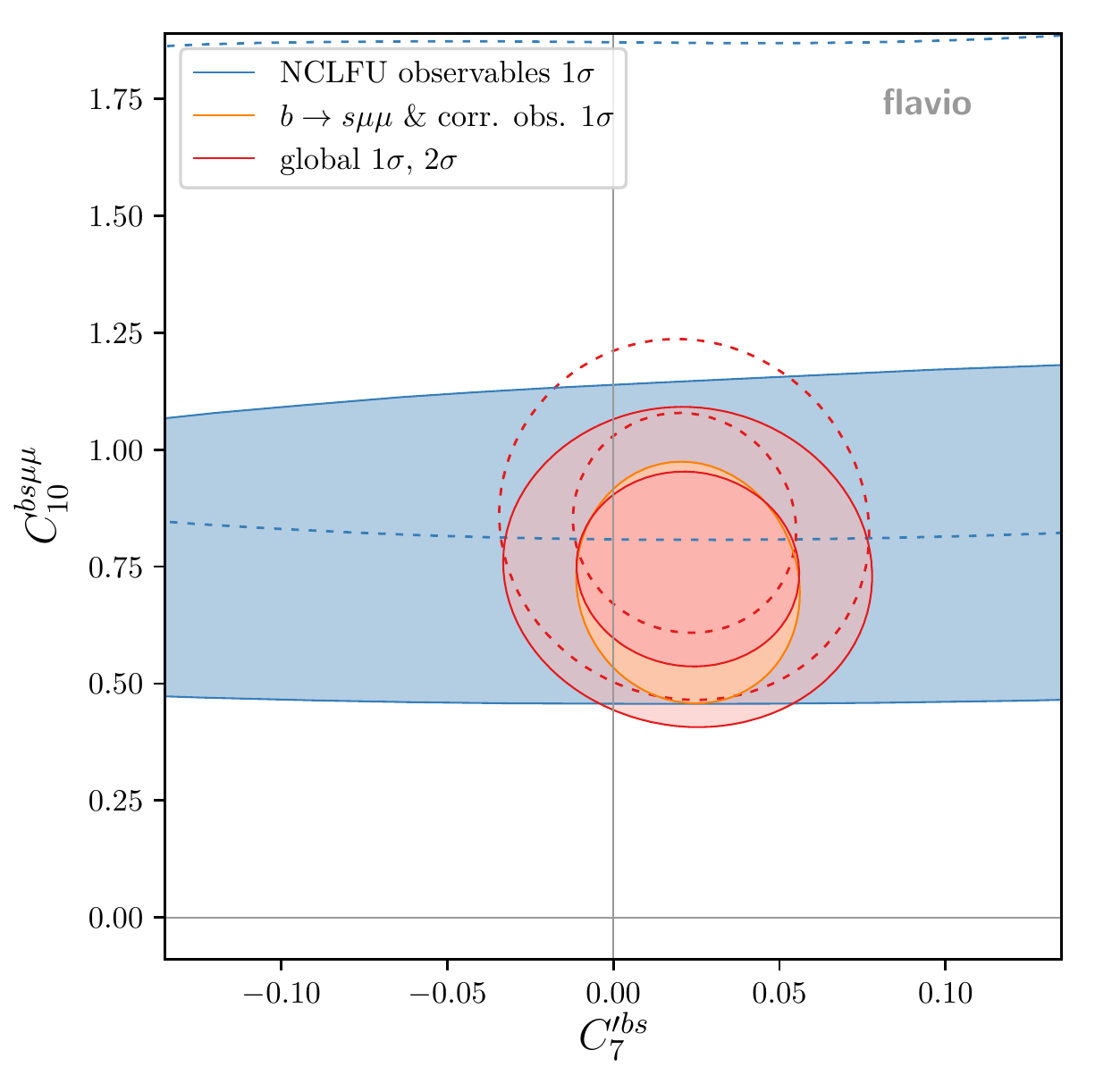}%
\caption{Likelihood contours from NCLFU observables ($R_{K^{(*)}}$ and $D_{P^\prime_{4,5}}$), $b\to s\mu\mu$ observables and the global likelihood
  in the space of $C_{7}^{bs}$ and $C_{10}^{bs\mu\mu}$ (left) and
  in the space of $C'{}_{7}^{bs}$ and $C_{10}^{bs\mu\mu}$ (right).
  All other Wilson coefficients are assumed to vanish.
  Solid (dashed) contours include (exclude) the Moriond-2019 results for $R_{K}$ and $R_{K^{*}}$.}
\label{fig:bsmumu_C7_C10}
\end{figure}

\begin{figure}[p]
\centering
\includegraphics[width=0.5\textwidth]{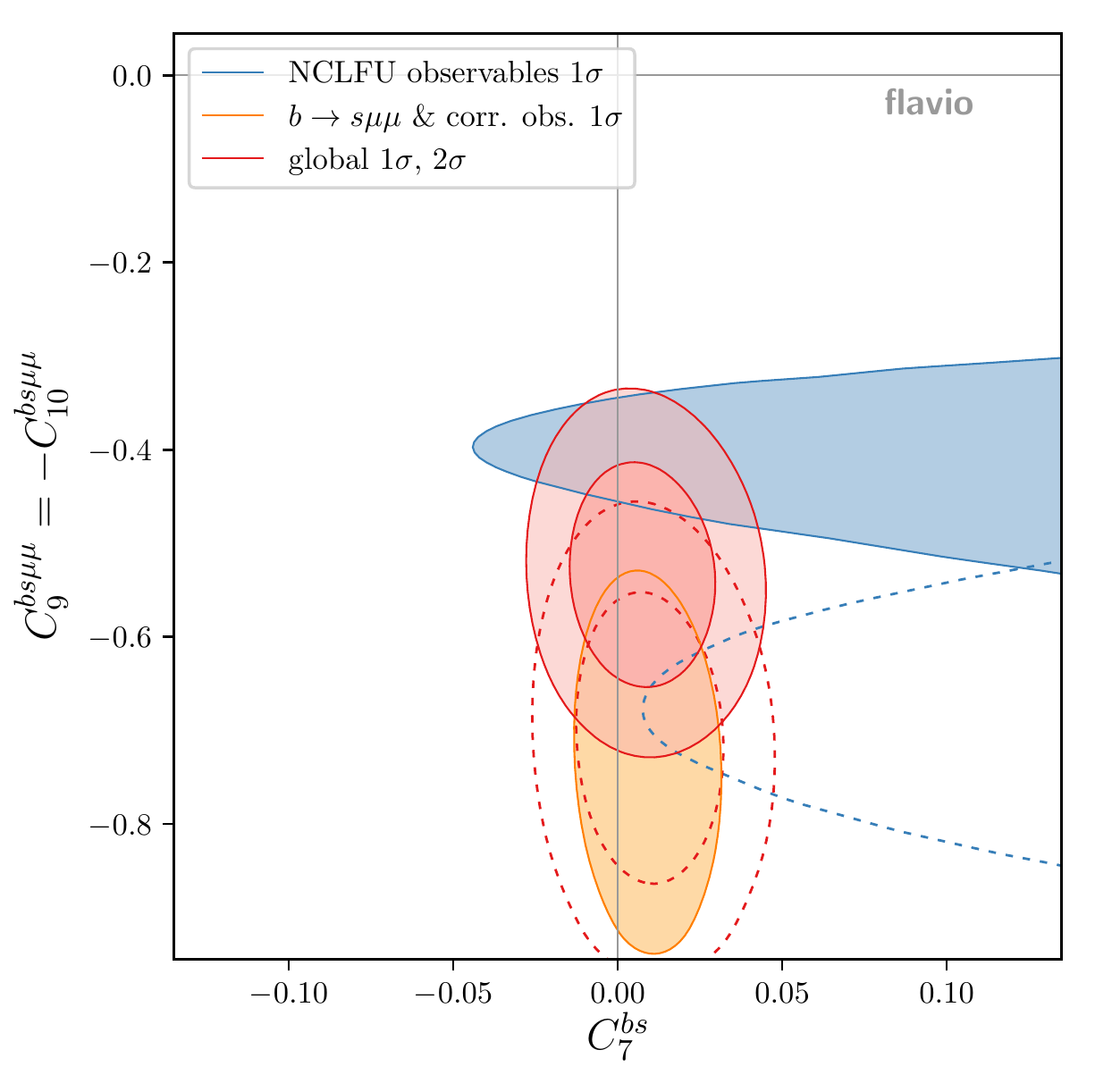}%
\includegraphics[width=0.5\textwidth]{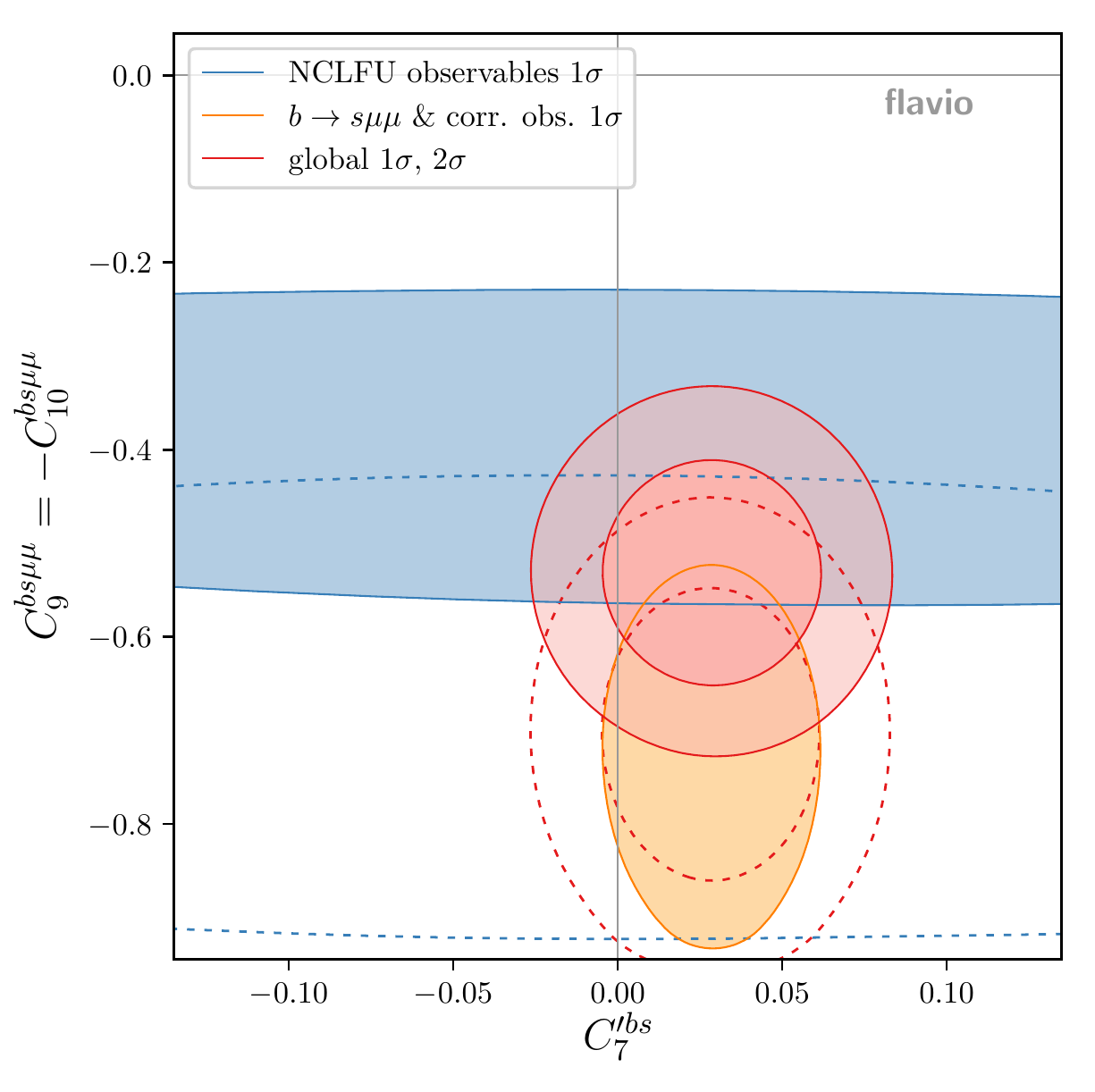}%
\caption{Likelihood contours from NCLFU observables ($R_{K^{(*)}}$ and $D_{P^\prime_{4,5}}$), $b\to s\mu\mu$ observables and the global likelihood
  in the space of $C_{7}^{bs}$ and $C_{9}^{bs\mu\mu}=-C_{10}^{bs\mu\mu}$ (left) and
  in the space of $C'{}_{7}^{bs}$ and $C_{9}^{bs\mu\mu}=-C_{10}^{bs\mu\mu}$ (right).
  All other Wilson coefficients are assumed to vanish.
  Solid (dashed) contours include (exclude) the Moriond-2019 results for $R_{K}$ and $R_{K^{*}}$.}
\label{fig:bsmumu_C7_C9-C10}
\end{figure}

\begin{figure}[p]
\centering
\includegraphics[width=0.5\textwidth]{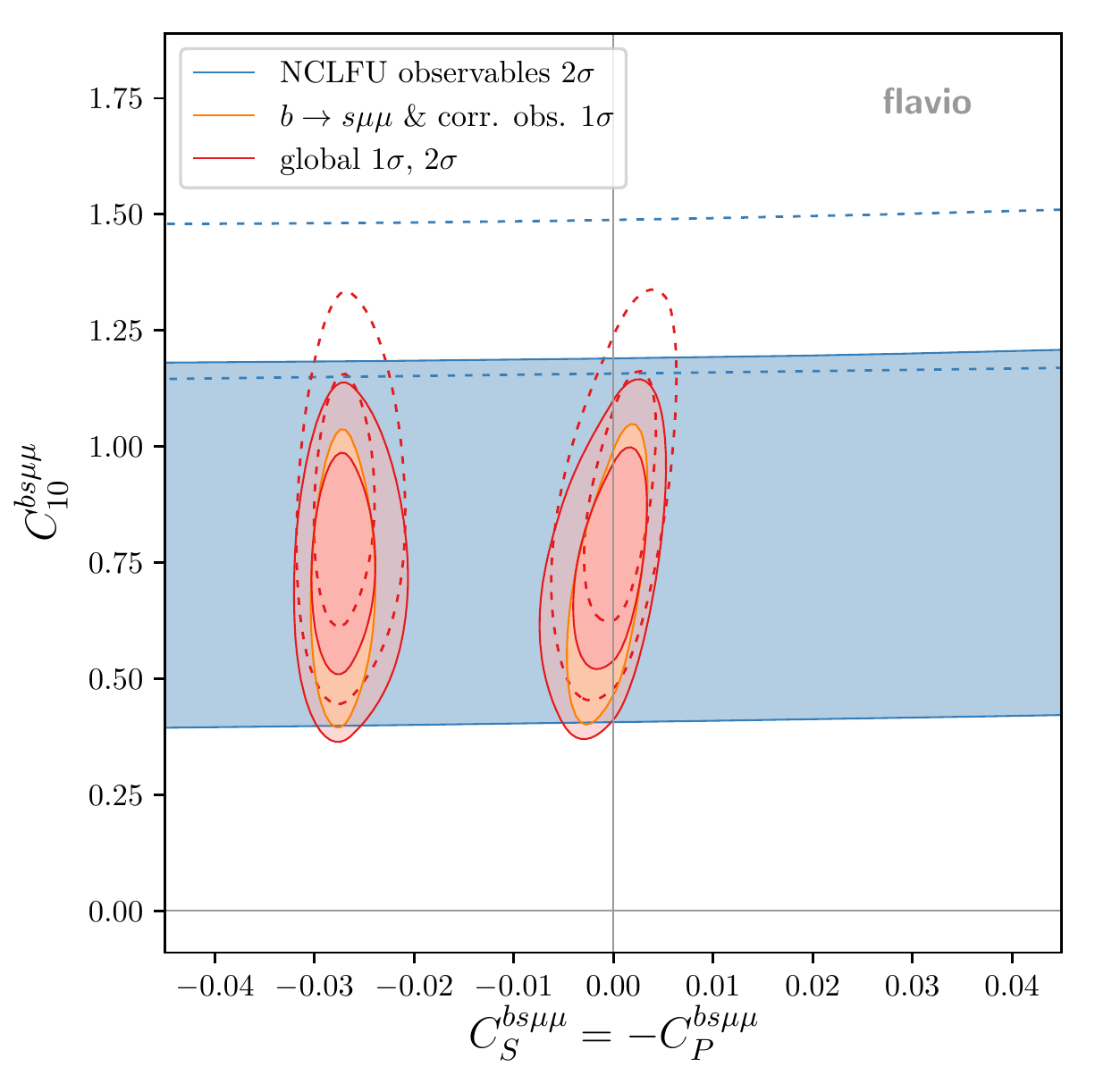}%
\includegraphics[width=0.5\textwidth]{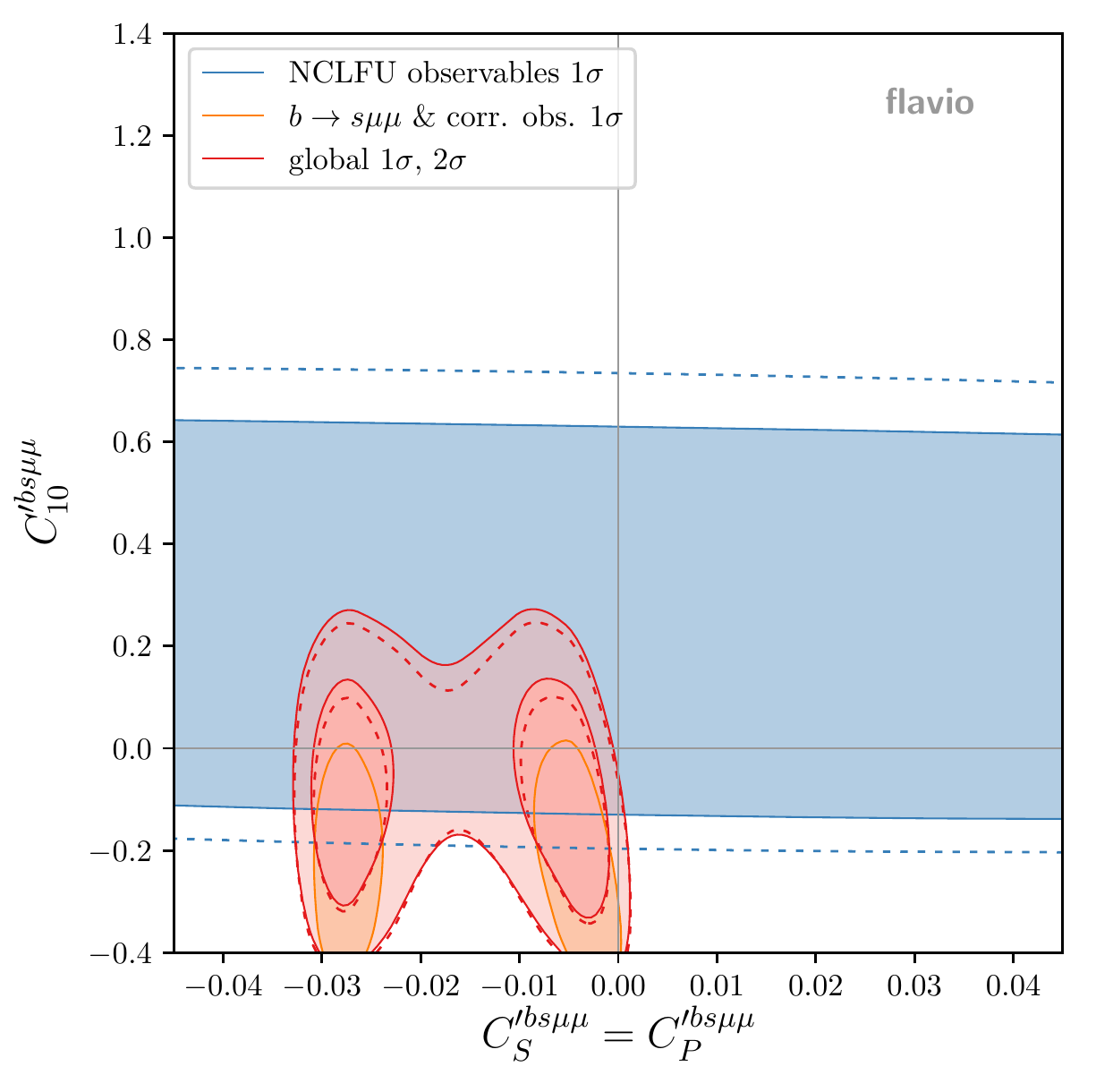}%
\caption{Likelihood contours from NCLFU observables ($R_{K^{(*)}}$ and $D_{P^\prime_{4,5}}$), $b\to s\mu\mu$ observables and the global likelihood
  in the space of $C_{S}^{bs\mu\mu}=-C_{P}^{bs\mu\mu}$ and $C_{10}^{bs\mu\mu}$ (left) and
  in the space of $C_{S}^{\prime bs\mu\mu}=C_{P}^{\prime bs\mu\mu}$ and $C_{10}^{\prime bs\mu\mu}$ (right).
  All other Wilson coefficients are assumed to vanish.
  Solid (dashed) contours include (exclude) the Moriond-2019 results for $R_{K}$ and $R_{K^{*}}$.}
\label{fig:bsmumu_CS_C10}
\end{figure}

\begin{figure}[p]
\centering
\includegraphics[width=0.5\textwidth]{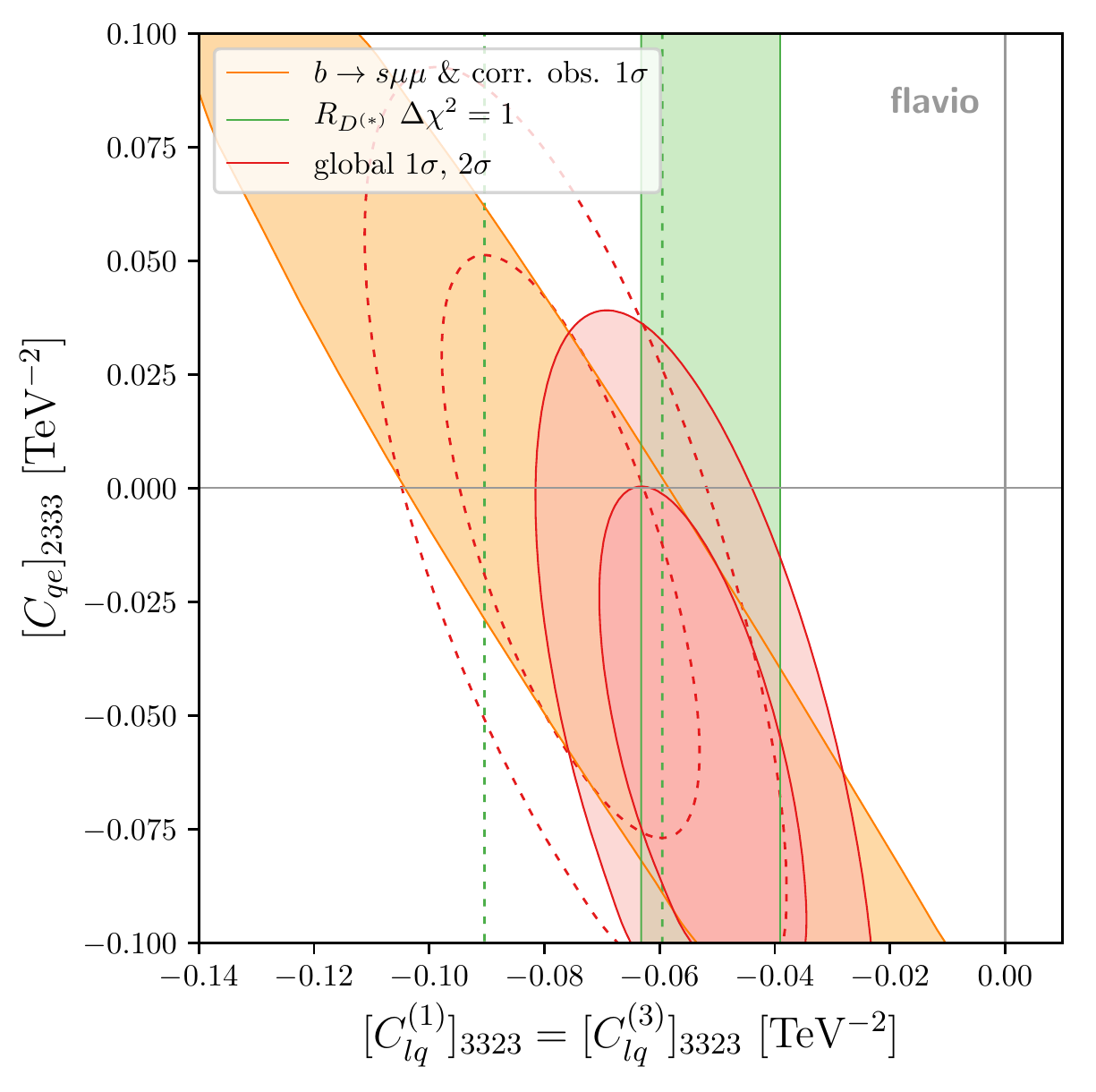}%
\includegraphics[width=0.5\textwidth]{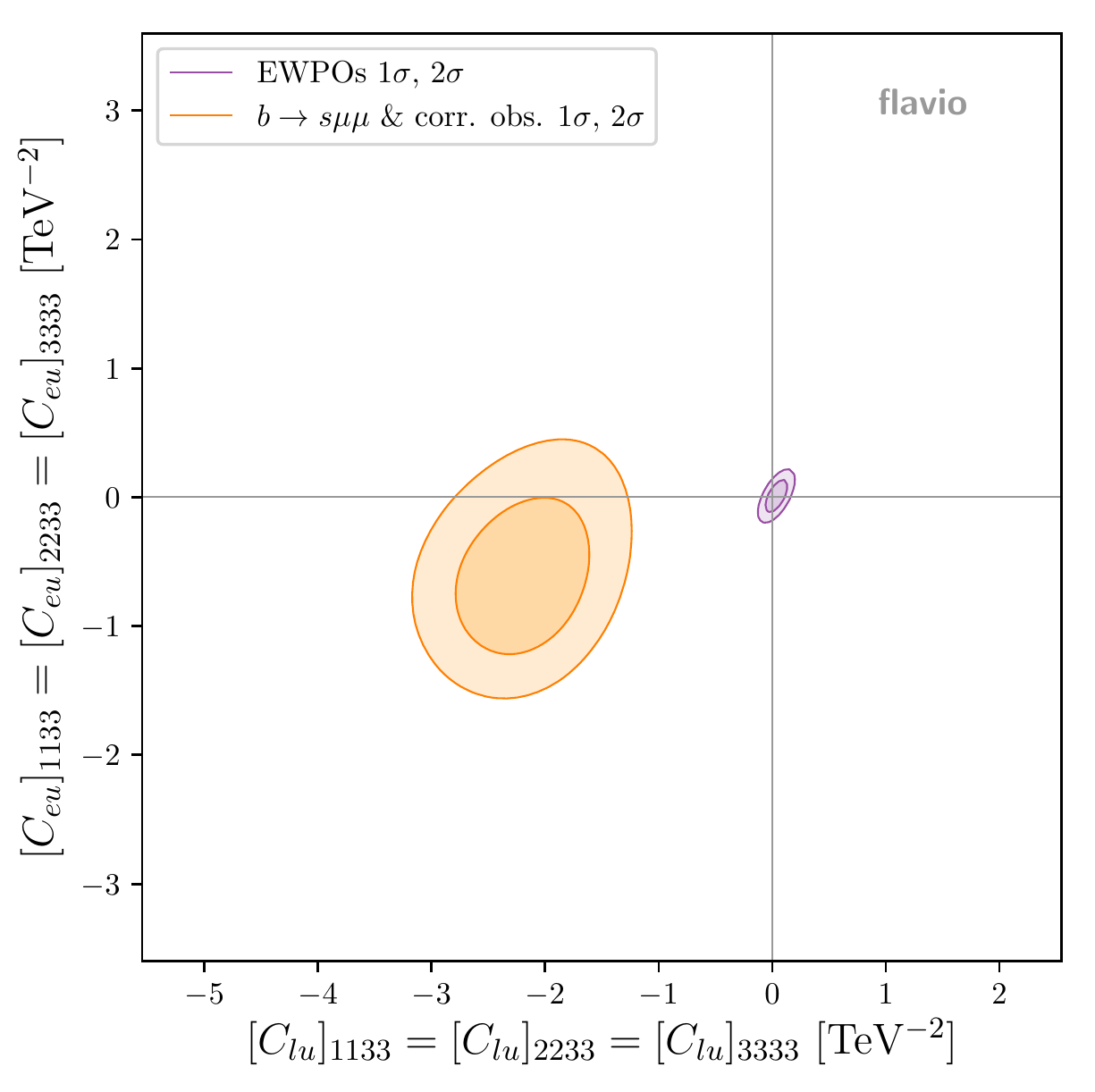}%
\caption{Likelihood contours from $R_{D^{(*)}}$, $b\to s\mu\mu$ observables and the global likelihood
  in the space of $[C_{lq}^{(1)}]_{3323}=[C_{lq}^{(3)}]_{3323}$ and $[C_{qe}]_{2333}$ (left) and
  in the space of $[C_{lu}]_{1133}=[C_{lu}]_{2233}=[C_{lu}]_{3333}$ and $[C_{eu}]_{1133}=[C_{eu}]_{2233}=[C_{eu}]_{3333}$ (right) at 2~TeV.
  All other Wilson coefficients are assumed to vanish at 2~TeV.}
\label{fig:qe2333andLFU}
\end{figure}

\begin{figure}[p]
\centering
\includegraphics[width=0.5\textwidth]{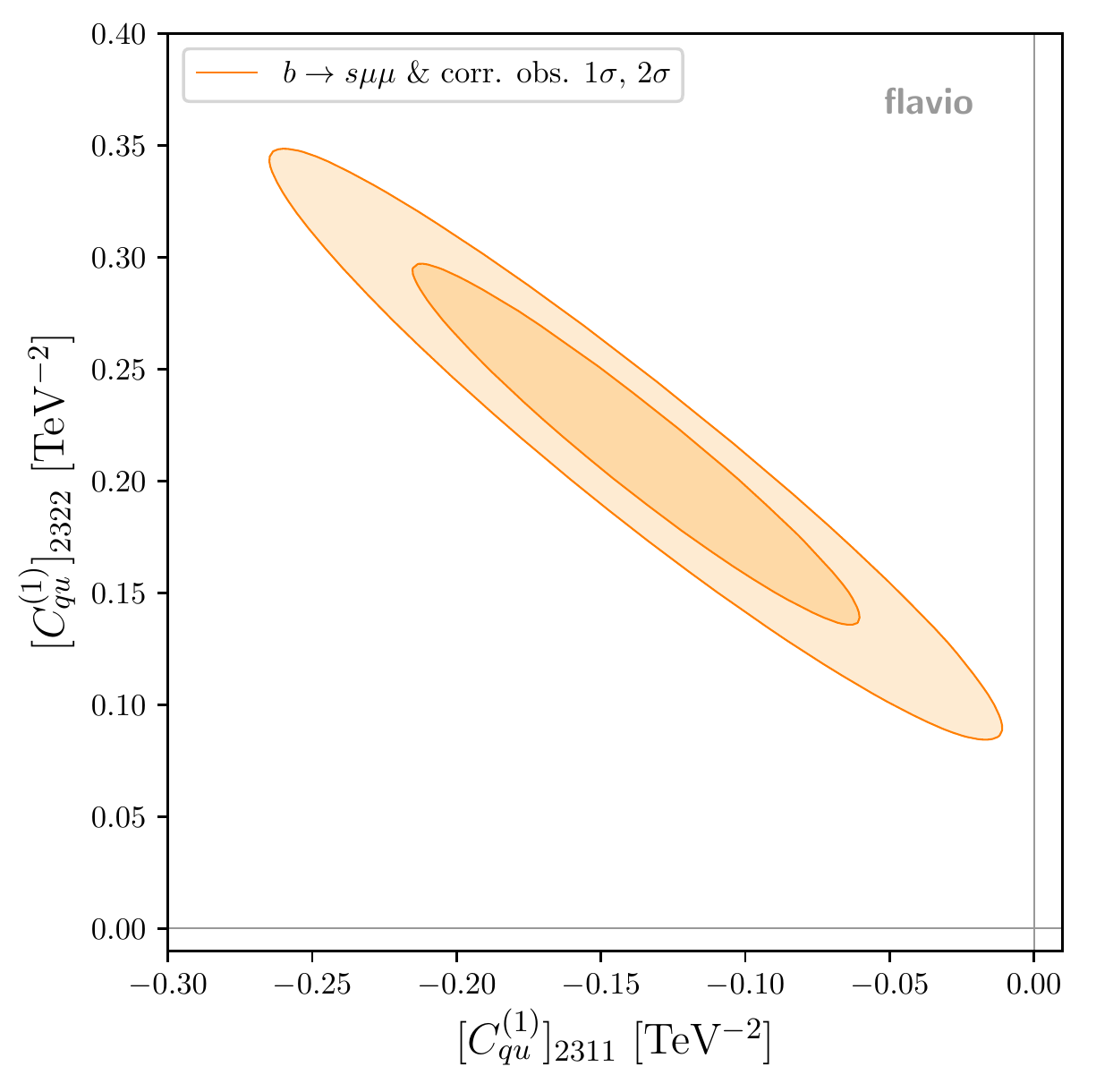}%
\caption{Likelihood contours from $b\to s\mu\mu$ observables
  in the space of $[C_{qu}^{(1)}]_{2311}$ and $[C_{qu}^{(1)}]_{2322}$ at 2~TeV.
  All other Wilson coefficients are assumed to vanish at 2~TeV.}
\label{fig:qu1_2311_qu1_2322}
\end{figure}

\begin{itemize}

\item The plots of Fig.~\ref{fig:axial} complement the ones in Fig.~\ref{fig:C9C10} and show Wilson coefficient spaces involving muonic axial-vector currents $C_{10}^{bs\mu\mu}$ and $C'{}_{10}^{bs\mu\mu}$.

\item Fig.~\ref{fig:bsmumu_C9e_C10e} shows the space of electronic Wilson coefficients $C_{9}^{bsee}$ and $C_{10}^{bsee}$. $R_K$ and $R_{K^*}$ can be explained simultaneously if both $C_{9}^{bsee}$ and $C_{10}^{bsee}$ are present.

\item The plots in Figs.~\ref{fig:bsmumu_C7_C9},~\ref{fig:bsmumu_C7_C10}, and~\ref{fig:bsmumu_C7_C9-C10} explore the impact of non-standard effects in the dipole coefficients $C_{7}^{bs}$ and $C'{}_{7}^{bs}$ that are switched on in addition to new physics in $C_{9}^{bs\mu\mu}$ (Fig.~\ref{fig:bsmumu_C7_C9}), $C_{10}^{bs\mu\mu}$ (Fig.~\ref{fig:bsmumu_C7_C10}), and $C_{9}^{bs\mu\mu}=-C_{10}^{bs\mu\mu}$ (Fig.~\ref{fig:bsmumu_C7_C9-C10}). The fits prefer $C_{7}^{bs}$ to be SM-like but show a slight ($\sim 1\sigma$) preference  for a positive shift in $C'{}_{7}^{bs}$.

\item In Fig.~\ref{fig:bsmumu_CS_C10} we consider the effects of non-zero scalar and pseudo-scalar Wilson coefficients that obey the SMEFT relations $C_{S}^{bs\mu\mu}=-C_{P}^{bs\mu\mu}$ (left) and $C_{S}^{\prime bs\mu\mu}=C_{P}^{\prime bs\mu\mu}$ (right). The most important constraint on these Wilson coefficients arises from the $B_s \to \mu^+\mu^-$ branching ratio. Mirror solutions in Wilson coefficient space that correspond to a mass eigenstate rate asymmetry $A_{\Delta \Gamma} \simeq -1 = - A_{\Delta \Gamma}^\text{SM}$ are allowed by present data.

\item The plots in Fig.~\ref{fig:qe2333andLFU} show scenarios with various combinatios of SMEFT Wilson coefficients. The left plot in Fig.~\ref{fig:qe2333andLFU} contains only tauonic Wilson coefficients, while the right plot shows a scenario with lepton flavour universal coefficients. The latter case is strongly constrained by electro-weak precision observables.

\item Finally, Fig.~\ref{fig:qu1_2311_qu1_2322} shows that the $b\to s \mu\mu$ anomalies can be explained by SMEFT 4-quark operators that enter the rare semi-leptonic decays at the loop level.

\end{itemize}


\FloatBarrier

\bibliographystyle{JHEP}
\bibliography{bibliography}

\end{document}